\documentclass[letterpaper,11pt,fleqn]{article}
\usepackage{jheppub}

\usepackage{graphicx}
\usepackage[figuresright]{rotating}
\usepackage{bm,amsmath,amssymb}
\usepackage[mathscr]{eucal}
\usepackage{color}
\usepackage{subcaption}
\usepackage[normalem]{ulem}
\usepackage{setspace}
\usepackage{multirow}

\long\def\comment#1{ }
\newcommand{\eqn}[1]{Eq.~\eqref{#1}}
\newcommand{\beq}{\begin{eqnarray}}
  \newcommand{\eeq}{\end{eqnarray}}

\newcommand{\dif}{{\rm d}}
\newcommand{\rmd}{{\rm d}}
\newcommand{\rme}{{\rm e}}
\newcommand{\rmi}{{\rm i}}

\newcommand{\rmJ}{{\rm J}}
\newcommand{\rmK}{{\rm K}}

\newcommand{\rmA}{{\rm A}}

\newcommand{\rmL}{{\rm L}}
\newcommand{\rmT}{{\rm T}}
\newcommand{\del}{\partial}

\newcommand{\mcal}{\mathcal}

\newcommand{\bl}{\bm{\ell}}
\newcommand{\bk}{\bm{k}}

\newcommand{\bq}{\bm{q}}
\newcommand{\bp}{\bm{p}}
\newcommand{\bx}{\bm{x}}
\newcommand{\by}{\bm{y}}

\newcommand{\bz}{\bm{z}}

\newcommand{\br}{\bm{r}}

\newcommand{\abar}{\bar{\alpha}_s}

\newcommand{\xbj}{x}

\newcommand{\minus}{\!-\!}

\title{\Large Saturation effects in SIDIS at very forward rapidities}

\author[a]{E.~Iancu,}

\author[b]{A.H.~Mueller,}

\author[c]{D.N.~Triantafyllopoulos,}

\author[c]{and S.Y.~Wei\,}

\affiliation[a]{Institut de physique th\'{e}orique, Universit\'{e} Paris Saclay, CNRS, CEA, F-91191 Gif-sur-Yvette, France}

\affiliation[b]{Department of Physics, Columbia University, New York, NY 10027, USA}

\affiliation[c]{European Centre for Theoretical Studies in Nuclear Physics and Related Areas (ECT*)\\and Fondazione Bruno Kessler, Strada delle Tabarelle 286, I-38123 Villazzano (TN), Italy}

\emailAdd{edmond.iancu@ipht.fr}
\emailAdd{ahm4@columbia.edu}
\emailAdd{trianta@ectstar.eu}
\emailAdd{swei@ectstar.eu}

\abstract{Using the dipole picture for electron-nucleus deep inelastic scattering at small
Bjorken $x$, we study the effects of gluon saturation in the nuclear target 
on the cross-section for SIDIS (single inclusive hadron, or jet, production). 
We argue that the sensitivity of this process to gluon saturation can be enhanced
by tagging on a hadron (or jet) which carries a large fraction $z \simeq 1$ of the longitudinal momentum
of the virtual photon. This opens the possibility to study gluon saturation in relatively hard processes, 
where the virtuality $Q^2$ is (much) larger than the target saturation momentum $Q_s^2$,
but such that $z(1-z)Q^2\lesssim Q_s^2$. Working in the limit $z(1-z)Q^2\ll Q_s^2$, we predict
new phenomena which would signal saturation in the SIDIS cross-section.  For sufficiently low 
transverse momenta $k_\perp\ll Q_s$ of the produced particle, 
the dominant contribution comes from elastic scattering in the black disk limit, which 
exposes the unintegrated quark distribution in the virtual photon. 
For larger momenta $k_\perp\gtrsim Q_s$, inelastic collisions take the leading role. They explore
gluon saturation via multiple scattering, leading to a Gaussian distribution in $k_\perp$ centred
around $Q_s$.
When $z(1-z)Q^2\ll Q^2$,
this results in a Cronin peak in the nuclear modification factor (the $R_{pA}$ ratio) 
at moderate values of $x$. With decreasing $x$, this peak is
washed out by the high-energy  evolution and replaced by nuclear suppression
($R_{pA}<1$) up to large momenta $k_\perp\gg Q_s$.
Still for  $z(1-z)Q^2\ll Q_s^2$, we also compute SIDIS cross-sections integrated over 
$k_\perp$. We find that both elastic and inelastic scattering are controlled by the black
disk limit, so they yield similar contributions, of zeroth order in the QCD coupling.
}
  
  \keywords{Perturbative QCD, Deep Inelastic Scattering, Gluon Saturation}

\begin{document}
\maketitle

\section{Introduction}
\label{sect:intro}

One of the main objectives of the future experimental program at the Electron-Ion
Collider (EIC) \cite{Accardi:2012qut,Aschenauer:2017jsk}
 is an in-depth study of the regime of high parton densities in the wavefunction of
a large nucleus accelerated to ultrarelativistic energies. The parton densities, and notably the
gluon distribution, are amplified both by the quantum evolution with increasing energy 
and by the presence of a large number $A\gg 1$ of nucleons in the structure of the 
nucleus (with $A\simeq 200$ for lead or gold). As a result of this evolution, 
the gluon distribution, as measured (at least, indirectly) in deep inelastic scattering (DIS),
rises rapidly with the energy --- roughly like an inverse power of the Bjorken variable $\xbj$,
which is small at high energy:  $\xbj\simeq Q^2/s\ll 1$ when $s\gg Q^2$. 
Here, $Q^2$ is the  virtuality of the space-like photon $\gamma^*$ 
exchanged between the electron and one nucleon from the nuclear target and $s$ 
is the center-of-mass energy squared for the $\gamma^*$--nucleon scattering.

One expects this rise to be eventually tamed by {\it gluon saturation}, i.e. by non-linear
effects associated with gluon self-interactions, which limit their occupation numbers
to values of order $1/\alpha_s$. The saturation effects should become manifest in the DIS
structure functions at sufficiently small values of $\xbj$ and for relatively low virtualities
$Q^2\lesssim Q_s^2(A,\xbj)$. The saturation momentum $Q_s(A,\xbj)$ is the 
typical transverse momentum of the saturated gluons. It increases with the energy, 
$Q_s^2\propto \xbj^{-\lambda_s}$ with $\lambda_s\simeq 0.2$, 
and also with the nucleon number $A$ ($Q_s^2\propto A^{1/3}$ when $A\gg 1$).
Yet, for the values of $\xbj$ that should be accessible at the EIC and for $A\simeq
200$, this scale $Q_s^2(A,\xbj)$ is still quite  low, in the ballpark of 1 to 2~GeV$^2$. Because of that,
the studies of gluon saturation in inclusive DIS at $Q^2\lesssim Q_s^2$ can be hindered 
by non-perturbative contamination: the perturbative quark-antiquark ($q\bar q$)
fluctuation of the virtual photon that one would like to use as a probe of the saturated gluon
distribution can mix with non-perturbative, hadronic (vector mesons), fluctuations. When
$Q^2\simeq 1\div 2$~GeV$^2$, the respective contributions to the inclusive DIS cross-section
may be difficult to separate from each other.

One common strategy to overcome such a difficulty is to consider less inclusive observables,
like multi-particle production, which involve several scales, some of which can be ``semi-hard''
(i.e. of order $Q_s$, or smaller) even for relatively large virtualities $Q^2\gg Q_s^2$
\cite{Gelis:2002nn,Marquet:2009ca,Dominguez:2011wm,Mantysaari:2019hkq}.
The ``golden probe'' which attracted most attention over the last years is the
production of a pair of hadrons (or jets)  in ``dilute-dense'' collisions ($eA$ or $pA$) at forward
rapidities (in the fragmentation region of the dilute projectile)
\cite{Marquet:2007vb,Albacete:2010pg,Dominguez:2011wm,Stasto:2011ru,Metz:2011wb,Lappi:2012nh,Iancu:2013dta,Kotko:2015ura,Dumitru:2015gaa,Altinoluk:2015dpi,Hatta:2016dxp,Marquet:2016cgx,vanHameren:2016ftb,Albacete:2018ruq,Dumitru:2018kuw,Salazar:2019ncp,Mantysaari:2019hkq,Kolbe:2020tlq,Mantysaari:2020lhf}. 
Even when the final particles/jets have relatively large transverse momenta ($k_{\perp}^2\sim Q^2\gg Q_s^2$), 
the physics of saturation is still visible in the
broadening of the back-to-back peak in their azimuthal angle distribution.
That said, this effect is not exempt from ambiguities either: it can hardly be distinguished 
from the broadening due to radiation in the final state (the ``Sudakov effect'') \cite{Mueller:2013wwa}.  
In order to reduce the latter, one must again
limit oneself to semi-hard values for $Q^2$ \cite{Zheng:2014vka}.


In this paper, we make the observation that saturation effects in relatively hard $eA$ collisions with 
$Q^2\gg Q_s^2$ could be observed at the EIC in an even simpler process, the semi-inclusive
production of a {\it single} hadron (or jet), a.k.a. SIDIS,
provided the measured particle/jet is {\it very forward} --- meaning that it
carries a large fraction $z\simeq 1$ of the longitudinal momentum of the virtual photon. 
The precise condition is that
$Q^2(1-z)\lesssim Q_s^2$, which for large virtualities  $Q^2\gg Q_s^2$ 
forces $z$ to be very close to one: $1-z\ll 1$. 
When this happens, saturation effects become important because the 
 $q\bar q$ fluctuation which mediates the interactions between the virtual 
photon and the nuclear target has a large transverse size, $r\gtrsim 1/Q_s$, hence it typically
scatters off saturated gluons with transverse momenta $p_\perp \sim Q_s$.

As implicit in the above discussion, throughout this paper we use the  ``colour-dipole'' picture for DIS
at small Bjorken $\xbj\ll 1$. This is the natural description in any Lorentz frame  --- such as
the rest frame of the hadronic target, or the center-of-mass frame for the $\gamma^*$--nucleon collision
 --- in which the virtual photon has a large longitudinal momentum, but zero 
 transverse momentum ($q_\perp=0$). 
In  such a frame and to leading order in pQCD 
(but including the high-energy evolution and the non-linear effects associated with gluon saturation), the
DIS process can be described as the interaction between a long-lived $q\bar q$ fluctuation of the virtual 
photon (a ``colour dipole'') and the small-$x$ gluons from the target.
The transverse size of this dipole\footnote{This is more precisely the dipole size
in the absence of any scattering; the dipole could be localised on a smaller scale by a
hard scattering with transferred momentum $k_\perp\gg \bar Q$.}
can be estimated as $r\sim 1/\bar Q$, where $\bar Q^2\equiv
z(1-z)Q^2$, with $z$ and $1-z$ the longitudinal fractions of the two quarks.

So long as $\bar Q^2\gg Q_s^2$, we are in the standard ``leading-twist'' regime, where the 
dipole-hadron scattering is weak and essentially ``counts'' the number of gluons in the target:
the dipole cross-section is proportional to the gluon distribution.
However, if $z(1-z)$ is sufficiently small, one can have $\bar Q^2\lesssim Q_s^2$ even
at high virtualities $Q^2\gg  Q_s^2$, and then the scattering is strong ---
that is, multiple scattering becomes  important and the dipole amplitude approaches the unitarity 
(or ``black disk'') limit, which is the dual description of gluon saturation in a dipole frame.
The impact of multiple scattering/gluon saturation on the final state also depends upon other
factors, like the transverse momentum $k_\perp$ of the produced hadron, the polarisation of
the virtual photon, and the type of process under consideration --- elastic\footnote{By ``elastic'',
we mean processes where also the hadronic target scatters coherently, that is,
where the CGC average is performed as the level of the amplitude. This is in line with the large-$N_c$
approximation that we shall generally use.} or inelastic.

In our subsequent analysis, we shall identify several physical regimes where saturation effects are expected
to be important for SIDIS and will provide leading order calculations for the relevant observables. 
Our general framework is the Colour Glass Condensate (CGC)
effective theory \cite{Iancu:2003xm,Gelis:2010nm,Kovchegov:2012mbw}, 
where the ``leading order approximation''
already includes all-order resummations of multiple scattering (via Wilson lines) and of soft gluon 
emissions (via the Balitsky--JIMWLK evolution equations \cite{Balitsky:1995ub,JalilianMarian:1997jx,JalilianMarian:1997gr,Kovner:2000pt,Iancu:2000hn,Iancu:2001ad,Ferreiro:2001qy}, 
or their BK truncation at large $N_c$  \cite{Balitsky:1995ub,Kovchegov:1999yj}).
The leading-order predictions of this theory for both single particle (quark) and two particle
(quark-antiquark) production\footnote{In applications to phenomenology, such partonic 
cross-sections should be convoluted with parton-to-hadron fragmentation functions.}
are well known \cite{Mueller:1999wm,Marquet:2009ca,Dominguez:2011wm} 
and will represent the starting point of our analysis.
Up to trivial factors, these cross-sections depend upon $Q^2$ and $z$ only via the {\it effective} 
virtuality $\bar Q^2\equiv z(1-z)Q^2$. We shall immediately specialise to the
interesting regime at $\bar Q^2\lesssim Q_s^2$, that we shall study via both analytic and
numerical calculations. As customary in the CGC framework, we shall describe the dipole-nucleus 
scattering by using the McLerran-Venugopalan (MV)  model \cite{McLerran:1993ni,McLerran:1994vd}
at low energies and the BK equation for the evolution with increasing energy.

In our analytic studies, we shall mostly focus on the strongly ordered case $\bar Q^2 \ll Q_s^2$;
besides allowing for controlled approximations, this also has the merit to be
conceptually clear: the interesting physical regimes are well separated and the physical interpretation
of the results is transparent. In this regime, the dipole-hadron scattering is typically strong 
and implies a transverse momentum transfer of order $Q_s$.
Accordingly, the $k_\perp$-distribution of the produced quark
is found to be a Gaussian with dispersion
$\langle k_\perp^2\rangle =Q_s^2$, which describes transverse momentum broadening via multiple scattering.
This Gaussian is surrounded by power-low distributions at both lower momenta and higher momenta.
At lower momenta $ k_\perp^2 \ll Q_s^2$, we find a $1/k_\perp^2$--distribution which is in fact
the distribution generated by the decay $\gamma^*_\rmT\to q\bar q$ of the transverse virtual photon
and which is preserved by the elastic scattering in the black disk limit. At larger momenta
$ k_\perp^2 \gg Q_s^2$, we find a power tail $\propto 1/k_\perp^4$ (up to logarithms), which is
produced via a single hard scattering with one of  the valence quarks in the nucleus.

A particularly suggestive way to exhibit the effects of saturation is by constructing the $R_{pA}$ ratio
for SIDIS, that is, the ratio of the SIDIS cross-sections in $eA$ and respectively $ep$ collisions,
with the latter scaled up by the number of participants. We find that this ratio behaves very 
similarly to the corresponding ratio  in d+Au collisions, as measured at
RHIC \cite{Arsene:2004ux,Adams:2006uz} and theoretically studied in several publications
\cite{Kharzeev:2002pc,Baier:2003hr,
Albacete:2003iq,Kharzeev:2003wz,Iancu:2004bx,Blaizot:2004wu}: that is, it shows nuclear suppression
at low momenta, $k_\perp <  Q_s$, and nuclear enhancement at larger momenta, $k_\perp \gtrsim  Q_s$, 
with the possible appearance of a (wide) Cronin peak near $k_\perp =  Q_s$.
This similarity is in fact natural: in both problems, the cross-section for
particle production is driven by the ``dipole'' unintegrated gluon distribution in the hadronic
target (a proton or a nucleus).

The results above mentioned strictly hold in the MV model, i.e. at relatively low energies (say, $\xbj\sim 10^{-2}$).
We will numerically study their high-energy evolution using the collinearly improved version of
the BK equation with running coupling recently proposed in \cite{Ducloue:2019ezk,Ducloue:2019jmy}
(see also the related works \cite{Beuf:2014uia,Iancu:2015vea,Iancu:2015joa,Lappi:2015fma,Lappi:2016fmu,Albacete:2015xza,Beuf:2020dxl}). 
We shall thus find that the effects of the evolution are quite mild up to moderately large 
rapidities (as relevant for the phenomenology at the EIC). The only qualitative changes are the emergence
of an ``anomalous'' dimension in the power-law distribution at high $k_\perp$  
and, especially,
a rapid suppression of the Cronin peak, which disappears after just one unit in rapidity --- once again, very similarly to what happens in
d+Au collisions at RHIC \cite{Arsene:2004ux,Adams:2006uz,Albacete:2003iq,Kharzeev:2003wz,Iancu:2004bx}.

Since the identification of very forward jets, or hadrons, with semi-hard $k_\perp\sim Q_s$
may be experimentally challenging, we also provide estimates for the SIDIS cross-sections integrated
over the transverse momentum $k_\perp$ --- that is, the cross-sections for producing a ``quark''
(physically, a hadron, or jet) with a given longitudinal fraction $z$, but any transverse momentum.
Once again, the most interesting regime for studies of saturation lies at $\bar Q^2\ll Q_s^2$. In this regime,
we find that the elastic and inelastic cross-sections are both dominated by the black disk limit and
hence they give identical contributions (at least, at leading order), as expected from the optical theorem.
This is especially interesting since the respective contributions come from integrating $k_\perp$--dependent
physics which is quite different in the two cases: the $\gamma^*_\rmT$--wavefunction in the case of the elastic
scattering and, respectively, the $k_\perp$--broadening of the measured quark for the inelastic one.
We also find it interesting to study the weak scattering regime at large effective virtuality $\bar Q^2\gg
Q_s^2$, where saturation is still visible, albeit rather indirectly, via the phenomenon of {\it geometric
scaling}  \cite{Stasto:2000er,Iancu:2002tr,Mueller:2002zm}. 

A striking property of the integrated cross-section is its rapid increase with $z$ when
 $z$  is large enough (say, $z\ge 0.9$). Such a rise should be easily
seen experimentally, although it could be more difficult to distinguish between the {\it power-law} increase 
expected in the dilute regime at $\bar Q^2\gg Q_s^2$ and, respectively, the {\it logarithmic} increase
in the black-disk regime at  $\bar Q^2\lesssim Q_s^2$. Once again, this distinction can become
sharper by constructing appropriate $R_{pA}$ ratios, that we shall introduce and study
in Sect.~\ref{sect:qllqs}.

Let us conclude this discussion with some remarks on the experimental
challenges to be faced when searching for saturation effects in SIDIS at forward rapidities.
First, these effects are associated with relatively rare events:  
the very asymmetric ($z\simeq 1\gg 1-z$) dipole
configurations of interest for us here represent only a small
fraction of the total DIS cross-section when $Q^2\gg Q_s^2$. 
Of course, asymmetric configurations {\it per se} are {\it not} that rare:
the {\it aligned jet} configurations, which dominate the DIS structure 
functions at small $x$ and very large $Q^2$, are asymmetric too
(when viewed in the dipole frame, they correspond to $q\bar q$ pairs with $z(1-z)\ll 1$).
Yet, the configurations that we consider in this paper are even {\it more} asymmetric:
they must obey $z(1-z)\lesssim Q_s^2/Q^2\ll 1$ (in order to be sensitive to gluon saturation), 
unlike the traditional aligned-jet configurations, for which $Q_s^2/Q^2\ll z(1-z) \ll 1$
(corresponding to weak scattering and dilute parton distributions; see the discussion
in Sect.~\ref{sect:sym-al}).

But albeit very asymmetric and rare when $Q^2\gg Q_s^2$, the configurations relevant for saturation
become more and more symmetric, and also typical, when decreasing $Q^2$ towards $Q_s^2$. 
So, in practice, one can use SIDIS to study the saturation physics by simultaneously 
varying $Q^2$ and $z$, provided the constraint $z(1-z)Q^2\lesssim Q_s^2$ remains satisfied. 
We believe that this leaves a sizeable phase-space to be explored by the experiments.

Another tricky issue, that we shall not explicitly address in this work, is the final-state evolution
of the leading parton and, related to that, the possibility to reconstruct its $z$ fraction from
the measured final state (a hadron or a jet). Clearly, this fraction can only be degraded
by the final state radiation. In order to keep a good control on it,
we foresee two possible strategies, depending upon the type of measurement.
If one measures a hadron in the final state, then one should focus on hadrons 
with a large longitudinal momentum fraction $z_h\simeq 1$ w.r.t. the virtual photon. Since
the longitudinal fraction of the leading parton is even larger, $z_h< z< 1$, we would already
have some control on the physical regime at work.
To have a more precise estimate for $z$, the data should be unfolded with the parton-to-hadron
fragmentation functions near $z=1$. Measuring jets instead of hadrons may look like a better
option, since a jet captures the quasi-totality of the longitudinal momentum of the leading
parton: $z_{\rm jet}$ is a good proxy for the $z$ of the quark. But this rises the question
about the possibility to reconstruct jets with semi-hard (total) transverse momenta $k_\perp
\sim Q_s$ and at very forward rapidities. 

The paper is organised as follows. Sect.~\ref{sect:dijet} presents a brief summary of known results
for particle production in DIS in the dipole picture and in the CGC formalism at leading order.
We start with the dijet cross-section, for which the physical and diagrammatic interpretations are
more transparent, and deduce the SIDIS cross-section by ``integrating out'' the kinematics of the 
unmeasured quark.  Sects.~\ref{sect:qllqs} and \ref{sect:intk} contain our main new results.
In Sect.~\ref{sect:qllqs} we consider the double-differential (in $z$ and $k_\perp$) SIDIS
cross-sections in the saturation regime at $\bar Q^2=z(1-z)Q^2< Q_s^2$. We present both
analytic results (valid in the strictly-ordered limit $\bar Q^2\ll Q_s^2$ and for 
different ranges in $k_\perp$) and numerical results, which include the effects of the 
high-energy evolution. In Sect.~\ref{sect:intk}, we discuss the SIDIS cross-section differential in $z$
but integrated over $k_\perp$. Once again, we combine analytic approximations (in the limiting
cases $\bar Q^2\ll Q_s^2$ and $\bar Q^2\gg Q_s^2$) and numerical results (for the whole kinematics).
In Sect.~\ref{sect:sym-al} we discuss the inclusive DIS cross-section (as obtained by integrating
the SIDIS cross-section over both $z$ and $k_\perp$), with the purpose to expose the signatures 
of saturation and to clarify the relative importance of the large-$z$ SIDIS configurations at high 
$Q^2\gg Q_s^2$. We notably explain the difference between these configurations and the
better known aligned jets. Sect.~\ref{sec:conc} contains our summary and conclusions.
We have relegated some of the calculations to five appendices, out of which one is devoted
to a brief review of the collinearly-improved BK equation.




\section{DIS in the dipole picture: from dijets to SIDIS}
\label{sect:dijet}

As mentioned in the Introduction, both the physical picture and the mathematical description of DIS 
at small Bjorken $x$ are greatly simplified if one describes this process in a {\it dipole frame}, 
by which one means a Lorentz frame in which the virtual photon ($\gamma^*$) has a large longitudinal 
momentum, $q_z^2\gg Q^2$, and zero transverse momentum, $q_\perp=0$. 
This choice is not unique: two such frames can be related
to each other by a boost  along the longitudinal direction.
As a matter of facts, most of the subsequent developments 
in this paper are boost-invariant, so they remain valid in any dipole frame\footnote{In 
applications to the phenomenology, these results must of course be translated to the
laboratory frame at the EIC.}. 
That said, both for the purposes of the physical discussion and for the description of
the high energy evolution\footnote{At leading order, the predictions of the BK equation are 
boost-invariant, but this symmetry is spoilt by the resummation of the higher-order
corrections amplified by large collinear logarithms. The collinear resummation  that
we shall employ here is adapted to the evolution of the target
 \cite{Ducloue:2019ezk,Ducloue:2019jmy} (see also Appendix~\ref{sec:BK} for a summary).}, 
 it is convenient to work in a special
frame where both $\gamma^*$ and the target are ultrarelativistic,
but such that most of the total energy is still carried by the target. In this frame, the 
high-energy evolution is fully encoded in the gluon distribution of the target, 
whereas the virtual photon has just enough energy to fluctuate
into a quark-antiquark ($q\bar q$) pair  --- a colour dipole ---, which then scatters off the target.
As a consequence of this scattering, the quark and the antiquark from the dipole can
lose their initial coherence and evolve into independent jets, or hadrons, in the final state.
%
%

Specifically, we chose the virtual photon to be the right mover, with light-cone (LC) 4-momentum
$q^\mu\equiv (q^+, q^-, \bq_\perp)= (q^+, -\frac{Q^2}{2q^+},
\bm{0}_\perp)$, whereas the nucleus is a left-mover with\footnote{We neglect the
proton mass $M$ which is much smaller than all the other scales in the
problem, $M^2\ll Q^2\ll 2P\cdot q$.} 4-momentum $P^\mu=\delta^{\mu-}P^-$ per nucleon.
In the high-energy or small Bjorken $x$ regime, with
\beq\label{xBj}
\xbj\equiv\,\frac{Q^2}{2P\cdot q}=\,\frac{Q^2}{2P^- q^+}\,\ll\,1\,,
\eeq
the coherence time $\Delta x^+\simeq 2q^+/Q^2$ of the virtual photon,
i.e.\ the  typical lifetime of its $q\bar q$
fluctuation, is much larger than the longitudinal extent $\sim A^{1/3}/P^-$ of
the nuclear target. This justifies the use of the eikonal approximation when computing the
dipole-target scattering.
%

\subsection{Dijet production in DIS at small $x$}

To leading order in perturbative QCD, the most detailed information about the final state
which is still inclusive with respect to the nucleus is contained in the cross-section for producing 
an on-shell $q\bar q$ pair. This depends upon the polarisation (transverse or longitudinal) state
of the virtual photon, with the following  results \cite{Gelis:2002nn,Dominguez:2011wm,Mantysaari:2019hkq}:
\begin{align}
\label{2j-sigmaL}
\frac{\dif\sigma^{\gamma^*_{\rmL} A \to q\bar q X}}{ \dif^3k_1\, \dif^3k_2} = &\, 8(2\pi)^2
 \frac{N_c \alpha_{\rm em} e_q^2}{q^+}\,\delta(q^+-k_1^+-k_2^+)\,
 z (1-z)
\bar{Q}^2
	\int \frac{\dif^2 \bx_1}{(2\pi)^2}\frac{\dif^2 \bx_1^\prime}{(2\pi)^2}
	\frac{\dif^2 \bx_2}{(2\pi)^2}\frac{\dif^2 \bx_2^\prime}{(2\pi)^2}
	 \nonumber\\*[0.2cm]
     &\times\rme^{-\rmi \bk_1 \cdot (\bx_1 - \bx_1^\prime)} \,
	\rme^{-\rmi \bk_2 \cdot (\bx_2 - \bx_2^\prime)}\,	\rmK_0(\bar{Q}r) \rmK_0(\bar{Q}r')\,
	\mathcal{O}^{(4)}_{x_g}(\bx_1, \bx_2, \bx_1^\prime, \bx_2^\prime),
\end{align}
and respectively:
\begin{align}
\label{2j-sigmaT}
\frac{\dif\sigma^{\gamma^*_{\rmT} A \to q\bar q X}}{ \dif^3k_1\, \dif^3k_2} = &\, 2(2\pi)^2
 \frac{N_c \alpha_{\rm em} e_q^2}{q^+}\,\delta(q^+-k_1^+-k_2^+)\,
 [z^2+ (1-z)^2]
\bar{Q}^2
	\int \frac{\dif^2 \bx_1}{(2\pi)^2}\frac{\dif^2 \bx_1^\prime}{(2\pi)^2}
	\frac{\dif^2 \bx_2}{(2\pi)^2}\frac{\dif^2 \bx_2^\prime}{(2\pi)^2}
	 \nonumber\\*[0.2cm]
	&\times	\rme^{-\rmi \bk_1 \cdot (\bx_1 - \bx_1^\prime)} \,
	\rme^{-\rmi \bk_2 \cdot (\bx_2 - \bx_2^\prime)}     \,
	\frac{\br \!\cdot \!\br'}{r\, r'}\,\rmK_1(\bar{Q}r) \rmK_1(\bar{Q}r')\,
     \mathcal{O}^{(4)}_{x_g}(\bx_1, \bx_2, \bx_1^\prime, \bx_1^\prime).
\end{align}
In writing these expressions, 
we used longitudinal momentum conservation, $q^+=k_1^++k_2^+$, to write
$k_1^+=zq^+$ and  $k_2^+=(1-z)q^+$, with $0< z<1$. 
Also, $\bar{Q}^2 \equiv z(1-z) Q^2$, while $\br\equiv\bx_1-\bx_2 $ 
and $\br'\equiv \bx_1^\prime - \bx_2^\prime$ are the transverse sizes of the dipole in the direct amplitude (DA) and in the complex-conjugate amplitude (CCA), respectively.  
The modified Bessel functions  $\rmK_0(\bar{Q}r) $ and $\rmK_1(\bar{Q}r) $
belong to the light-cone wavefunctions describing the $q\bar q$ fluctuation of the virtual photon
for the two possible polarisations. These functions exponentially vanish
when $\bar{Q}r\gtrsim 1$, showing that the transverse size of the dipole cannot be much larger than
$1/\bar Q$.
Via the uncertainty principle, this in turn implies that the quark and the antiquark are produced by the
decay of the photon with typical transverse momenta $k_{i\perp}^{\rm prod}\sim \bar Q$. Of course, these 
{\it production} momenta can be subsequently modified by the scattering,
so their {\it measured} values $k_{1\perp}$ and $k_{2\perp}$ can be very
different from $\bar Q$ (typically, larger than it).

The QCD scattering between the dipole and the gluons from the nuclear target
is encoded in the following $S$-matrix structure,
\beq\label{O4}
\mathcal{O}^{(4)}_{x_g}(\bx_1, \bx_2, \bx_1^\prime, \bx_2^\prime)
\equiv     1 + S^{(4)}_{x_g}(\bx_1, \bx_2 ; \bx_2^\prime, \bx_1^\prime)-
     S^{(2)}_{x_g}(\bx_1, \bx_2) - S^{(2)}_{x_g}(\bx_2^\prime, \bx_1^\prime),
     \eeq
where the two-point (dipole) and four-point (quadrupole) functions are defined as
\begin{equation}\label{dipole}
S^{(2)}_{x_g}(\bx_1, \bx_2)
\,\equiv\,\frac{1}{N_{c}}\left\langle \mathrm{tr} \,\big(
V(\bm{x}_1)V^{\dagger}(\bm{x}_2)\big)\right\rangle_{x_g},
\end{equation}
and, respectively,
\begin{equation}
\label{quadrupole}
S^{(4)}_{x_g}(\bm{x}_1,\bm{x}_2;\bx_2^\prime,\bx_1^\prime)\,\equiv\,\frac{1}{N_{c}}\,\left\langle
\mathrm{tr}\big(V(\bm{x}_1)V^{\dagger}(\bm{x}_2)V(\bx_2^\prime)
V^{\dagger}(\bx_1^\prime)\big)\right\rangle_{x_g}.
\end{equation}
Here, $V(\bm{x})$ and $V^{\dagger}(\bm{x})$
are Wilson lines in the fundamental representation, which describe multiple scattering for
the quark and the anti-quark, in the eikonal approximation; explicitly,
\beq\label{Wilson}
  V(\bm{x})\,=\,{\rm T}\exp\left\{ ig\int dx^{+}\, t^{a}A^{-}_a(x^{+},\, \bm{x})\right\}.
   \eeq
The colour field  $A^{-}_a$ is a random quantity whose correlations (computed
within the CGC effective theory) encode the information about the gluon distribution in the target.
Via the (non-linear) B-JIMWLK evolution, these correlations depend upon the value $x_g$ of the longitudinal momentum fraction of the gluons which participate in the collision. To the leading order approximation at hand, this value is determined by energy-momentum conservation
together with the condition that the produced ``jets'' (the $q\bar q$ pair) be on mass-shell: 
the total light-cone energy in the final state, that is, $k_1^-+k_2^-$ with $k_i^-=k^2_{i\perp}/2k_i^+$,
must be equal to the sum of the (negative) LC energy of the virtual photon, $q^-=-Q^2/2q^+$,
 and the (positive) LC energy $p^-=x_gP^-$ transferred by the gluons from the target, via the collision. This
 condition yields
 \beq\label{xg}
 x_g = \frac{1}{2P\cdot q}\left(\frac{k^2_{1\perp}}{z} + \frac{k^2_{2\perp}}{1-z} + Q^2\right).
 \eeq
Note that this effective value $x_g$ can be significantly larger than the Bjorken $x$ variable \eqref{xBj}
if the final transverse momenta $k_{i\perp}$ are considerably larger than $\bar Q$. For what follows,
it is useful to keep in mind that the {\it typical} transverse momentum that can be transferred
by the collision is of the order $Q_s\equiv Q_s(A, x_g)$ --- the nuclear saturation momentum at the longitudinal
scale $x_g$. So, we typically expect  $k_{i\perp}^2\sim {\rm max}\big(\bar Q^2, \,Q_s^2\big)$.
 The total transverse momentum ${\bm \Delta}
\equiv \bk_1 + \bk_2$ of the produced pair is generated via multiple scattering, hence its
magnitude is of order $Q_s$.


\begin{figure}
  \centerline{
    \includegraphics[width=0.85\textwidth]{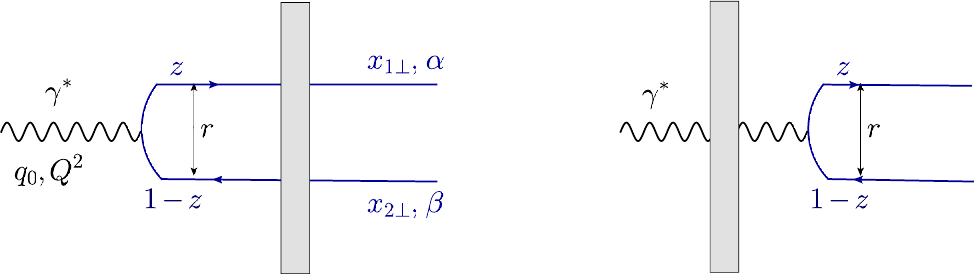}}
  \caption{\small Left: The virtual photon splits before crossing the  nuclear shockwave;
  the originally colourless $q \bar q$ pair acquires colour via the scattering off the gluons 
in the target. Right: The virtual photon splits
  after crossing the target, thus giving rise to a colourless  $q \bar q$ pair in the final state.}
  \label{fig:2jets-ampli}
\end{figure} 

Notice that the non-linear effects associated with the high gluon density in the target
enter the cross-sections \eqref{2j-sigmaL}--\eqref{2j-sigmaT} at two levels: on one hand, via
the multiple scattering of the dipole projectile, as encoded in the Wilson lines, cf. \eqn{Wilson};
on the other hand, via the non-linear effects in the B-JIMWLK evolution of the 
multi-partonic $S$-matrices which appear in Eqs.~\eqref{O4}--\eqref{quadrupole}.

For what follows, it is useful to have a more detailed microscopic understanding of this $S$-matrix 
structure. To that aim, let us move one step
backwards and study the amplitudes underlying the cross-sections \eqref{2j-sigmaL}--\eqref{2j-sigmaT}.
Within the time-ordered LC perturbation theory, there are two contributing amplitudes,
illustrated in Fig.~\ref{fig:2jets-ampli}: the virtual photon
can split into the  $q\bar q$ pair either long before, or long after, it crosses the nuclear ``shockwave''.
In both cases, the quark and the antiquark are originally produced by the decay $\gamma^*\to
q\bar q$ in the same colour state. For the process in Fig.~\ref{fig:2jets-ampli}.~left, this colour state
can be rotated by the scattering off the nucleus, separately for the quark and the antiquark;
hence, the 2 fermions will generally end up in different colours states, say $\alpha$ and 
$\beta$, with transition amplitudes measured by the matrix elements of the Wilson lines.
For the other process, cf. Fig.~\ref{fig:2jets-ampli}.~right,  one clearly has $\alpha=\beta$.
 This discussion motivates the following expression for the total amplitude \cite{Gelis:2002nn},
\beq\label{dijet-ampl}
\mathcal{A}_{\alpha\beta}(z, \bk_1, \bk_2; q^+, Q^2)
=
\int_{\bx_1,\bx_2}
\rme^{-\rmi(\bx_1\cdot\bk_1+\bx_2\cdot\bk_2)}\,\Psi(r, z; q^+, Q^2)
\, \big(V({\bx_1}) V^\dagger({\bx_2})-1\big)_{\alpha\beta},
\eeq
where $\Psi(r, z; q^+, Q^2)$ generically denotes the  $q\bar q$ component of the $\gamma^*$
wavefunction, for either transverse or longitudinal polarisations (see Eqs.~(19)--(20) in
 \cite{Dominguez:2011wm} for explicit expressions). The negative sign in front of the unit matrix in \eqn{dijet-ampl} can be viewed as a consequence of probability conservation. The total amplitude
vanishes in the absence of scattering, as expected: the space-like photon cannot decay into a
pair of {\it on-shell} fermions without additional interactions.
After taking the modulus squared of this amplitude and performing the CGC average
over the target field, one indeed generates the $S$-matrix  structure in Eq.~\eqref{O4}. 
This discussion also shows that the two dipole $S$-matrices which enter
 \eqn{O4} with negative signs are interference terms between the two processes shown
 in Fig.~\ref{fig:2jets-ampli}.
 
 It will be also interesting to consider dijet production via {\it elastic} (or ``coherent'') scattering. 
 The ``elastic'' version of   \eqn{dijet-ampl} is
 obtained by demanding the $q\bar q$ pair to remain a colour singlet  in the final state and by
 performing the CGC averaging already at the level of the amplitude (thus enforcing elastic
 scattering also for the target); this yields
  \beq\label{dijet-elas}
  \hspace*{-.4cm}
\mathcal{A}_{\rm el}(z, \bk_1, \bk_2; q^+, Q^2) =
\int_{\bx_1,\bx_2}
\rme^{-\rmi(\bx_1\cdot\bk_1+\bx_2\cdot\bk_2)}\,\Psi(r, z; q^+, Q^2)
\, \frac{1}{N_{c}}\left\langle \mathrm{tr} \,\big(
V(\bm{x}_1)V^{\dagger}(\bm{x}_2)-1\big)\right\rangle_{x_g}.
\eeq
Clearly the respective cross-sections take the general form in Eqs.~\eqref{2j-sigmaL}--\eqref{2j-sigmaT},
but with a new $S$-matrix structure: $\mathcal{O}^{(4)}_{x_g}\to \mathcal{O}^{\,\rm el}_{x_g}$, with\footnote{One can also introduce a generalisation of \eqn{Oel} in which the CGC average
is performed only at the level of the cross-section and which describes a part of the inelastic diffraction
\cite{Mantysaari:2019hkq}. However the difference w.r.t. our ``fully elastic'' cross-section vanishes
in the large-$N_c$ limit, which is the limit that we shall use in practice anyway.}:
\beq\label{Oel}
\mathcal{O}^{\,\rm el}_{x_g}(\bx_1, \bx_2, \bx_1^\prime, \bx_2^\prime)
\equiv     1 + S^{(2)}_{x_g}(\bx_1, \bx_2)  S^{(2)}_{x_g}(\bx_2^\prime, \bx_1^\prime)-
     S^{(2)}_{x_g}(\bx_1, \bx_2) - S^{(2)}_{x_g}(\bx_2^\prime, \bx_1^\prime).
     \eeq
Finally, the cross-section for inelastic dijet production is  the difference between
the total cross-section and its elastic component, so the associated $S$-matrix structure reads
\beq\label{Oinel}
\mathcal{O}^{\,\rm inel}_{x_g}=\mathcal{O}^{(4)}_{x_g}- \mathcal{O}^{\,\rm el}_{x_g}=
S^{(4)}_{x_g}(\bx_1, \bx_2 ; \bx_2^\prime, \bx_1^\prime)-
S^{(2)}_{x_g}(\bx_1, \bx_2)  S^{(2)}_{x_g}(\bx_2^\prime, \bx_1^\prime)\,.\eeq

\subsection{Single inclusive jet production in DIS at small $x$}
\label{sect:general}

Given the dijet results in the previous section, it is easy to deduce the corresponding
results for the quantity in which we are primarily interested in this work: the cross-section
for the inclusive production of a single ``jet'' (a quark or an antiquark), or SIDIS.
Assume that we measure the quark, for definiteness. Then we have to integrate out
the kinematics of the unmeasured antiquark in Eqs.~\eqref{2j-sigmaL}--\eqref{2j-sigmaT}.
The integral over  $k_2^+$ simply removes the $\delta$-functions for longitudinal 
momentum conservation. That over  $\bk_2$ identifies the transverse coordinates
of the antiquark in the DA and in the CCA: $\bx_2=\bx_2^\prime$. In turn, this introduces an
important simplification in the colour structure: the original quadrupole in
\eqn{quadrupole} reduces to an effective dipole, $S^{(2)}_{x_g}(\bx_1, \bx'_1)$, built with the
transverse coordinates of the measured quark  in the DA and in the CCA, respectively.
By also assuming that the target is homogeneous in the transverse plane (a disk with radius $R_A$), a further 2-dimensional integration can be performed to give a factor $\pi R^2_A$. Assembling everything together,
we find the following expressions for the longitudinal and transverse SIDIS cross sections:
\begin{align}
\label{sigmaL}
	\frac{\dif\sigma^{\gamma^*_{\rmL} A \to q X}}{\dif z\, \dif^2\bk} = 
	8 R^2_A N_c \alpha_{\rm em} e_q^2 z (1-z) \mcal{J}_\rmL,
\end{align}
\begin{align}
\label{sigmaT}
	\frac{\dif\sigma^{\gamma^*_{\rmT} A \to q X}}{\dif z\, \dif^2\bk} = 
	2 R^2_A N_c \alpha_{\rm em} e_q^2 \big[z^2 +(1-z)^2\big]
	\mcal{J}_\rmT.
\end{align}
We have introduced here the reduced cross-sections,
\begin{align}
\label{JL}
	\mcal{J}_\rmL(\bk,\bar Q) \equiv
	\pi \bar{Q}^2
	\int \frac{\dif^2 \br}{(2\pi)^2}\frac{\dif^2 \br'}{(2\pi)^2}\,
	\rme^{-\rmi \bk \cdot (\br - \br')} 
	\rmK_0(\bar{Q}r) \rmK_0(\bar{Q}r')
	[T(\br) + T(-\br') - T(\br \!-\! \br')] ,
\end{align}
\begin{align}
\label{JT}
	\mcal{J}_\rmT(\bk,\bar Q)\equiv
	\pi \bar{Q}^2
	\int \frac{\dif^2 \br}{(2\pi)^2}\frac{\dif^2 \br'}{(2\pi)^2}
	\rme^{-\rmi \bk \cdot (\br - \br')}
	\frac{\br \!\cdot \!\br'}{r\, r'}\, \rmK_1(\bar{Q}r) \rmK_1(\bar{Q}r') 
	[T(\br) + T(-\br') - T(\br \!-\! \br')],
\end{align}
which encode all the non-trivial dynamics. As suggested by their notations, $\mcal{J}_\rmL$
and $\mcal{J}_\rmT$ depend upon the virtuality of the
initial photon $Q^2$ and upon the splitting fraction $z$ for the $\gamma^*\to q\bar q$
decay only via the product $\bar Q^2=z(1-z)Q^2$ (the ``effective virtuality''). 

In writing Eqs.~\eqref{JL}--\eqref{JT}, we found it useful
to replace the dipole $S$-matrix by the corresponding scattering amplitude, 
$T(\br,x_g)\equiv  1-S^{(2)}_{x_g}(\bx_1, \bx_2)$, with $\br=\bx_1- \bx_2$. Also,
the dependence of $T$ upon the longitudinal fraction $x_g$ of the gluons from the
target was kept implicit,  to simplify the notations. 
The relevant values for $x_g$ will be discussed in a moment. But before that,
it is useful to separate the above cross sections into elastic and inelastic pieces, following
Eqs.~\eqref{Oel}--\eqref{Oinel}. For the transverse photon, one has
\begin{align}
\label{JTel}
	\mcal{J}_{\rmT,\rm{el}} = 
	\pi \bar{Q}^2
	\int \frac{\dif^2 \br}{(2\pi)^2}\frac{\dif^2 \br'}{(2\pi)^2}
	\rme^{-\rmi \bk \cdot (\br - \br')}
	\frac{\br \!\cdot \!\br'}{r\, r'}\, \rmK_1(\bar{Q}r) \rmK_1(\bar{Q}r') T(\br) T(-\br'),
\end{align}
\begin{align}
\label{JTinel}
\hspace*{-0.8cm}
	\mcal{J}_{\rmT,\rm{in}}= 
	\pi \bar{Q}^2\!
	\int \frac{\dif^2 \br}{(2\pi)^2}\frac{\dif^2 \br'}{(2\pi)^2}
	\rme^{-\rmi \bk \cdot (\br - \br')}
	\frac{\br \!\cdot \!\br'}{r\, r'}\, \rmK_1(\bar{Q}r) \rmK_1(\bar{Q}r') [T(\br) + T(-\br') - T(\br \!-\! \br')-T(\br) T(-\br')],
\end{align}
and similarly for the longitudinal one.

As promised, let us now specify the values for $x_g$ (the longitudinal momentum fraction of the gluons
from the target which are involved in the scattering) that should be used in evaluating the above SIDIS
cross-sections. By inspection of \eqn{xg}, one sees that, in the case of two-particle production,
 $x_g$ depends upon the transverse momenta $\bk_1$ and $\bk_2$ of both final partons.
 So, our above treatment of the integral over $\bk_2$, which ignored this additional dependence,
 may look as just an approximation. This is indeed true in general, but not also for the {\it elastic}
 scattering: one can easily check that, for the case of a homogeneous target, the r.h.s of \eqn{JTel}
 is proportional to the delta-function $\delta^{(2)}(\bk_1+\bk_2)$. (In the elastic scattering, there is
 no net transfer of transverse momentum from the target to the produced partons.)
 Accordingly, the integral over $\bk_2$ simply identifies $\bk_2=-\bk_1$, 
 and the version of  \eqn{xg} for SIDIS in the elastic channel reads
  \beq\label{xgel}
 x_g = \frac{1}{2P\cdot q}\left(\frac{k^2_{\perp}}{z(1-z)} + Q^2\right)\qquad\mbox{for elastic SIDIS},
 \eeq
 without any approximation. As for the inelastic channel, it is easy to identify what should be
 the right approximation to \eqn{xg}: the integral over $\bk_2$ is controlled
 by the {\it typical} transverse momenta of the unmeasured parton, which as discussed after
  \eqn{xg} are of the order of the largest among $\bar Q^2$ and $Q_s^2$. So in practice
  we shall use  \beq\label{xginel}
 x_g = \frac{1}{2P\cdot q}\left(\frac{k^2_{\perp}}{z} + 
 \frac{{\rm max}\big(\bar Q^2, \,Q_s^2(A,x_g)\big)}{1-z} +
 Q^2\right)\qquad\mbox{for inelastic SIDIS}.
 \eeq
  
 The integrations over $\br$ and $\br'$ in Eqs.~\eqref{JL}--\eqref{JTinel} are controlled by a product
of 3 factors: the modified Bessel functions $\rmK_0(\bar{Q}r) $ and $\rmK_1(\bar{Q}r) $ describing
the wavefunction of the virtual photon, the amplitude $T(\br)$ for the elastic dipole-nucleus scattering, and the exponential phase for the Fourier transform. Each of these factors introduces its
own characteristic scale: the effective virtuality $\bar Q^2=z(1-z)Q^2$, the target
saturation momentum $Q_s(A,x_g)$, and, respectively, the transverse momentum $k_\perp\equiv
|\bk|$ of the produced quark. As already mentioned, the Bessel functions effectively restrict the
integrations to values $r\lesssim 1/\bar Q$ (larger values are exponentially suppressed). Also,
the Fourier phase is rapidly oscillating for dipole sizes $r \gg 1/k_\perp$. Finally, the dipole
amplitude is small, but rapidly increasing with $r$, for small dipoles,
$T(x_g,\br)\sim r^2Q_s^2\ll 1$ when $r\ll 1/Q_s$ (``colour transparency''), 
but this growth is tamed by multiple scattering when $r\gtrsim 1/Q_s$. 
For larger dipole sizes, $r\gg  1/Q_s$, the amplitude saturates the unitarity limit $T=1$.
So, clearly, the result of these integrations depends upon the competition between the 3 scales
aforementioned.

In this work, we would like to explore the regime of strong scattering where $T\sim 1$. This is interesting 
since multiple scattering it the reflexion of gluon saturation when DIS is viewed in the dipole frame.
(The partonic picture of saturation in terms of gluon occupation numbers becomes manifest only in an infinite momentum frame for the target, such as the Bjorken frame \cite{Mueller:2001fv}.) 
Hence, we would like to choose the external scales $k_\perp$, $z$, and $Q^2$ in such a way that the integrations
favour dipole sizes $r,\,r'\gtrsim 1/Q_s$. For this to be true, we clearly need the two following conditions:
\begin{equation}
	\label{cons}
	\bar{Q}^2 \lesssim Q_s^2 \quad \textrm{and} \quad k_{\perp}^2 \lesssim Q_s^2.
\end{equation}
(Otherwise the integrations will be heavily suppressed either due to the exponential fall-off of the Bessel function, or due to the oscillations introduced by the Fourier transform.)
Notice that these conditions may be relaxed a bit, since the weak scattering regime at $r Q_s \ll 1$ is affected
too by saturation, via the phenomenon of geometric scaling. Accordingly, in what follows we shall also
study the high-momentum tails at $k_{\perp}^2 \gg Q_s^2$. Yet, as we shall see, the most interesting
signatures of multiple scattering and gluon saturation come indeed from the regime
where the scattering is strong.

\section{Saturation effects in SIDIS at $z(1-z)Q^2\ll Q_s^2$}
\label{sect:qllqs}

The fact that the longitudinal momentum fraction $z$ of the produced particle can be measured
in SIDIS, at least in principle, opens a very interesting possibility for studies of saturation: one can
select processes such that $Q^2$ is sufficiently high,  $Q^2\gtrsim Q_s^2$, for the perturbative treatment
of the $\gamma^*$ decay to be justified, while at the same time $z$ is sufficiently close to 1 
for $\bar Q^2=z(1-z)Q^2$ to be (much) smaller than $Q_s^2$ --- meaning that the dipole scattering
is sensitive to gluon saturation in the nuclear target. In this section we shall analytically 
study the strongly ordered case $\bar Q^2 \ll Q_s^2$, via suitable approximations. 
The analytical results will be supported by numerical calculations which also cover 
more general situations, where the two scales are comparable to each other: $\bar Q^2 \lesssim Q_s^2$.
In the subsequent analysis, it will be convenient to distinguish between three regimes for  the 
transverse momentum $k_{\perp}$ of the produced quark: 
{\ttfamily (i)} $k_{\perp}^2 \ll Q_s^2$, {\ttfamily (ii)} 
$k_{\perp}^2 \sim Q_s^2$, and  {\ttfamily (iii)}  $k_{\perp}^2 \gg Q_s^2$. 

%

\subsection{$k_{\perp}^2 \ll Q_s^2$:  probing the virtual photon wavefunction} 
\label{sect:LCWV}

When both $\bar Q^2$ and $k_{\perp}^2$ are much smaller than $Q_s^2$, 
the scattering is predominantly elastic. This will be confirmed by the
subsequent calculations, but can  also be understood via the following argument. 
In this regime, the integrals in \eqn{JTel} for $\mathcal{J}_{\rmT,\rm{el}}$ 
are dominated by relatively large dipole sizes, which obey $1/Q_s \lesssim 
r,\,r'\ll 1/\bar Q$, for which the scattering is as strong as possible:  $T\simeq 1$. 
(The respective contributions from smaller dipoles are suppressed by colour transparency: 
  $T(r)\!\propto\! r^2$ for $r\ll 1/Q_s$.)  On the contrary, for the inelastic cross-section, 
 the would-be dominant,  ``black disk'' ($T=1$), contributions to the
 individual terms in  \eqn{JTinel} mutually cancel in their sum. Accordingly, the net
 contribution to $\mathcal{J}_{\rmT,\rm{in}}$ 
 comes from smaller dipoles $r,\,r'\lesssim 1/Q_s$ and thus is parametrically suppressed
 compared to $\mathcal{J}_{\rmT,\rm{el}}$. Clearly, the presence of the last term 
 in the r.h.s. of  \eqn{JTinel},  which is quadratic in $T$ and becomes
 important at strong scattering, was essential for the above argument.
 
 \subsubsection{Elastic scattering near the unitarity limit}
 
 So let us focus on the elastic cross-section in this section.
As visible in \eqn{JTel}, this has a factorised structure 
--- the cross-section is the modulus squared of the elastic amplitude ---, hence it is
enough to discuss one of the two integrations, say the one over $\br$. As just mentioned,
the dominant contribution when $\bar{Q}^2, \,k_{\perp}^2 \ll Q_s^2$ can be computed
by replacing $T\to 1$ in  \eqn{JTel}. After this replacement, the elastic
amplitude reduces to the Fourier transform (FT) of the $\gamma^*$ wavefunction
(in particular, it becomes independent of the QCD coupling $\alpha_s$). By also using the
integrals \eqref{int1} and \eqref{int2} from Appendix~\ref{sec:app1}, one finds
\begin{align}
\label{JLlow}
\mathcal{J}_{\rmL,\rm{el}} (\bk) = \frac{1}{4\pi} \frac{\bar{Q}^2}{(k_{\perp}^2 +\bar{Q}^2)^2}
\quad \textrm{for} \quad \bar{Q}^2, k_{\perp}^2 \ll Q_s^2,
\end{align}
and
\begin{align}
\label{JTlow}
\mathcal{J}_{\rmT,\rm{el}}(\bk)  = \frac{1}{4\pi} \frac{k_{\perp}^2}{(k_{\perp}^2 +\bar{Q}^2)^2}
\quad \textrm{for} \quad \bar{Q}^2, k_{\perp}^2 \ll Q_s^2,
\end{align}
for longitudinal and transverse polarisations, respectively.
By definition, these expressions represent the  transverse momentum distributions of the
quark and the antiquark generated by the decay of the virtual photon.  For a virtual photon
with longitudinal polarisation, this distribution is peaked near $k_\perp\sim \bar Q$ (in particular,
it vanishes in the photo-production limit $\bar Q^2\to 0$), whereas for a transverse photon, 
it is logarithmically distributed within the range $ \bar{Q}^2 \ll k_{\perp}^2 \ll Q_s^2$, where
$\mathcal{J}_{\rmT,\rm{el}} \simeq 1/(4\pi k_{\perp}^2)$.


It is quite
remarkable that a DIS measurement in the regime where the QCD scattering is as strong as possible
($T=1$) carries no information about the target, but merely unveils the structure
of the virtual photon. This is due to the fact that, in this strong scattering regime, the internal
structure of the target is not visible for the dipole projectile: the target looks ``black'', 
i.e. totally absorptive, for any $r\gg 1/Q_s$. 
In terms of Feynman graphs, the contributions shown in Eqs.~\eqref{JLlow}--\eqref{JTlow}
fully correspond to the second graph in Fig.~\ref{fig:2jets-ampli} --- that is, the decay of $\gamma^*$
occurs {\it after} crossing the nuclear target. 
The contribution of the first graph to the elastic scattering is proportional to the dipole $S$-matrix 
(recall \eqn{dijet-elas}), which vanishes in the  ``black disk'' limit. 
But in the second graph, the dipole does not ``see'' the target at all,
so in that case $S(r)=1$ independently of the value of $r$.

To summarise, in the black disk/deeply saturated regime at $\bar{Q}^2\ll k_{\perp}^2 \ll Q_s^2$, 
the cross-section for SIDIS is controlled by the elastic scattering of the $q\bar q$ fluctuation of
a virtual photon with transverse polarisation and can be approximated as 
\begin{align}
\label{JTellow}
\mathcal{J}_{\rmT} \simeq \mathcal{J}_{\rmT,{\rm el}} \simeq  
\frac{1}{4\pi k_{\perp}^2}
\quad \textrm{for} \quad \bar{Q}^2 \ll k_{\perp}^2 \ll Q_s^2.	
\end{align}

Although not manifest in this result for $\mathcal{J}_{\rmT}$,
the details of the QCD scattering (in particular, the nature
of the target) are of course important, as they determine its validity range (notably, the upper
limit $Q_s^2\equiv Q_s^2(A,x_g)$ on $k_{\perp}^2$), as well as the corrections beyond this universal 
form. We shall shortly evaluate these corrections, both analytically (within the context of the 
McLerran-Venugopalan model)
and numerically (from solutions to the BK equation). 
To that aim, we also need the relevant value of $x_g$ --- 
the longitudinal momentum fraction of the participating gluons from the target  ---, which fixes
the rapidity phase-space for the high-energy evolution. For elastic processes, this is
generally given by \eqn{xgel}, which for the present kinematics reduces to
  \beq\label{xgel2}
 x_g \simeq \frac{1}{2P\cdot q}\,\frac{k^2_{\perp}}{z(1-z)}\,=\xbj\,
 \frac{k^2_{\perp}}{\bar Q^2}\,\gg\,\xbj\,.
 \eeq
As indicated by the last inequality, this value $x_g$ is typically much larger than the Bjorken $x$
variable relevant for inclusive DIS, meaning that the effects of the high energy evolution are less
important than one might naively expect. But before we study this evolution in more detail,
let us focus on the situation at lower energies, say $x_g\sim 10^{-2}$, where the MV model applies.

\subsubsection{Elastic scattering in the MV model}

The McLerran-Venugopalan (MV) model \cite{McLerran:1993ni,McLerran:1994vd} is
a semi-classical approximation for the gluon distribution in a large nucleus, valid at weak coupling and for
moderately high energies --- namely,  so long as one can neglect the effects of the high-energy 
evolution. In this model, the nuclear gluon distribution is assumed to be
generated via incoherent, classical, radiation by the valence quarks. This leads to a Gaussian 
distribution for the target colour fields $A^{-}_a$ (recall \eqn{Wilson}), for which it is possible to analytically
compute the relevant Wilson-lines correlators. In particular, the dipole $S$-matrix is found as
(see e.g. \cite{Iancu:2002xk,Iancu:2003xm})
\beq\label{SMV} 
S_0(r)=\exp\left\{-\frac{r^2Q_{A}^2}{4}\ln\frac{1}{r^2\Lambda^2}\right\}\,,
\eeq
where the quantity in the exponent is the dipole amplitude in the single scattering approximation
(and in the MV model). This quantity $Q_{A}^2$, which has the dimension of a transverse momentum
squared, is proportional to the colour charge density squared of the valence quarks per unit transverse
area. It scales with the nucleon number $A$
and with the QCD coupling $\alpha_s$ like $Q_{A}^2\propto \alpha_s^2 A^{1/3}$. (One factor
of $\alpha_s$ is inherent in the colour charge density, the other one comes from the coupling
to the dipole.) Furthermore,  $\Lambda$ is the QCD confinement scale and the
logarithm has been generated by integrating over Coulomb exchanges with transverse
momenta $q_\perp$ within the range $\Lambda^2 \ll q_\perp^2 \ll 1/r^2$.  \eqn{SMV} is derived
under the assumption that both $Q_A^2$ and  $1/r^2$ are much larger than $\Lambda^2$
and is valid to leading logarithmic accuracy w.r.t. to the Coulomb log.

Our present interest in \eqn{SMV} is twofold. First, as already mentioned, 
this will be used as the initial condition for the BK evolution with decreasing $x_g$.
Second, due to its relative simplicity, \eqn{SMV} allows us to deduce relatively accurate, analytic,
approximations for the SIDIS cross-sections of interest for us here. In particular, in this section
we shall construct an approximation to the elastic cross-section $\mathcal{J}_{\rmT,{\rm el}}$,
which extends the previous result in \eqn{JTellow} to larger transverse momenta,  $k_\perp\sim Q_s$.
The corresponding approximation for the inelastic cross-section $\mathcal{J}_{\rmT,{\rm in}}$ will
be discussed in the next section.

Before we proceed, let us specify the definition of the saturation momentum $Q_s(A)$ that we shall
use in this context\footnote{Notice that the definition of $Q_s(A)$ to be used in the analytic calculations
based on the MV model is slightly different from the one that we shall generally use, for practical
convenience, in the numerical calculations.}: this is the value $Q_s(A)=2/r$ for which the exponent 
in \eqn{SMV} becomes equal to one; this condition yields
\beq\label{QsMV}
Q_s^2(A)\,\simeq\,Q_{A}^2 \ln\frac{Q_s^2(A)}{\Lambda^2}\,.\eeq

We now return to \eqn{JTel} for $\mathcal{J}_{\rmT,{\rm el}}$ and construct an analytic approximation
valid within the MV model and within the extended range
$\bar{Q}^2 \ll k_{\perp}^2 \lesssim Q_s^2$.  Since the interesting dipole sizes  
obey $r \ll 1/\bar Q$, one can keep only the leading order term, 
$ \rmK_1(\bar{Q}r) \simeq 1/(\bar{Q}r)$, in the small-argument
expansion of the Bessel function; then \eqn{JTel} reduces to\footnote{Incidentally,
the quantity $ \mathcal{W}(\bk)$ is formally the same as the so-called Weisz\"acker-Williams
unintegrated gluon distribution as computed in the MV model \cite{Iancu:2003xm}.}
\begin{align}
\label{JTelMV}
	\mcal{J}_{\rmT,\rm{el}} \simeq
	\pi \big |\nabla^i_{\bk} \mathcal{W}\big |^2,\qquad\mbox{with}\qquad
	 \mathcal{W}(\bk)\equiv  
	\int \frac{\dif^2 \br}{(2\pi)^2}\,
	\rme^{-\rmi \bk \cdot \br}\,
	\frac{T(\br)}{r^2}\,.
\end{align}
Since independent of $\bar Q$, this result is formally the same as the limit of \eqn{JTel} 
when $\bar Q\to 0$. As such, it holds independently of the precise form of
the dipole amplitude. But the use of the MV model, i.e.  $T(r)= 1-S_0(r)$ with $S_0(r)$ 
shown in \eqn{SMV},  enables us to almost exactly compute the integral in \eqn{JTelMV}.
Indeed, one can easily see that the piece of the integral involving $S_0$ is
dominated by $r=2/Q_s$ (the largest value of $r$ which is allowed by the decay of $S_0$).
Hence, in evaluating that piece, one can replace $r\to 2/Q_s$ within the slowly varying 
logarithmic factor in the exponent of \eqn{SMV}. That is, one can effectively replace 
the dipole $S$-matrix by the Gaussian $S_0(r)\simeq \exp(-{r^2Q_{s}^2}/{4})$, where we have also
used the definition \eqref{QsMV} for  $Q_{s}^2$.
The ensuing integral over $\br$ can be exactly evaluated, as explained
in Appendix~\ref{sec:elastic}; one thus finds
\begin{align}
\label{JTelGBW}
 \mathcal{J}_{\rmT,{\rm el}} \simeq  
\frac{1}{4\pi k_{\perp}^2}\,\rme^{-\frac{2k_\perp^2}{Q_s^2}}
\qquad \textrm{for} \qquad \bar{Q}^2 \ll k_{\perp}^2 \lesssim Q_s^2.
\end{align}
This result looks indeed as a natural extension of our previous estimate  in \eqn{JTellow}
 to larger transverse momenta  $k_\perp\sim Q_s$.  It has a natural physical interpretation.
 The power $1/ k_{\perp}^2$ is the transverse momentum distribution produced by the decay
 of the transverse virtual photon, as already discussed. The Gaussian $\exp(-{2k_\perp^2}/{Q_s^2})$
 describes the subsequent  broadening of this distribution via {\it elastic}
multiple scattering between the $q\bar q$ dipole and the gluons from the nucleus. 
 \eqn{JTelGBW}  shows that the deviations from the pure power law should become
important  for $k_\perp^2 \sim Q_s^2/2$, where the factor 1/2 reflects the elastic nature
of the scattering process.

\subsubsection{Numerical results for $k_{\perp}^2 \lesssim Q_s^2$}

\begin{figure}
  \centering
  \begin{subfigure}[t]{0.48\textwidth}
    \includegraphics[width=\textwidth]{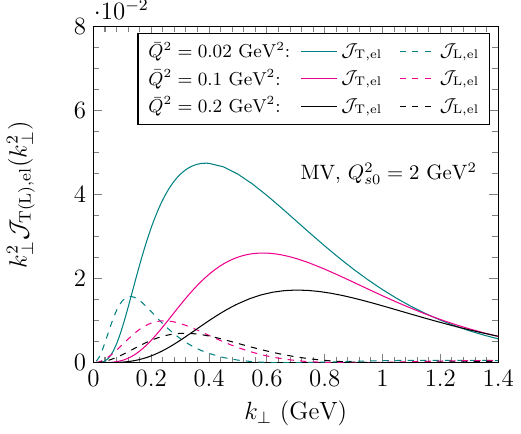}
    \caption{}\label{fig:lowkT_a}
  \end{subfigure}
  \hfill
  \begin{subfigure}[t]{0.48\textwidth}
    \includegraphics[width=\textwidth]{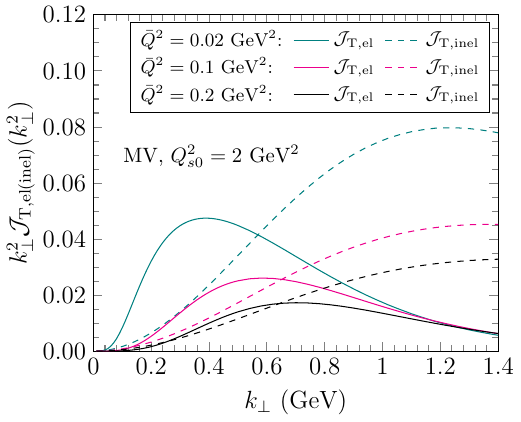}
    \caption{}\label{fig:lowkT_b}
  \end{subfigure}
  \caption{\small (a) 
  Comparison between the elastic cross-sections in the longitudinal and the transverse sector,
  respectively.
 The (reduced) cross-sections multiplied by $k_{\perp}^2$ are shown as functions of $k_\perp$, for
 $k_\perp\le  Q_{s0}$ and several
 values $\bar{Q}^2 \ll Q_{s0}^2$. Here, $Q_{s0}^2=2$~GeV$^2$ is the saturation momentum
 in the MV model. (b)  Comparison between the elastic and inelastic cross-sections in the transverse sector.
 Elastic scattering dominates for low enough $k_\perp$,
shows a maximum around $k_{\perp}^2 \sim \bar{Q} Q_{s0}$, and then decreases. Inelastic scattering keeps increasing until $k_{\perp} \sim Q_{s0}$; it becomes the dominant contribution already below $Q_{s0}$
   (see also the discussion in Sect.~\ref{sec:medkT}).}
    \label{fig:lowkT}
    \end{figure}

As promised, we shall now compare the  analytic approximations that we previously developed
for the kinematical range at $\bar{Q}^2 \ll Q_s^2$ and $k_{\perp}^2 \lesssim Q_s^2$ 
with numerical calculations.

We start with the MV model, for which we shall use a slightly modified version of \eqn{SMV},
which is well defined for any value of $r$, including $r> 1/\Lambda$, and thus allows
for unrestricted numerical integrations; this reads (below, $e=2.718...$ is the Euler number)
\beq\label{SMVnum} 
S_0(r)=\exp\left\{-\frac{r^2Q_{A}^2}{2}\ln\left(\frac{1}{r\Lambda}+e\right)\right\}\,.
\eeq
Also, in all the numerical simulations, we shall universally define the saturation momentum
as the value $Q_s=2/r$ for which the dipole amplitude is equal to 1/2: $T(r=2/Q_s)=1/2$. This is
slightly different from its previous definition \eqref{QsMV}, but the difference is irrelevant for our purposes.
In practice, we have chosen $Q_{A}^2$ and $\Lambda^2$ such as
$Q_{s0}^2=2$~GeV$^2$ for the MV model $S$-matrix in \eqn{SMVnum}.
(For more clarity, in all the plots and the associated discussions
we use the specific notation $Q_{s0}^2$ for the saturation momentum in the MV model.)

{Our numerical results corresponding to the MV model are shown
in the two plots in Fig.~\ref{fig:lowkT}. For convenience, we display results for
the reduced cross-sections multiplied by $k_\perp^2$. That is, we plot
dimensionless quantities like  $k_\perp^2 \mathcal{J}_{\rmT,{\rm el}}(k_\perp)$
as functions of $k_\perp$, in the saturation region at $k_\perp\le Q_{s0}$.
We use $Q_{s0}^2=2$~GeV$^2$ together with 3 values for the effective virtuality $\bar Q^2$,
namely $\bar Q^2=0.02,\, 0.1,$ and 0.2~GeV$^2$, which are well separated from each
other and also from $Q_{s0}^2$, and thus are useful for indicating theoretical trends.
These values for $\bar Q^2$ are quite low and should be taken with a grain of salt.
The largest among them,  $\bar Q^2= 0.2~{\rm GeV}^2$,
corresponds to a transverse momentum $k_\perp=\bar Q\simeq 450~{\rm MeV}$, which is a reasonable
scale for the transition from non-perturbative to perturbative regimes. The even lower values,
like $\bar Q^2= 0.02~{\rm GeV}^2$, are strictly speaking non-perturbative, but they are
only used to illustrate the interesting phenomena, outside any phenomenological context.}

Specifically, in Fig.~\ref{fig:lowkT_a} we compare the elastic cross-sections in the transverse and in
the longitudinal sector, respectively, whereas in Fig.~\ref{fig:lowkT_b} we compare
the elastic and the inelastic cross-sections, both in the transverse sector.
The curves in Fig.~\ref{fig:lowkT_a} have the expected shapes in light of the previous
discussion in this paper: $k_\perp^2 \mathcal{J}_{\rmL,{\rm el}}$ is peaked around 
$k_\perp=\bar Q$, whereas  $k_\perp^2 \mathcal{J}_{\rmT,{\rm el}}$ shows a wide
maximum (``plateau'') at, roughly, $\bar Q< k_\perp < Q_{s0}/\sqrt{2}$. When 
increasing $\bar Q$, the height of this maximum is decreasing and its position
gets displaced towards larger values of $k_\perp$. This can be understood
as follows: when $\bar Q\to 0$,  \eqn{JTelGBW} shows that the maximum of
$k_\perp^2 \mathcal{J}_{\rmT,{\rm el}}$ lies at $k_\perp=0$ and its height is equal to
$1/(4\pi)\simeq 0.08$. The effect of increasing $\bar Q$ can be appreciated by using
the following approximation, which interpolates between  \eqn{JTlow} at low $k_\perp\sim \bar Q\ll Q_{s0}$ 
and  \eqn{JTelGBW} at larger momenta $\bar Q\ll k_\perp\sim Q_{s0}$:
 \begin{align}
\label{JTlow-inter}
\mathcal{J}_{\rmT,\rm{el}}(\bk)  \simeq \frac{1}{4\pi} \frac{k_{\perp}^2}{(k_{\perp}^2 +\bar{Q}^2)^2}
\,\rme^{-\frac{2k_\perp^2}{Q_s^2}}
\qquad \textrm{for} \quad \bar{Q}^2 \ll Q_s^2 \quad\mbox{and}\quad  k_{\perp}^2\lesssim Q_s^2.
\end{align}
It is not difficult to check that the function $k_\perp^2 \mathcal{J}_{\rmT,{\rm el}}$
develops a maximum at $k_\perp^2\sim \bar Q Q_{s0}$ and that the height of this maximum
is proportional to $\exp(-2\bar Q/{Q_{s0}})$ --- in qualitative agreement with Fig.~\ref{fig:lowkT_a}.

Fig.~\ref{fig:lowkT_a} also shows that
the longitudinal piece $\mathcal{J}_{\rmL,{\rm el}}$ dominates at very low momenta
$k_\perp\lesssim \bar Q$, but it is much smaller than the transverse piece 
$\mathcal{J}_{\rmT,{\rm el}}$ at any $k_\perp\gg \bar Q$. Of course, one should be 
careful when comparing numerical results for  $\mathcal{J}_{\rmT}$ and  $\mathcal{J}_{\rmL}$,
as these quantities are weighted by different factors in the respective {\it physical} 
cross-sections, cf. Eqs.~\eqref{sigmaL}--\eqref{sigmaT}.  Notably, the longitudinal cross-section 
has an additional suppression when $z\to 1$, due to the prefactor $z(1-z)$ in \eqref{sigmaL}.
Altogether, we conclude that the longitudinal contribution is tiny and can
be safely neglected in the physical regime under consideration.

Concerning Fig.~\ref{fig:lowkT_b}, we notice that the elastic cross-section $\mathcal{J}_{\rmT,{\rm el}}$
is indeed larger than the inelastic one  $\mathcal{J}_{\rmT,{\rm in}}$ at sufficiently small 
$k_\perp\ll Q_s$, as expected, but this hierarchy is changing quite fast when increasing $k_\perp$. 
The reasons for this behaviour
should become clear in the next section, where the inelastic contribution will be explicitly
computed.

\begin{figure}
 \centering
  \includegraphics[width=0.5\textwidth,page=1]{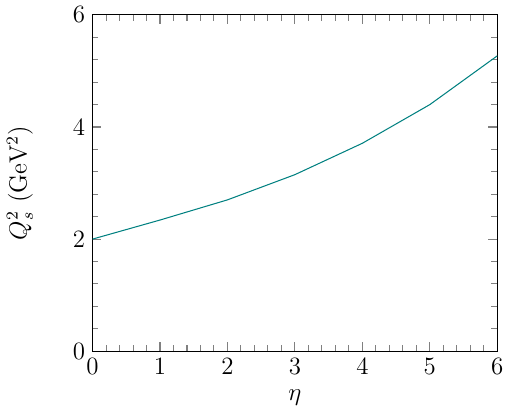}    
 \caption{\small Rapidity dependence of the saturation momentum squared,
 as obtained by solving the collinearly resummed BK equation described in Appendix \ref{sec:BK} and with MV model initial condition with $Q_{s0}^2=2$~GeV$^2$.}\label{fig:Qseta}
\end{figure}

\begin{figure}
  \centering
  \begin{subfigure}[t]{0.48\textwidth}
    \includegraphics[width=\textwidth]{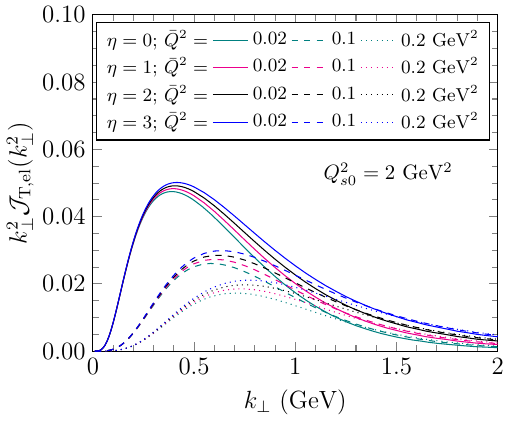}
    \caption{} \label{fig:BKlowkT_a}
  \end{subfigure}
  \hfill
  \begin{subfigure}[t]{0.48\textwidth}
    \includegraphics[width=\textwidth]{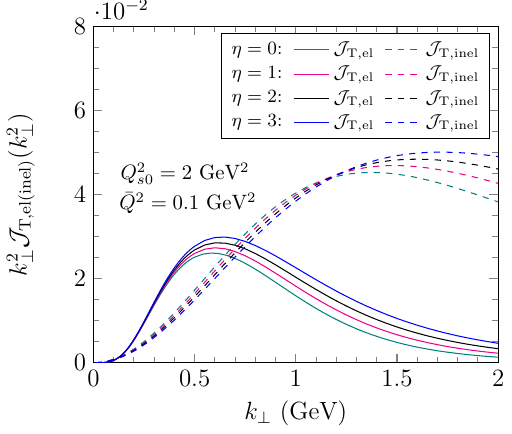}
    \caption{} 
    \label{fig:BKlowkT_b}
  \end{subfigure}
  \caption{\small (a) Rapidity evolution of the  transverse elastic cross-section multiplied by $k_{\perp}^2$ for $Q_{s0}^2=2$~GeV$^2$, for 3 different values for $\bar{Q}^2$ and 4 different rapidities ($\eta=0,\,1,\,2,\,3$);
  see the main text for details.
    (b) Rapidity evolution of the transverse elastic and inelastic cross-sections multiplied by $k_{\perp}^2$, for 
    fixed $\bar{Q}^2=0.1$~GeV$^2$ and the same 4 values for $\eta$.}
  \label{fig:BKlowkT}
    \end{figure}

Furthermore, Fig.~\ref{fig:BKlowkT} shows the results of the high-energy evolution
for the elastic and inelastic (reduced) cross-sections, in the transverse sector alone.
These results have been obtained by solving an improved version of the 
BK equation with running coupling (rcBK), which also includes all-order resummations 
of large radiative corrections enhanced by double and single collinear logarithms
 \cite{Ducloue:2019ezk,Ducloue:2019jmy}. (For convenience, 
this equation is briefly summarised in Appendix~\ref{sec:BK}.)
The evolution ``time'' is the target rapidity\footnote{The use of the target rapidity as opposed
 to the projectile rapidity not only is natural in
the context of DIS, but also has the advantage to automatically ensure the proper
time-ordering (with ``time'' = the LC time $x^+$ of the projectile) of the successive, soft,
gluon emissions, thus avoiding large higher-order corrections --- the {\it anti-collinear}
double logarithms --- which would lead to an unstable evolution \cite{Lappi:2015fma}; see the
discussion in  \cite{Ducloue:2019ezk}.}
 $\eta=\ln(x_0/x_g)$, related to the ``minus''  longitudinal momentum ($p^-=x_g P^-$) 
 of the gluons from the target which are involved in the scattering. 
Specifically, we use the $S$-matrix in the MV model, \eqn{SMVnum}, as an initial condition at $\eta=0$
and evolve the collinearly-improved rcBK equation up to $\eta=3$. 
The value $x_0$ of $x_g$ at which one starts this evolution is irrelevant for our purposes;
in applications to phenomenology, one often takes $x_0=10^{-2}$.
For the subsequent discussion, it is also interesting to know what are the values of the target
saturation momentum for the rapidities to be explored in this paper ($\eta\le 6$). These are
numerically extracted from the solution to the (improved) BK equation, with the results
 shown in Fig.~\ref{fig:Qseta}.

In Fig.~\ref{fig:BKlowkT_a} we show the elastic cross-section  
$k_\perp^2 \mathcal{J}_{\rmT,{\rm el}}$ for $\eta\le 3$ and the
same 3 values for the ratio $\bar Q^2/Q_{s0}^2$ as in  Fig.~\ref{fig:lowkT}. The results are plotted 
as functions of $k_\perp$, up to $k_\perp= 2$~GeV. (For comparison, we note that
 $Q_s(\eta=3)\simeq 1.77$~GeV, cf.  Fig.~\ref{fig:Qseta}.)
In Fig.~\ref{fig:BKlowkT_b}, we fix $\bar Q^2/Q_{s0}^2=0.05$ and display the functions 
$k_\perp^2 \mathcal{J}_{\rmT,{\rm el}}$ and $k_\perp^2 \mathcal{J}_{\rmT,{\rm in}}$ for the same
values of $\eta$ and in the same range in $k_\perp$ as in the left plot.  By inspection
of these curves, one sees that the effects of the evolution are rather mild (at least,
for the moderate values of $\eta$ under consideration). This is a consequence of the 
aforementioned resummations (the running coupling corrections and the collinear logarithms), 
which considerably slow down the evolution as compared to the leading-order BK equation.

Note finally a subtle point that we have glossed over in the above discussion: 
strictly speaking, the relevant values for $x_g$ may
depend upon the measured transverse momentum $k_\perp$, as shown in \eqn{xgel2} for the case
of elastic scattering. In practice, we have neglected this dependence, since it would greatly complicate the numerical solutions to the BK equation. However, given the smallness of the evolution effects,
as visible in Fig.~\ref{fig:BKlowkT}, and the logarithmic dependence of $\eta$ upon $x_g$,
 it should be quite clear that such a $k_\perp$--dependent change in the value of $x_g$ cannot
 change our qualitative discussion. As for the inelastic sector, we shall shortly discover (in the next section)
 that this ambiguity is even less important.
%

\subsection{$k_{\perp}^2 \sim Q_s^2$:  transverse momentum broadening} 
\label{sec:medkT}

In the previous section we have seen that, when increasing $k_\perp^2$ above $\bar Q^2$, 
the elastic contribution to SIDIS is rapidly decreasing, according to \eqn{JTellow}.
On the other hand, the inelastic contribution is expected to be most important when 
$k_\perp \sim Q_s$, since this is the typical transverse momentum transferred by the target to the produced quark via incoherent multiple scattering. This expectation is confirmed by the numerical results in
Figs.~\ref{fig:lowkT} and \ref{fig:BKlowkT}, which show that $k_\perp^2\mathcal{J}_{\rmT,{\rm in}}$ keeps
increasing when $k_\perp$ increases towards $Q_s$. In this section, we shall construct an analytic
approximation for $\mathcal{J}_{\rmT,{\rm in}}$ for transverse momenta in the range 
$\bar{Q}^2\ll k_{\perp}^2 \sim Q_s^2$ (by which we mean that $k_{\perp}$ can be both smaller,
or larger, than $Q_s$, but not {\it much} larger; the tail of the cross-section at 
$k_{\perp} \gg Q_s$ will be studied in Sect.~\ref{sec:highkT}). 
This will allow us to physically understand the respective numerical 
results in Figs.~\ref{fig:lowkT} and \ref{fig:BKlowkT}, as well as their extension to larger values
$k_{\perp} \gtrsim Q_s$.

\subsubsection{Inelastic scattering off the saturated gluons}

To study the inelastic cross-section \eqref{JTinel} for $k_\perp \sim Q_s$, it is convenient to return
to its representation in terms of dipole $S$-matrices (recall \eqn{Oinel}), 
rather than scattering amplitudes. One has
\begin{align}
\label{JTinel2}
\hspace*{-0.1cm}
	\mcal{J}_{\rmT,\rm{in}}= 
	\pi \bar{Q}^2
	\int \frac{\dif^2 \br}{(2\pi)^2}\frac{\dif^2 \br'}{(2\pi)^2}
	\rme^{-\rmi \bk \cdot (\br - \br')}
	\frac{\br \!\cdot \!\br'}{r\, r'}\, \rmK_1(\bar{Q}r) \rmK_1(\bar{Q}r') [S(\br \!-\! \br')-S(\br) S(-\br')],
\end{align}
where we recall that $\bar Q^2\ll k_\perp^2\sim Q_s^2$. In this regime, the double
integral is dominated by dipole sizes within the range $1/Q_s\lesssim  r,\,r'\, \ll 1/\bar Q$, which 
 (at least,  marginally)  correspond to
strong scattering: the dipole $S$-matrices are significantly smaller than one, albeit
their actual values cannot be neglected, i.e. one cannot work in the unitarity limit $S\to 0$
(or $T\to 1$), since the net result would vanish in that limit. Yet, when $S\ll 1$, the second term 
within the square brackets in \eqn{JTinel2}, which is quadratic in $S$, can be neglected next
to the first term there, which is linear in $S$.

 To evaluate the dominant contribution, we first observe that the product $\rme^{-\rmi \bk \cdot (\br - \br')}
 S(\br \!-\! \br')$ restricts the double integration to values $\br$ and $\br'$ such that $|\br- \br'|
 \lesssim 1/k_\perp \sim 1/Q_s\ll 1/\bar Q$.  This makes it convenient to change
one integration variable, say $\br'\to \bm{\rho} \equiv \br - \br'$. Then, after also using the small-argument
 expansion of the Bessel function,  $ \rmK_1(z)\simeq 1/z$ for $z\ll 1$, one sees
 that the integral over $\br$ at fixed $\bm{\rho}$ shows a logarithmic enhancement over the range 
 $\rho \ll r\ll 1/\bar Q$, with $\rho=|\bm{\rho}|$. Indeed, one can write
 \beq\label{logint}
\frac{\br \!\cdot \!\br'}{r\, r'}\, \rmK_1(\bar{Q}r) \rmK_1(\bar{Q}r')
\,\simeq\,\frac{1}{\bar{Q}^2r^2}\quad\mbox{for}\quad 1/\bar Q \gg r\gg\rho\,,\eeq
and the ensuing integral over $\br$ yields the logarithm $\ln(1/\rho^2 \bar Q^2)$, which is 
large when $1/\rho\sim Q_s$. To leading logarithmic accuracy with respect to this transverse
log, one can further write
\begin{align}
\label{JTinlow}
\mathcal{J}_{\rmT,{\rm in}} 
\simeq  &\,
\frac{1}{16\pi^2}
\int \dif^2\bm{\rho}\, \rme^{-\rmi \bk \cdot \bm{\rho}} S(\bm{\rho}) \,\ln \frac{1}{\rho^2\bar{Q}^2}
\nonumber\\*[0.2cm]
\simeq  &\,
\frac{1}{16\pi^2}\ln \frac{Q_s^2}{\bar{Q}^2}
\int \dif^2\bm{\rho}\, \rme^{-\rmi \bk \cdot \bm{\rho}} S(\bm{\rho}) 
\quad \textrm{for} \quad \bar{Q}^2 \ll k_{\perp}^2 \sim Q_s^2,
\end{align}
where the second line follows from the fact that the integral over $\rho$
is controlled by values $\rho\sim 1/Q_s$, as we shall shortly check; hence,
to the leading-log accuracy of interest, it is legitimate to replace  $\rho\to 1/Q_s$
within the argument of the logarithm from the first line.

One can verify that the inelastic contribution $\mathcal{J}_{\rmL,{\rm in}}$ of
the longitudinal photon has no logarithmic enhancement\footnote{Using
$ \rmK_0(z)\simeq \ln(1/z)$ for $z\ll 1$, one can check that, in the kinematical
regime at study, the integrals in \eqref{JTinel} are dominated by very large dipoles,
with $r,\,r'\sim 1/\bar Q$, but such that $\rho=|\br - \br'|\lesssim 1/k_\perp$. Accordingly,
there is no logarithmic phase-space for the integral over $r$ at fixed $\rho$.}
and hence it is parametrically smaller than the transverse contribution in \eqn{JTbroad}.

The final result in \eqn{JTinlow} has a simple physical interpretation. The 
logarithm $\ln({Q_s^2}/{\bar{Q}^2})$ 
comes from integrating over the {\it initial} transverse momentum distribution 
of the quark, as generated by the decay of the virtual photon $\gamma^*_\rmT$ 
and shown in \eqn{JTellow}. The second factor in
the r.h.s. of \eqn{JTinlow} is recognised as the Fourier transform of the dipole $S$-matrix.
It describes the {\it final} transverse momentum distribution of the quark, as generated via 
(multiple) scattering off the nuclear target.

The dipole $S$-matrix appearing in \eqn{JTellow} should be evaluated at a rapidity
$\eta=\ln(x_0/x_g)$ with $x_g$ given by \eqn{xginel}; within the present kinematics,
that is,  $k_{\perp}^2 \sim Q_s^2\gg \bar{Q}^2$,  this yields
\beq\label{xgQs}
 x_g = \frac{1}{2P\cdot q}
\, \frac{Q_s^2}{z(1-z)} \ \Longrightarrow \ \eta
  = \ln \frac{x_0}{\xbj} - \ln  \frac{Q_s^2}{\bar Q^2}\,,
  \eeq
with $\xbj$ as defined in \eqn{xBj}.
In the regime of interest here, the rapidity  $\eta$ can be considerably smaller than the 
value $\ln(x_0/\xbj)$ that would be normally used for studies of inclusive  DIS. 

\subsubsection{Inelastic scattering in the MV model}

To obtain a more explicit expression for $\mathcal{J}_{\rmT,{\rm in}} $, let us evaluate 
 \eqn{JTinlow} within the MV model, that is, by using the dipole $S$-matrix in \eqn{SMV}.
This calculation can be simplified via the same argument as discussed in relation
with \eqn{JTelMV}:
in the interesting range at $k_\perp\lesssim Q_s$, the FT in \eqn{JTinlow} is 
dominated by the largest value of $\rho$ that is allowed by the dipole $S$-matrix,
namely $\rho\simeq 2/Q_s$. To the accuracy of interest, one can replace the $S$-matrix
in \eqn{SMV} by the Gaussian $S_0(r)\simeq \exp\{-{r^2Q_{s}^2}/{4}\}$, with $Q_{s}^2$
defined in \eqn{QsMV}. The one easily finds
\begin{align}
\label{JTbroad}
\mathcal{J}_{\rmT,{\rm in}} 
\simeq  
\frac{1}{4\pi}\ln \frac{Q_s^2}{\bar{Q}^2}\,\frac{1}{Q_s^2}\,\rme^{-{k_\perp^2}/{Q_s^2}}
\quad \textrm{for} \quad \bar{Q}^2 \ll k_{\perp}^2 \sim Q_s^2.	
\end{align}
This Gaussian distribution in $k_\perp$ is the hallmark of a diffusive process:
it describes transverse momentum broadening via soft ($\Lambda^2 \ll q_\perp^2 \lesssim Q_s^2$) 
multiple scattering between the produced quark and the gluons in the nucleus. 

At this point, we have analytic estimates for the predictions of the MV model 
for both the elastic ($\mathcal{J}_{\rmT,{\rm el}}$)
and the inelastic ($\mathcal{J}_{\rmT,{\rm in}}$) cross-sections, and for transverse
momenta within the range $\bar Q^2\ll k_\perp^2\sim Q_s^2$. By comparing
the respective results in Eqs.~\eqref{JTelGBW} and \eqref{JTbroad}, one finds that the
elastic scattering dominates over the inelastic one so long as $k_\perp$ is small enough for
$Q_s^2/ k_\perp^2 > \ln ({Q_s^2}/{\bar{Q}^2})$. Using $Q_s^2\gg \bar Q^2$, it is not difficult
to check that the ``critical'' value $k_\perp^2\simeq Q_s^2/\ln ({Q_s^2}/{\bar{Q}^2})$ at which
the elastic and the inelastic contributions are roughly equal to each other lies in between the
respective maxima\footnote{More precisely, these maxima refer to the reduced cross-sections
multiplied by $k_\perp^2$, as in Figs.~\ref{fig:lowkT} and \ref{fig:BKlowkT}.}, 
namely $k_\perp^2\simeq Q_s\bar Q$ for 
the elastic piece and $k_\perp^2\simeq Q_s^2$ for the inelastic one. 
This is in agreement with the numerical results shown in
Figs.~\ref{fig:lowkT_b} and \ref{fig:BKlowkT_b}. More results
will be presented in the next subsection.
 
 \subsubsection{Numerical results for $k_{\perp}^2 \sim Q_s^2$}

\begin{figure}
  \centering
  \begin{subfigure}[t]{0.48\textwidth}
    \includegraphics[width=\textwidth]{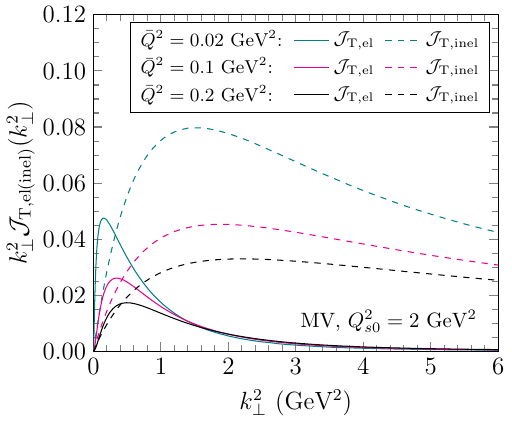}
    \caption{}\label{fig:el-inel_a}
  \end{subfigure}
  \hfill
  \begin{subfigure}[t]{0.48\textwidth}
    \includegraphics[width=\textwidth]{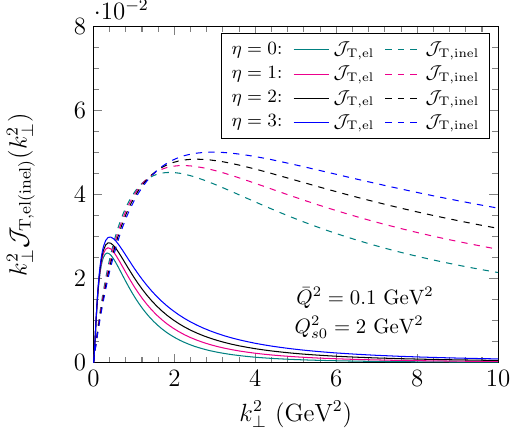}
    \caption{}\label{fig:el-inel_b}
  \end{subfigure}
  \caption{\small (a) The transverse elastic and inelastic cross-sections multiplied by $k_{\perp}^2$, 
  plotted as functions of $k_{\perp}^2$ up to $k_\perp^2= 3Q_{s0}^2=6$~GeV$^2$ and
  for 3 different values of $\bar{Q}^2$. 
  (b) Rapidity evolution up to $\eta=3$ of the results from the left plot corresponding to
   $\bar{Q}^2=0.1$~GeV$^2$. We note that $Q_s^2(\eta=3)\simeq 3.13$~GeV$^2$.}\label{fig:el-inel}
\end{figure}


The two plots shown in Fig.~\ref{fig:el-inel} extend the previous ones in
Fig.~\ref{fig:lowkT_b} and, respectively, Fig.~\ref{fig:BKlowkT_b} 
to larger values for the transverse momenta, namely up to  $k_\perp^2= 3Q_{s}^2$.
Notice also that, as compared to Figs.~\ref{fig:lowkT} and \ref{fig:BKlowkT}, the
rescaled cross-sections  $k_\perp^2 \mathcal{J}_{\rmT,{\rm in}}$ and 
$k_\perp^2 \mathcal{J}_{\rmT,{\rm el}}$ are now plotted as functions of $k_\perp^2$,
and not of $k_\perp$.
Fig.~\ref{fig:el-inel_a} shows the respective predictions of the MV model 
for $Q_{s0}^2=2$~GeV$^2$ and 3 values different for the effective virtuality $\bar Q^2$, namely
$\bar Q^2/Q_{s0}^2=0.01,\, 0.05,\,0.1$.  Fig.~\ref{fig:el-inel_a} illustrates
the effects of the high-energy evolution up to $\eta =3$, for 
$k_\perp^2\lesssim 3Q_s^2(\eta=3)$ and  a fixed ratio $\bar Q^2/Q_{s0}^2=0.05$.

We have already discussed the results for the elastic cross-section in some detail,
in relation with Figs.~\ref{fig:lowkT} and \ref{fig:BKlowkT}. So, let us now focus on the
inelastic contributions. The respective MV-model curves in Fig.~\ref{fig:el-inel_a} show the 
same qualitative behaviour as the analytic estimate in \eqn{JTbroad} (multiplied by
$k_{\perp}^2$): a rapid
increase at low $k_{\perp}^2\ll Q_s^2$, followed by a wide peak around $Q_s^2$, and finally a
slow decrease at larger values $k_{\perp}^2>Q_s^2$.   Also, when increasing
$\bar Q^2$, the position of the peak remains unchanged --- as it should, since this is located at
$k_\perp^2=Q_s^2$ --- but its height is decreasing, due to the reduction in the logarithmic 
factor $\ln({Q_s^2}/{\bar{Q}^2})$ in \eqn{JTbroad}. Notice the difference between
this behaviour and that of the elastic cross-section, for which both the height and the position
of the peak are changing with $\bar{Q}$, as discussed in relation with 
Fig.~\ref{fig:lowkT} (and also visible in Fig.~\ref{fig:el-inel_a}).

This general trend is preserved by the BK evolution up to $\eta=3$, as shown in  Fig.~\ref{fig:el-inel_b}.
Due to the increase in the saturation momentum with $\eta$, cf. Fig.~\ref{fig:Qseta}, the
position of the peak is (slightly) moving towards larger values of $k_\perp^2$, for both the
elastic and the inelastic cross-sections. Also, the magnitude of the cross-sections is
growing with $\eta$, albeit quite slowly and only at sufficiently large transverse momenta
 $k_\perp^2\gtrsim Q_s^2(\eta)$. This growth is more important
 for the inelastic channel, where it can be attributed
to the corresponding increase in the logarithm $\ln({Q_s^2(\eta)}/{\bar{Q}^2})$, cf. \eqn{JTinlow}.

%

\subsection{$k_{\perp}^2 \gg Q_s^2$: hard scattering and BK evolution}
\label{sec:highkT}

%

When the transverse momentum $k_\perp$ of the produced quark is much larger than 
the target saturation momentum,  $k_\perp^2\gg Q_s^2$, the physical picture of SIDIS  
is quite different: the (elastic and inelastic) cross-sections are controlled by
relatively small dipoles, with transverse sizes $r,\,r'\ll 1/Q_s$, which scatter only weakly off the
hadronic target: $T(r)\ll 1$. An immediate consequence of that is that the elastic cross-section,
which is quadratic in $T$ (cf. \eqn{JTel}), is even more strongly suppressed relative to the 
inelastic one, which in turn is dominated by the first 3 terms, linear in $T$, in \eqn{JTinel}. 
That said, gluon saturation may still influence SIDIS even for such high transverse momenta, 
via its consequences 
(like {\it geometric scaling}  \cite{Stasto:2000er,Iancu:2002tr,Mueller:2002zm}) 
for the high-energy evolution of the dipole scattering amplitude.

In this section, we shall consider the situation where $k_\perp$ is much larger than both
$Q_s$ and $\bar Q$, without any strong ordering between the two latter scales 
(this ordering becomes irrelevant when $k_\perp^2\gg Q_s^2,\,\bar Q^2$). 
We shall provide analytic estimates in two limiting situations: for relatively small
energies, where we shall rely on the MV model, and for very high energies, 
where we shall use an analytic approximation to the solution to the BK equation. 
These estimates will be completed with
numerical solutions to the BK equation, which also apply to intermediate energies.

\subsubsection{Hard scattering in the MV model}

To compute the total (or inelastic) cross-section for $k_\perp^2\gg Q_s^2,\,\bar Q^2$ within
the MV model, one can use the single scattering approximation to the dipole $S$-matrix \eqref{SMV}, that is, 
 \beq\label{MVT0}
 T_0\equiv 1-S_0\simeq \frac{r^2Q_{A}^2}{4}\ln\frac{1}{r^2\Lambda^2}\,,
 \eeq
together with the small-argument expansion of
the Bessel function: $ \rmK_1(z)\simeq 1/z$ for $z\ll1$.  Under these assumptions, 
the dominant contribution to  $\mathcal{J}_{\rmT}$  
(the only one to be enhanced by a large transverse logarithm)
is that generated by the last term in
\eqn{JT}, i.e. $T(\br \!-\! \br')$, which refers to the scattering of the 
produced quark. (This term enters \eqn{JT} with a negative sign, but it actually gives a positive
contribution to the FT; see below.) Indeed, for this term, the dipole sizes $r$ and $r'$ 
can be as large as $1/\bar Q$, since it is only their difference $\bm{\rho}=\br \!-\! \br'$
which is constrained by the Fourier phase to small values $\rho\lesssim 1/k_\perp$.  
Changing integration variables from $\br,\,\br'$ to  $\br,\,\bm{\rho}$, the integral 
over $r$ is logarithmic within the interval $\rho\ll r\ll 1/\bar Q$ (recall \eqn{logint}),
and one finds
\begin{align}
\label{JTsingle}
\mathcal{J}_{\rmT} 
& \simeq  
\frac{1}{16\pi^2}
\int \dif^2\bm{\rho}\, \rme^{-\rmi \bk \cdot \bm{\rho}} \,\ln \frac{1}{\rho^2\bar{Q}^2}
\left(- \frac{\rho^2Q_{A}^2}{4}\ln\frac{1}{\rho^2\Lambda^2}
\right)\nonumber\\*[.2cm]
&\simeq \,\frac{1}{4\pi}\,\frac{Q_{A}^2}{k_\perp^4}\left(\ln \frac{k_\perp^2}{\Lambda^2}
+\ln \frac{k_\perp^2}{\bar{Q}^2}\right)\,
\quad \textrm{for} \quad  k_{\perp}^2\gg \bar{Q}^2, Q_s^2.
\end{align}
The expression in the first line is recognised as the single-scattering
approximation to the first line of  \eqn{JTinlow}. The obtention of the final result
in the second line
is a bit subtle and requires the more elaborated calculations presented in Appendix~\ref{sec:mv}.
Here, we shall rather focus on its physical origin. 

Notice first that, for the kinematics
at hand, the integral over $\rho$ is controlled by the Fourier phase, which selects
$\rho\sim 1/k_\perp$. One may naively think that the dominant contribution can be
obtained by replacing $\rho^2\to 1/k_\perp^2$ within both logarithms inside the integrand:
such a contribution would be {\it quadratic} in the transverse logarithms and hence 
parametrically larger than the {\it linear} combination shown
in the second line of  \eqn{JTinlow}.  However, if one does so, then the ensuing double 
logarithm gets multiplied by an integral over $\bm{\rho}$ which vanishes
when $k_\perp > 0$ (since its result is proportional to $\nabla^2_{\bk}\delta^{(2)}(\bk)$).
So, it is natural to find a result which is only {\it linear} in the transverse logs, and also 
{\it symmetric} under the exchange $\bar Q^2\leftrightarrow \Lambda^2$, like the integrand
in the first line.

Consider now the physical interpretation of the two terms in the final result. We start
with the second one, whose origin is quite transparent. 
As in the case of \eqn{JTinlow}, the $\ln({k_\perp^2}/{\bar{Q}^2})$ 
comes from integrating over the initial $k_\perp$--distribution 
of the quark produced by the decay of $\gamma_\rmT^*$,
 \eqn{JTellow}. As for the associated power tail $\propto 1/k_\perp^4$, this
is generated by a hard single scattering between the quark and a nucleon from the
target, with a transferred momentum $q_\perp\simeq k_\perp$. 

Concerning the first term in \eqn{JTsingle},
it is quite clear that the logarithm $\ln({k_\perp^2}/{\Lambda^2})$ is directly inherited from
the dipole amplitude \eqref{MVT0}: this is a {\it Coulomb} logarithm, generated via a relatively
soft exchange, within the range $\Lambda^2\ll q_\perp^2\ll k_\perp^2$. Accordingly, the power-like
tail $\propto 1/k_\perp^4$ for this term should somehow be associated with the decay of the virtual
photon. This may look surprising since, as discussed in Sect.~\ref{sect:LCWV},
the $\gamma_\rmT^*$ decay is expected to produce a much softer tail $\propto 1/k_\perp^2$, cf. \eqn{JTellow}.
Yet, as shown by the calculation in Appendix~\ref{sec:mv}, a $1/k_\perp^4$ 
tail is in fact generated by the {\it difference}
$\frac{1}{(\bk\minus\bq)^2}-\frac{1}{k_\perp^2}$ between the two such distributions,
 with and without soft rescattering in the final state.

 In principle, \eqn{JTsingle} should be used in the case where
 $\bar Q^2\gg \Lambda^2$, to justify the use of a perturbative (partonic) description
for the $\gamma^*$ wavefunction; in such a case, the first logarithm in the final result would dominate
over the second one. In practice though, the difference between $\bar Q^2$ and  $\Lambda^2$ may 
not be that important, which explains why we have kept both logarithms in  \eqn{JTsingle}. 

For completeness, let us also observe that the elastic cross-section $\mathcal{J}_{\rmT, {\rm el}}$
in the MV model and in the single scattering approximation can be easily estimated from
Eqs.~\eqref{JTel} and \eqref{MVT0}. One thus finds $\mathcal{J}_{\rmT, {\rm el}}\sim Q_{A}^4/{k_\perp^6}$
(see Appendix~\ref{sec:elastic} for an explicit calculation and \eqn{Wlim1} for the final result).
This is suppressed  by a power of $ Q_{A}^2/k_\perp^2\ll 1$
compared to the inelastic piece (a ``higher twist effect''), as expected.

\subsubsection{Hard scattering at very high energies}

We now turn to the high-energy limit, where the asymptotic solution to the BK equation in the regime 
$r Q_s \ll 1$ is known as \cite{Mueller:2002zm}
\begin{align}
	\label{tbkdil}
	T(r) \simeq (r^2 Q_s^2)^\gamma \ln 
	\frac{1}{r^2 Q_s^2}
	\quad \textrm{for} \quad r Q_s \ll 1,
\end{align}
where the ``anomalous dimension''  $\gamma \simeq 0.63$ is significantly smaller than 1.  It is here understood
that $Q_s\equiv Q_s(A,\eta)$ depends upon the rapidity $\eta=\ln(x_0/x_g)$, assumed to be large:
$\abar\eta \gg 1$. \eqn{tbkdil} shows {\it geometric scaling} \cite{Stasto:2000er,Iancu:2002tr,Mueller:2002zm}: 
it depends upon the 2 independent kinematical 
variables $r$ and $\eta$ (and also upon the nucleon number $A$) only via the product $r^2 Q_s^2(A,\eta)$.
This property is an imprint of the physics of saturation on the behaviour of the amplitude in the weak
scattering regime at $rQ_s\ll 1$.
 
Clearly, the asymptotic behaviour at high energies is not precisely the regime to be explored via
DIS at EIC, where only moderate values $\eta\lesssim 3$ should be accessible. Here, we shall
rather use  \eqn{tbkdil} as a convenient {\it Ansatz} which interpolates 
(when decreasing $\gamma$ from 1 to 0.63) between the MV model at low 
energy, cf.  \eqn{MVT0}, and the prediction of the BK equation at very high energy.

Given this formal similarity between the BK amplitude in \eqn{tbkdil} and the MV amplitude in \eqn{MVT0},
it is clear that the argument previously used in relation with \eqn{JTsingle}  can be taken
over to the present situation.  The proper generalisation of the first line in  \eqn{JTsingle}  reads
\begin{align}
\label{JTasymp}
\mathcal{J}_{\rmT} 
\simeq  
\frac{1}{16\pi^2}
\int \dif^2\bm{\rho}\, \rme^{-\rmi \bk \cdot \bm{\rho}} \,\ln \frac{1}{\rho^2\bar{Q}^2}
\left(- (\rho^2Q_s^2)^\gamma\,
\ln\frac{1}{\rho^2Q_s^2}
\right).
\end{align}
However, since the power $\gamma$ is strictly smaller than one, the mathematical
treatment of \eqn{JTasymp} is considerably simpler. Namely, it is now legitimate to replace
$\rho^2\to 1/k_\perp^2$ within the arguments of both logarithms inside the integrand,
since the large contribution thus obtained 
has a non-zero coefficient (given by the integral \eqref{int4} in Appendix~\ref{sec:int}).
One thus finds
\begin{align}
\label{JTbkhigh}
\mathcal{J}_{\rmT} \simeq 
\mathcal{J}_{\rmT,{\rm in}} \simeq 
\frac{4^{\gamma} \gamma  \Gamma(1+\gamma)}{\Gamma(1-\gamma)} 
\frac{1} {8\pi k_{\perp}^2}
\left(\frac{Q_s^2}{k_{\perp}^2}\right)^\gamma 
\ln \frac{k_{\perp}^2}{\bar{Q}^2}
\ln \frac{k_{\perp}^2}{Q_s^2}
\quad \textrm{for} \quad k_{\perp}^2\gg \bar{Q}^2, Q_s^2.
\end{align}
The longitudinal cross section $\mathcal{J}_{\rmL,{\rm in}}$ can be similarly estimated: one finds the same power-like behaviour as in Eq.~\eqref{JTbkhigh}, but with a single logarithmic enhancement.
Hence,  Eq.~\eqref{JTbkhigh} provides the dominant contribution ---
in the sense of a leading logarithmic accuracy ---  to SIDIS in this high-$k_\perp$ regime,
Comparing this result to the low energy behaviour shown in \eqn{JTsingle}, we observe that the
the high-energy evolution in presence of saturation produces a softer tail at large transverse momenta.
Also,  Eq.~\eqref{JTbkhigh} shows no special scaling, like geometrical scaling, due to the presence
of {\it two} external scales, $\bar Q^2$ and $Q_s^2$, in the final result
(rather than only one such a scale in the dipole amplitude \eqref{tbkdil}). We shall
return to this point in the numerical analysis to follow.


For completeness, let us also show the corresponding result for the 
elastic cross section, as obtained via similar manipulations starting with Eq.~\eqref{JTel}:
\begin{align}
\label{JTelbkhigh} 
\mathcal{J}_{\rmT,{\rm el}} \simeq 
\frac{4^{2\gamma} \Gamma^2(1+\gamma)}{\Gamma^2(1-\gamma)} 
\frac{1} {4\pi k_{\perp}^2}
\left(\frac{Q_s^2}{k_{\perp}^2}\right)^{2\gamma}
\ln^2 \frac{k_{\perp}^2}{Q_s^2}
\quad \textrm{for} \quad k_{\perp}^2\gg \bar{Q}^2, Q_s^2.
\end{align}
As anticipated, this is strongly suppressed as  compared to the inelastic part in Eq.~\eqref{JTbkhigh},
due to the larger power of ${Q_s^2}/{k_{\perp}^2}$: $2\gamma$ instead of $\gamma$.

One interesting conclusion which emerges from the calculations in Sects.~\ref{sec:medkT} and 
\ref{sec:highkT} is that, for all transverse momenta $k_\perp\gtrsim
Q_s$ and for any energy (from $\eta\simeq 0$, where the MV model applies, 
up the very large $\eta\gg 1/\abar$, where one can use the asymptotic solution to the BK equation),
the dominant contribution to SIDIS at low virtuality ($\bar Q^2\ll Q_s^2$) is generated
by the scattering of the measured quark {\it alone} (as opposed to the scattering of the quark-antiquark
dipole).  Of course, the scattering of the whole dipole does matter {\it in general} --- this is represented
by the terms $T(\br)$ and $T(\br')$ in \eqn{JT} ---, but its contribution turns out to be (parametrically) 
less important than that from the measured quark alone --- the piece involving $T(\br \!-\! \br')$ in \eqn{JT}.

\begin{figure}
  \centering
  \begin{subfigure}[t]{0.48\textwidth}
    \includegraphics[width=\textwidth]{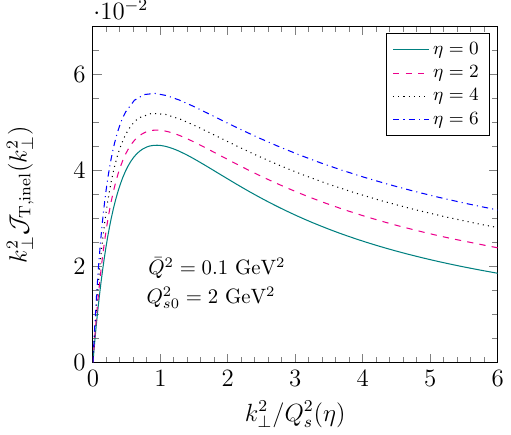}
    \caption{} \label{fig:highkt_a}
  \end{subfigure}
  \hfill
  \begin{subfigure}[t]{0.48\textwidth}
    \includegraphics[width=\textwidth]{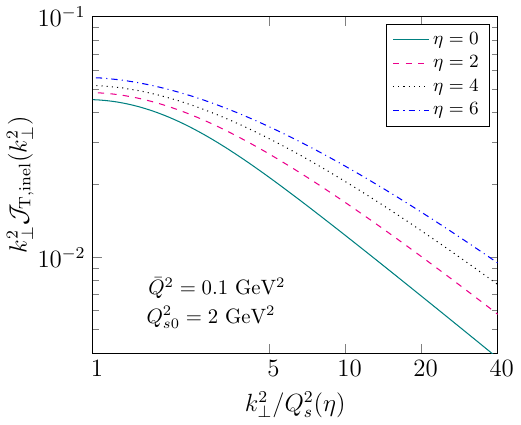}
    \caption{}\label{fig:highkt_b}
  \end{subfigure}
  \caption{\small Approximate geometric scaling behaviour of the transverse inelastic cross section multiplied by $k_\perp^2$ at (a) low and intermediate transverse momenta and (b) high transverse momenta. With increasing $\eta$ the high-$k_{\perp}$ tail becomes slightly softer.}\label{fig:highkt}
\end{figure}

 \subsubsection{Numerical results for $k_{\perp}^2 \gtrsim Q_s^2$} 

In Fig.~\ref{fig:highkt} we show numerical results for the quantity $k_{\perp}^2\mathcal{J}_{\rmT,{\rm in}}$
at different rapidities $\eta\le 6$ in terms of the dimensionless variable $k_{\perp}^2/Q_s^2(\eta)$, up
to relatively large values $k_{\perp}^2\gg Q_s^2(\eta)$. In the left plot, we use a linear scale and
show all the transverse momenta up to $k_{\perp}^2/Q_s^2(\eta) = 6$, in order to exhibit
the global shape of the
function $k_{\perp}^2\mathcal{J}_{\rmT,{\rm in}}$. In the right plot, we use a logarithmic
scale and show only the high momentum regime at $1 < k_{\perp}^2/Q_s^2(\eta) < 40$.
Note that after multiplication by $k_{\perp}^2$, the inelastic cross-section $\mathcal{J}_{\rmT,{\rm in}}$
is very close to a {\it scaling function} (see e.g. Eqs.~\eqref{JTbroad}, \eqref{JTsingle} and
\eqref{JTbkhigh}) --- that is, a function of the ratio  $k_{\perp}^2/Q_s^2(\eta)$. This scaling behaviour
is however not exact (not even in our analytic approximations): it is broken by logarithmic factors.
For instance, \eqn{JTinlow}, which we recall is valid when $k_\perp\sim Q_s$,
 involves an overall factor $\ln ({Q_s^2(\eta)}/{\bar{Q}^2})$, which increases with $\eta$,
 due to the respective behaviour of the saturation momentum. Similarly, at larger momenta
 $k_\perp^2\gg Q_s^2$, one can rely on \eqn{JTbkhigh} to deduce that the function
$k_{\perp}^2\mathcal{J}_{\rmT,{\rm in}}$ can be written as the sum of a scaling
function and a scaling-violating term proportional to $\ln ({Q_s^2(\eta)}/{\bar{Q}^2})$.
(To that aim, one should rewrite one of the logarithms in \eqn{JTbkhigh} as
$\ln ({k_{\perp}^2}/{\bar{Q}^2}) = \ln ({k_{\perp}^2}/{{Q}_s^2}) + \ln({Q_s^2}/{\bar{Q}^2})$.)
These analytic arguments suggest that, when plotted in terms of the scaling
variable $k_{\perp}^2/Q_s^2(\eta)$, the function
$k_{\perp}^2\mathcal{J}_{\rmT,{\rm in}}$ should still increase with $\eta$, 
albeit only slightly. This is in agreement with the numerical findings in Fig.~\ref{fig:highkt_a}.

Another interesting feature of the numerical results, better seen in Fig.~\ref{fig:highkt_b}, is the
emergence of an anomalous dimension $\gamma < 1$ as a consequence of the high-energy
evolution: indeed, the high-$k_\perp$ tails corresponding to $\eta\gtrsim 4$ are clearly seen
to be softer (i.e. to show a slower decrease with increasing $k_{\perp}^2$) then the MV model
curve, for which $\gamma=1$. That said, the rapidities $\eta\le 6$ considered here are still too small
to feature the asymptotic expectation $\gamma\simeq 0.63$. Besides, the numerical extraction
of this power $\gamma$ is further hindered by the additional logarithmic dependences, as visible
in \eqn{JTbkhigh}.

%

\subsection{Cronin peak}
\label{sec:cronin} 

 As an application of the previous developments in this section, let us discuss
 an interesting physical consequence of the multiple
 scattering and the associated transverse momentum broadening, that might be measured at the
 EIC. Namely, let us compute the nuclear modification factor $R_{pA}$ for the SIDIS process,
 that is, the ratio between particle production (with a given $z$ and $k_\perp$) in $eA$ and
 respectively $ep$ collisions, with the latter scaled up by the number of participants. 
 In the approximations at hand, this is obtained as 
 \beq\label{RpA}
 R_{pA}(z,\bk)\,=\,\frac{\frac{\dif\sigma^{\gamma^*\!A \to q X}}{\dif z\, \dif^2\bk}}
 {A \frac{\dif\sigma^{\gamma^*p \to q X}}{\dif z\, \dif^2\bk}}\,
 \simeq\,\frac{\mcal{J}_\rmT^A(z,\bk)}{A^{1/3}\mcal{J}_\rmT^p(z,\bk)}\,,
 \eeq
 where in the second equality we have restricted ourselves to the transverse sector, since this
 dominates in the interesting kinematical regime. Namely we are interested
 in transverse momenta of the order of the saturation momentum in the nucleus: $k_\perp\sim Q_s(A)$.

\begin{figure}
  \centering
  \begin{subfigure}[t]{0.48\textwidth}
    \includegraphics[width=\textwidth]{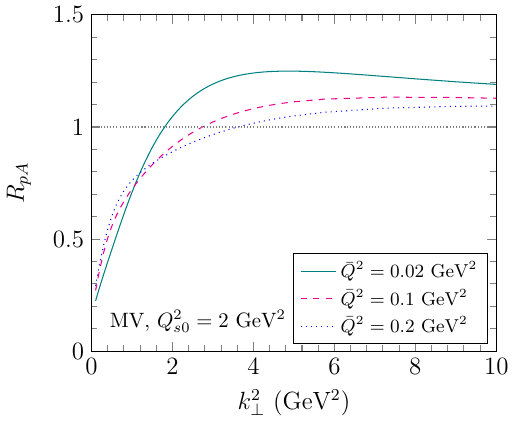}
    \caption{}\label{fig:Cronin_a} 
  \end{subfigure}
  \hfill
  \begin{subfigure}[t]{0.48\textwidth}
    \includegraphics[width=\textwidth]{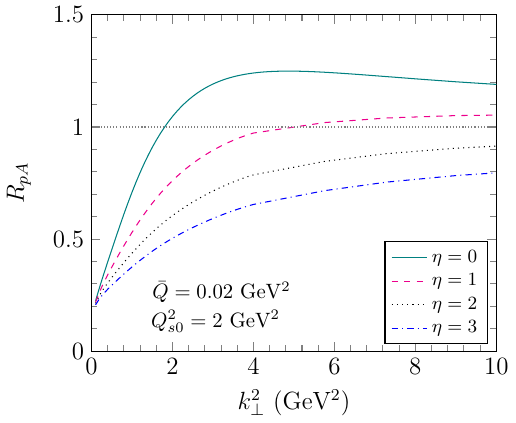}
    \caption{}\label{fig:Cronin_b}
  \end{subfigure}
  \caption{\small The ratio $R_{pA}$ defined in Eq.~\eqref{RpA}, plotted as a function of $k_{\perp}^2$.
  (a) The respective MV model predictions
   for 3 different values of $\bar{Q}^2$. For sufficiently small $\bar{Q}^2$ the ratio goes well above 1 at mid transverse momenta. (b) Rapidity evolution of the results from the left plot corresponding to
   $\bar{Q}^2=0.02$~GeV$^2$. 
   After a couple of rapidity steps the ratio stays below 1 for all values of $k_\perp^2$ under
   consideration.}\label{fig:Cronin}
\end{figure}

 For an analytic estimate of \eqn{RpA} at relatively low energies (say, $x_g\sim 0.01$), we can use
 the MV model for both types of targets, with $Q_A^2= A^{1/3}Q_p^2$. Specifically, for 
  $k_\perp\sim Q_s(A)$, the saturation effects are important for the nuclear target, for which one
  should use the full MV result in \eqn{JTbroad}, but they are not important for the proton target,
for which one should rather use the high-$k_\perp$ approximation in \eqn{JTsingle}.
To make the argument clearer, let us focus on the physically interesting regime at $\bar Q^2\gg
\Lambda^2$; in that case, one can neglect the second logarithm in the final result in \eqn{JTsingle},
which allows us to write
 \beq\label{RpAMV}
 R_{pA}(k_\perp)\,\simeq\,\frac{\ln ({Q_s^2(A)}/{\bar{Q}^2})}{\ln ({k_\perp^2}/\Lambda^2)}\,
 \frac{ k_\perp^4 \,\rme^{-{k_\perp^2}/{Q_s^2(A)}}}{A^{1/3} Q_{p}^2\, Q_s^2(A)}\
  \underset{k_\perp\sim Q_s(A)}{\longrightarrow}\ \ln\frac{Q_s^2(A)}{\bar Q^2}  \,>\,1,
 \eeq
 where the last estimate holds for $k_\perp\sim Q_s(A)$. (We have also used $Q_A^2= A^{1/3}Q_p^2$
 together with \eqn{QsMV} for the nuclear saturation momentum.)
 The first equality in \eqn{RpAMV} also shows that $R_{pA}(k_\perp)$ increases with  $k_\perp$
 so long $k_\perp< Q_s(A)$, but it decreases at larger momenta $k_\perp\gg Q_s(A)$.
 Hence, \eqn{RpAMV} shows a peak at $k_\perp\simeq Q_s(A)$ and the height of this peak
 is expected to be larger than one in the limit where $Q_s^2(A)\gg  \bar Q^2$. This is similar
to the Cronin peak observed in hadron production at midrapidities in d+Au collisions at RHIC
\cite{Arsene:2004ux,Adams:2006uz}. In that context too, this enhancement  
can be explained by gluon saturation in the nuclear wavefunction and its consequences in terms
of multiple scattering \cite{Kharzeev:2002pc,Baier:2003hr,
Albacete:2003iq,Kharzeev:2003wz,Iancu:2004bx,Blaizot:2004wu}.

The existence of such a peak at the level of the MV model is numerically studied in
Fig.~\ref{fig:Cronin_a}. This figure shows the $R_{pA}$ ratio in \eqn{RpA} as a function of $k_\perp^2$
(up to $5Q_s^2=10$~GeV$^2$) and for 3 values,  $\bar Q^2/Q_s^2=0.01,\, 0.05,\,0.1$,
for the ratio between the relevant scales. A mild peak, centered at $k_\perp^2$ slightly above
 $Q_s^2$, is indeed seen in the case $\bar Q^2/Q_s^2=0.01$, but not also in the 2 other cases.
A perhaps more robust prediction, valid for all the 3 cases under investigation, is that the ratio 
$R_{pA}(k_\perp)$ increases quite fast at low momenta until it reaches a value of order 1 for
$k_\perp\sim Q_s$. For larger $k_\perp> Q_s$, it shows either a mild peak, or a plateau with height
$R_{pA} > 1$. The rapid increase at low $k_\perp\ll Q_s$ is a consequence of gluon saturation in
the nucleus, as quite clear by inspection of the calculation in \eqn{RpAMV}.

\begin{figure}
  \centering
  \begin{subfigure}[t]{0.48\textwidth}
    \includegraphics[width=\textwidth]{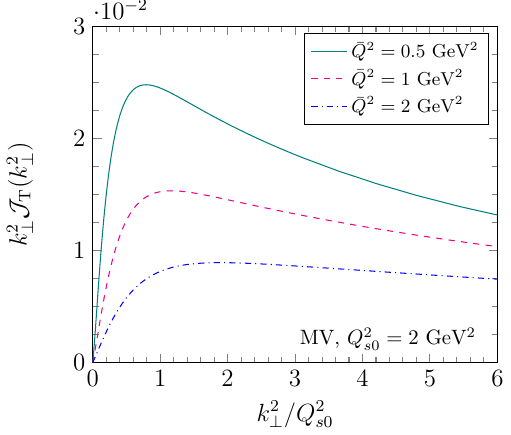}
    \caption{}\label{fig:largeQbar_a} 
  \end{subfigure}
  \hfill
  \begin{subfigure}[t]{0.48\textwidth}
    \includegraphics[width=\textwidth]{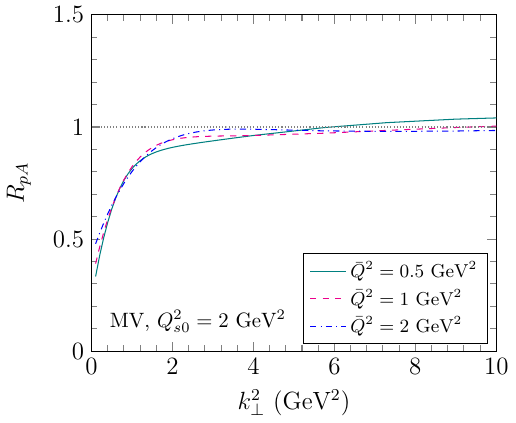}
    \caption{}\label{fig:largeQbar_b}
  \end{subfigure}
  \caption{\small Extension of some of the previous results to larger values of $\bar{Q}^2$
  for fixed $ Q_{s0}^2=2$~GeV$^2$, namely
   $\bar{Q}^2/Q_{s0}^2=0.25,\,0.5$ and 1 (within the framework of the MV model).
  (a) The rescaled cross-section $k_\perp^2 \mathcal{J}_{\rmT}$ 
   plotted as a function of $k_{\perp}^2/Q_{s0}^2$.
   (b)The ratio $R_{pA}$ defined in Eq.~\eqref{RpA}, plotted as a function of $k_{\perp}^2$.}
   \label{fig:largeQbar}
\end{figure}

From the lessons at RHIC, we also know that it is interesting to study the evolution of  the $R_{pA}$ ratio 
with increasing rapidity. Fig.~\ref{fig:Cronin_b} shows our respective predictions for the most
favourable case $\bar Q^2/Q_s^2=0.01$, as obtained from 
numerical solutions to the BK equation up to $\eta=3$. Remarkably, one can see that the peak
flattens out already after one step in rapidity and that the ratio  $R_{pA}$ remains strictly below 1
 at any  $k_\perp^2\le 10$~GeV$^2$ for $\eta\ge 2$. A similar behaviour was experimentally observed
 in d+Au collisions at RHIC \cite{Arsene:2004ux,Adams:2006uz}
 and theoretically explained (within the CGC framework)
 in Refs.~\cite{Albacete:2003iq,Kharzeev:2003wz,Iancu:2004bx}. Clearly, the same
 physical mechanism is working in the present set-up: for the values of $k_\perp$ considered
 in Fig.~\ref{fig:Cronin_b}, the nucleus is still close to saturation, hence it evolves only slowly, unlike
 the proton, which is in a dilute regime, so its gluon distribution rises much faster with increasing $\eta$.
 This difference transmits to the respective dipole amplitudes and hence to the two cross-sections
 in the numerator and the denominator of  \eqn{RpA}, respectively.

{
So far in this section, we have focused on the well-separated regime at $\bar Q^2\ll Q_s^2$, 
where the saturation effects are more striking and which allowed for 
analytic approximations.  In view of the phenomenology, it is however important to notice
that the saturation effects should remain important when the two scales are comparable to each other,
$\bar Q^2\sim Q_s^2$,  since the corresponding dipole sizes are large ($r\sim 1/Q_s$ 
when $k_\perp^2 \lesssim  Q_s^2$), hence the unitarity 
corrections are of $\mathcal{O}(1)$. To illustrate this point, we have repeated some of our calculations
for larger values of $\bar Q^2$, namely $\bar Q^2/Q_s^2=0.25,\,0.5$ and 1, with the results shown
in Fig.~\ref{fig:largeQbar} (for the MV model, for definiteness).
As visible in these plots, the saturation effects --- a maximum in the rescaled cross-section  $k_\perp^2 \mathcal{J}_{\rmT}$ near $k_\perp^2 =  Q_s^2$ (cf. Fig.~\ref{fig:largeQbar_a}) and 
a modest enhancement ($R_{pA}\gtrsim 1$) in the $R_{pA}$ ratio at $k_\perp^2 >  Q_s^2$ (cf. Fig.~\ref{fig:largeQbar_b}) --- are clearly seen for $\bar Q^2=0.25Q_s^2$, but smoothly
disappear when further increasing  $\bar Q^2$. Such a behaviour is indeed consistent with
our previous results, notably those appearing in Eqs.~\eqref{JTbroad} and \eqref{RpAMV}.
}

\section{$k_{\perp}$-integrated SIDIS and saturation}
\label{sect:intk}

In the previous section, we have studied the differential cross-section 
$\frac{\dif\sigma^{\gamma^* \!A \to q X}}{\dif z\, \dif^2\bk}$ for producing a quark with a given
longitudinal momentum fraction $z$ and a given transverse momentum $\bk$ in DIS processes with
relatively low effective virtualities $\bar Q^2=z(1-z)Q^2\ll Q_s^2$. 
Since this double differential quantity may be difficult to
measure in the actual experiments, in what follows we shall discuss its more inclusive version
${\dif\sigma}/{\dif z}$ where the transverse momentum distribution of the
produced quark is integrated out --- only its longitudinal fraction $z$ is measured. We shall study in more
detail two kinematical limits: \texttt{(i)} $\bar Q^2\ll Q_s^2$, which is the most interesting regime
for a study of saturation, and  \texttt{(ii)} $\bar Q^2\gg Q_s^2$, where saturation can manifest itself
{\it indirectly}, via the phenomenon of geometric scaling. But before that, let us briefly list the
general formulae valid for any kinematics.

\subsection{General expressions}

Within our current formalism, the fact of integrating out $\bk$ simply amounts to identifying the transverse coordinates, $\br$ and $\br'$, of the measured quark in the DA and in the CCA, respectively. So, 
it is very easy to deduce the general respective expressions for the elastic, inelastic and total cross-sections.
To avoid a proliferation of formulae, we consider the transverse sector alone, for which we find 
(for the reduced cross-sections defined in Eqs.~\eqref{JL}--\eqref{JT})
\begin{align}
\label{intJel} 
	\int \dif^2 \bk\, \mcal{J}_{\rmT,{\rm el}} = 
	\bar{Q}^2
	\int \frac{\dif^2 \br}{4\pi} \,
	\rmK_1^2(\bar{Q}r) \,|T(\br)|^2,
\end{align}
for the elastic piece,
\begin{align}
\label{intJin} 
	\int \dif^2 \bk\, \mcal{J}_{\rmT,{\rm in}} = 
	\bar{Q}^2
	\int \frac{\dif^2 \br}{4\pi} \,
	\rmK_1^2(\bar{Q}r) \left[T(\br) + T(-\br)- |T(\br)|^2\right],
\end{align}
for the inelastic one, and, finally,
\begin{align}
\label{intJT}
	\int \dif^2 \bk\, \mcal{J}_\rmT = 
	\bar{Q}^2
	\int \frac{\dif^2 \br}{4\pi} \,
	\rmK_1^2(\bar{Q}r)  \left[T(\br) + T(-\br)\right],
\end{align}
for the total cross-section. We have also used the fact that $T(-\br)=T^*(\br)$, as it can be
easily checked by using $T=1-S$ together with the definition \eqref{dipole}
of the dipole $S$-matrix\footnote{$T$ and $S$ are real for all the explicit calculations in this paper,
but in general they can include an imaginary part describing the exchange of an object
with negative charge parity (``odderon''). In pQCD, such $C$-odd exchanges require al least 
three gluons; see e.g. the discussions in \cite{Kovchegov:2003dm,Hatta:2005as}.}
. The $T$-matrix structure in \eqn{intJin}  can be equivalently written as
\beq
T(\br) + T(-\br)- |T(\br)|^2 \,=\,1 - |S(\br)|^2\,,
\eeq
where the r.h.s. is recognised as the probability for the dipole with size $\br$ to suffer
an inelastic scattering (recall that $|S(\br)|^2$ is the respective survival probability).
Similarly, \eqn{intJT} involves $T(\br) + T(-\br)=2 {\rm Re}\,T(\br)$, which is the expected
prediction of the optical theorem for the dipole total cross-section per unit impact parameter.

As  a further check, let us observe that by integrating \eqn{intJT} over $z$ one
recovers the expected, leading-order, result for the fully inclusive DIS cross-section in the dipole
factorisation  (here, for a transverse photon and for one flavour of massless quark):
\begin{equation}
\label{sigmat}
	\sigma_{\gamma_{\rmT}^*A} (x,Q^2)= e_q^2 \frac{\alpha_{\rm em} N_c}{2\pi^2}\,
			\int_0^1 \dif z\int \dif^2\br\,
	 \left[z^2+(1-z)^2\right] \bar{Q}^2
\rmK_1^2\big(\bar{Q}r\big)\,\sigma_{\rm dipole}(\eta, r),
	\end{equation}
with $\sigma_{\rm dipole}(\eta, r)=2\pi R_A^2T(\eta, r)$ when $T$ is real (as generally in this paper).

\subsection{Low virtuality $z(1-z)Q^2\ll Q_s^2$: the black disk limit}

It should be clear by now that the physics of strong scattering or gluon saturation can be most
directly investigated by focusing on the low-virtuality regime at $\bar Q^2\ll Q_s^2$. In this regime, 
all the cross-sections shown in Eqs.~\eqref{intJel}--\eqref{intJT} are dominated by large dipoles 
such that $1/ Q_s \ll r \ll 1/ \bar{Q}$, for which one can use the black disk limit $T(r)=1$ together
with the expansion of the Bessel function for small values of its argument. Indeed, this intermediate
range in $r$ yields a logarithmically enhanced contribution, which is easily found as 
\begin{align}
\label{intJlow}
	\int \dif^2 \bk\, \mcal{J}_\rmT \simeq 
	2 \int \dif^2 \bk\, \mcal{J}_{\rmT,{\rm el}} \simeq
	2 \int \dif^2 \bk\, \mcal{J}_{\rmT,{\rm in}} \simeq  
	\frac{1}{2} \ln \frac{Q_s^2}{\bar{Q}^2}
	\quad \textrm{for} \quad \bar{Q}^2 \ll Q_s^2.
\end{align}
Since obtained by invoking the black disk limit, these results are almost
insensitive to the details of the QCD scattering. In particular, they are independent
of the QCD coupling $\alpha_s$, except for the weak (logarithmic) dependence introduced
by the scale $Q_s^2$ in the argument of the logarithm (recall that $Q_s^2\sim \alpha_s^2$). 
To better understand the physical origin of these results, 
it is instructive to re-derive them by integrating over $\bk$ the double-differential 
cross-sections computed in Sect.~\ref{sect:qllqs}.

Concerning the elastic cross-section, it is quite clear that the respective logarithm in \eqn{intJlow}
is generated by integrating the $k_\perp$--distribution in \eqn{JTellow} over
the range $\bar{Q}^2 \ll k_{\perp}^2 \ll Q_s^2$. More precisely, the MV estimate in \eqn{JTelGBW} 
shows that the logarithmic integration over $k_\perp^2$ is cut off by multiple scattering already
at $k_\perp^2\sim Q_s^2/2$. Hence the corresponding  estimate in  \eqn{intJlow}, although correct
to leading logarithmic accuracy, is expected to be larger than the actual result.

Consider now the inelastic cross-section: in this case, the logarithm  $\ln({Q_s^2}/{\bar{Q}^2})$ 
is already visible in the differential distribution in the second line of  \eqn{JTinlow}, or in its MV
model estimate \eqref{JTbroad}. By integrating any of these two expressions over all $k_\perp$, 
one finds exactly the result displayed in \eqn{intJlow}. Of course, the Gaussian broadening
in  \eqn{JTbroad} only holds within a limited range of momenta around $Q_s$, but this is precisely
the range which dominates the $k_\perp$--integration to the accuracy of interest\footnote{One can
easily check that the power-like tail of the distribution at $k_\perp\gg Q_s$, e.g. $\mathcal{J}_{\rmT} \propto
Q_s^2/k_\perp^4$ in the MV model, merely yields a contribution of order 1 to the integrated cross-section.}.

%

\begin{figure}
  \centering
  \begin{subfigure}[t]{0.48\textwidth}
    \includegraphics[width=\textwidth]{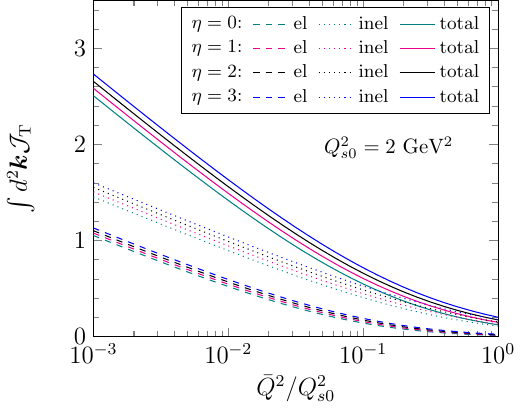}
    \caption{}\label{fig:int_a}
  \end{subfigure}
  \hfill
  \begin{subfigure}[t]{0.48\textwidth}
    \includegraphics[width=\textwidth]{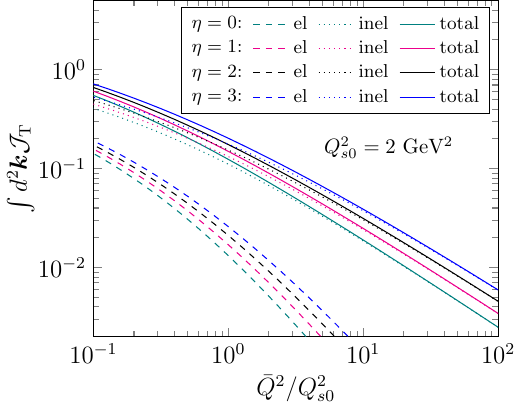}
    \caption{}\label{fig:int_b}
  \end{subfigure}
  \caption{\small SIDIS cross-sections integrated over $\bk$ and their high-energy evolution up to $\eta=3$:
    the elastic, inelastic and total cross-sections are plotted in terms of $\bar{Q}^2/Q_s^2$ and
  for 4 values of the rapidity $\eta$. (a) Logarithmic increase at low $\bar{Q}^2$: elastic and inelastic
  cross-sections are still comparable with each other. (b) Power law behaviour at high $\bar{Q}^2$: inelastic scattering strongly dominates the total cross-section.}\label{fig:int}
\end{figure}

These considerations are corroborated by the numerical results in Fig.~\ref{fig:int_a}, which show
the integrated elastic, inelastic, and total cross-sections in terms of  $\ln(\bar{Q}^2/ Q_s^2)$
in the saturation regime at $\bar{Q}^2 \ll Q_s^2$. Together with the predictions
of the MV model (the curves with $\eta=0$), 
we also exhibit the results of the BK evolution up to $\eta=3$. For $\bar{Q}^2/Q_s^2 \lesssim 0.1$,
all the curves are roughly straight lines, with slopes which are consistent with \eqn{intJlow}. 
As expected, the curves corresponding to elastic scattering lie below the inelastic ones,
but they are still comparable in so far as orders of magnitude are concerned.

\subsection{High virtuality $z(1-z)Q^2\gg Q_s^2$: geometric scaling}
\label{sec:highbarQ}

Even though not so directly relevant for studies of saturation (since controlled by weak scattering; see below),
the case $\bar{Q}^2 \gg Q_s^2$ is still interesting, notably for studies of the high-energy evolution
in the presence of saturation. Also, it features a strong $z$ -dependence of the cross-section, to be
discussed in Sect.~\ref{sec:zdep}.

In this regime, the dominant contributions to the integrals in Eqs.~\eqref{intJel}--\eqref{intJT} come 
from very small dipoles with $r \ll 1/Q_s$, for which one can use the single
scattering approximation. Consider first the low energy case, where one can use 
the MV model, cf. Eq.~\eqref{MVT0}, to deduce
\begin{align}
\label{intJhighMV}
        \int \dif^2 \bk\ \mcal{J}_{\rmT} \simeq
       \bar{Q}^2
      \int 
	\frac{\dif^2 \br}{2\pi} \,
	\rmK_1^2(\bar{Q}r)\,  
	 \frac{r^2 Q_{\rmA}^2}{4}
	 \ln \frac{1}{r^2\Lambda^2}\,.
\end{align}
In principle, one should cut the above integration at $r \sim 1/Q_s$, but since the Bessel function
is exponentially decaying for $\bar{Q}r\gg Q_sr\gg 1$, we can safely extend the integration back to infinity. The angular integration is trivial and  keeping only the logarithmically enhanced term, we find 
\begin{align}
\label{intJhighMV2}
        \int \dif^2 \bk\ \mcal{J}_{\rmT} \simeq
\frac{1}{6}\, 
        \frac{Q_\rmA^2}{\bar{Q}^2}\,
        \ln \frac{\bar{Q}^2}{\Lambda^2}
        \qquad \textrm{for} \quad \bar{Q}^2 \gg Q_s^2. 
\end{align}
It is easy to check that this dominant contribution arises from dipole sizes $r\sim 1/\bar Q\ll 1/Q_s$,
thus justifying our above use of the weak scattering approximation. Up to a numerical factor,
the quantity $(Q_{A}^2/\alpha_s)\ln({\bar{Q}^2}/{\Lambda^2})$ represents the nuclear gluon distribution 
per unit transverse area
generated by the valence quarks, as probed on a transverse resolution scale $\bar{Q}^2$.
So, \eqn{intJhighMV2} has a simple physical interpretation: the small projectile dipole with transverse size
$r\sim 1/\bar Q\ll 1/Q_s$  scatters with a probability $\sim\alpha_s$ off all the gluons with 
transverse sizes similar to its own and that it encounters while crossing the nucleus --- those
within a longitudinal tube with transverse area $1/\bar Q^2$.

The same exercise can be repeated for the amplitude given the BK solution at  asymptotically high energies,
shown in Eq.~\eqref{tbkdil}. One finds
\begin{align}
\label{intJhigh}
	\int \dif^2 \bk\, \mcal{J}_\rmT \simeq 
	\int \dif^2 \bk\, \mcal{J}_{\rmT,{\rm in}} \approx 
	\left(\frac{Q_s^2}{\bar{Q}^2}\right)^\gamma 
	\ln \frac{\bar{Q}^2}{Q_s^2}
	\qquad \textrm{for} \quad \bar{Q}^2 \gg Q_s^2,
\end{align}
where we omitted the overall  proportionality constant since this is anyway not under control when writing
the BK amplitude as in \eqn{tbkdil}.  We have here anticipated that the elastic cross section is more
strongly suppressed in this limit. One can indeed check that
\begin{align}
\label{intJelhigh} 
	\int \dif^2 \bk\, \mcal{J}_{\rmT,{\rm el}} \approx 
	\left(\frac{Q_s^2}{\bar{Q}^2}\right)^{2\gamma} 
	\ln^2 \frac{\bar{Q}^2}{Q_s^2}
	\qquad \textrm{for} \quad \bar{Q}^2 \gg Q_s^2.
\end{align}

The high-energy and large-virtuality 
results in Eqs.~\eqref{intJhigh}--\eqref{intJelhigh}  show two remarkable features:
\texttt{(i)}  the  ``anomalous'' dimension $\gamma <1$ 
specific to the high-energy evolution (BK/JIMWLK) 
in the presence of saturation, and \texttt{(ii)} geometric scaling --- the fact that these expressions
depend upon the (effective) virtuality $\bar Q^2=z(1-z) Q^2$ and upon the COM energy of the scattering
only via the ratio ${\bar{Q}^2}/{Q_s^2(\eta)}$. Notice that geometric scaling also holds at low virtualities
${\bar{Q}^2}\ll {Q_s^2(\eta)}$, at least within the black disk approximation in \eqn{intJlow}.  So, for
sufficiently high energies (or large $\eta$), this is expected to be a robust feature of the
integrated SIDIS cross-sections at any $\bar{Q}^2$.

\begin{figure}
  \centering
  \begin{subfigure}[t]{0.48\textwidth}
    \includegraphics[width=\textwidth]{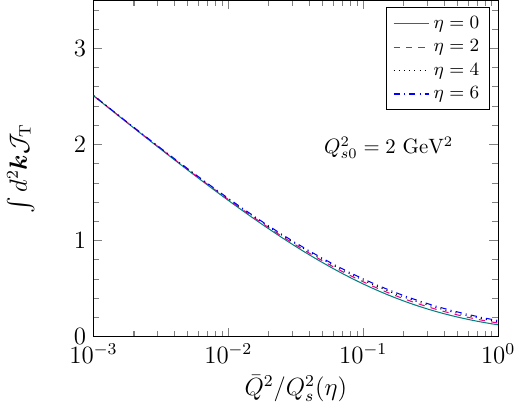}
    \caption{}\label{fig:scaling_a}
  \end{subfigure}
  \hfill
  \begin{subfigure}[t]{0.48\textwidth}
    \includegraphics[width=\textwidth]{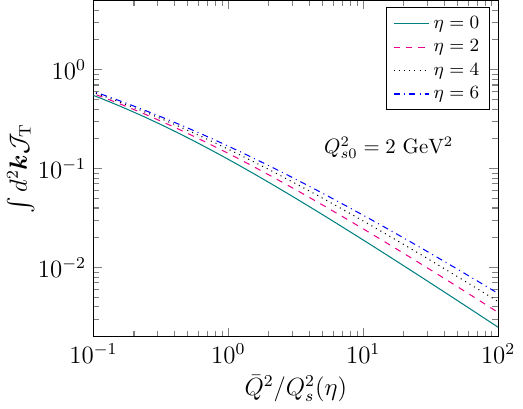}
    \caption{}\label{fig:scaling_b}
  \end{subfigure}
  \caption{\small Testing geometric scaling for the SIDIS transverse cross-section integrated over $\bk$,
  from solutions to the collinearly improved rcBK up to $\eta=6$.
  (a) Low $\bar{Q}^2$: the scaling is quasi-perfect. (b) High $\bar{Q}^2$:
   with increasing $\eta$, the power law tail softens and the scaling becomes more accurate.}\label{fig:intscaling}
\end{figure}

We have checked these properties against numerical solutions to the BK equation,
 with the results shown in  Fig.~\ref{fig:int_b}. With increasing $\eta$, one clearly sees 
a softening of the distribution at high $\bar Q^2$ as compared to the low-energy (MV model) result
\eqn{intJhighMV2}. This softening reflects the emergence of the  anomalous dimension, although 
 the values of $\eta\le 3$ considered there are still to small to observe the asymptotic value 
 $\gamma \simeq 0.63$ expected at very high energies.
 
 To also check geometric scaling, we have plotted in
 Fig.~\ref{fig:intscaling} the results for the integrated cross-section
 $\int \dif^2 \bk\ \mcal{J}_{\rmT} $ in terms of the scaling
variable $\bar{Q}^2/ Q_s^2(\eta)$, up to the larger rapidity $\eta=6$.
In the low virtuality regime at ${\bar{Q}^2}\ll {Q_s^2(\eta)}$, one finds a quasi-perfect scaling
(see Fig.~\ref{fig:scaling_a}),
in agreement with the leading-order estimate in  Eq.~\eqref{intJlow}.
At higher virtualities, ${\bar{Q}^2}\gtrsim {Q_s^2(\eta)}$, the scaling becomes better
with increasing rapidity (see Fig.~\ref{fig:scaling_b}) --- again, as expected.

\subsection{The $z$-dependence of the integrated cross-sections}
\label{sec:zdep}

\begin{figure}
  \centering
  \begin{subfigure}[t]{0.48\textwidth}
    \includegraphics[width=\textwidth]{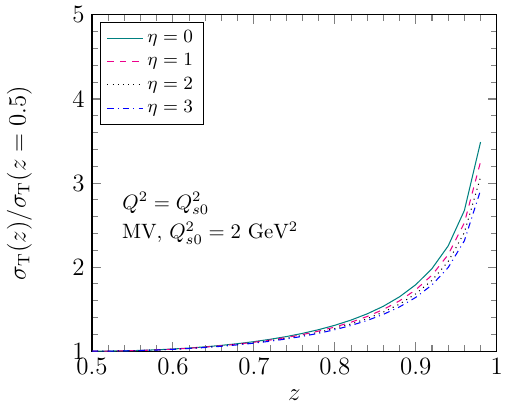}
    \caption{}\label{fig:zdep_a}
  \end{subfigure}
  \hfill
  \begin{subfigure}[t]{0.48\textwidth}
    \includegraphics[width=\textwidth]{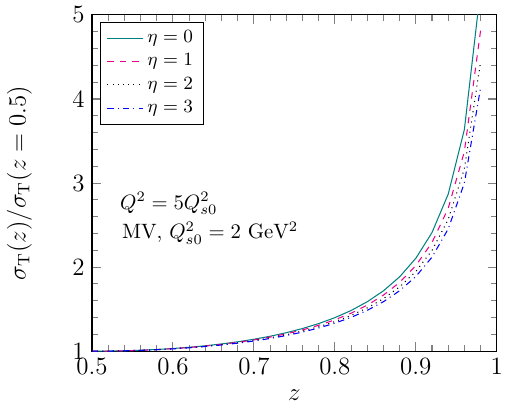}
    \caption{}\label{fig:zdep_b}
  \end{subfigure}
  \caption{\small The normalized $\bk$-integrated transverse cross section-defined in \eqn{sigmaz} plotted
   as a function of $z$ for various values of $\eta$ at (a) $Q^2=Q_{s0}^2$ (``low $\bar{Q}^2$'') 
   and (b) $Q^2=5Q_{s0}^2$ (``high $\bar{Q}^2$''). When $z$ approaches 1, one sees
   a sharper increase in Fig.~\ref{fig:zdep_b} than in Fig.~\ref{fig:zdep_a}.}\label{fig:zdep}
\end{figure}

Yet a different way of looking at the results for the integrated cross-sections 
is to plot them in terms of the longitudinal momentum fraction $z$ of the measured quark, for
fixed values of the virtuality $Q^2$ and of the target saturation momentum ${Q_s^2(\eta)}$. 
In  Fig.~\ref{fig:zdep} we show our respective numerical results for two choices for the virtuality,
 $Q^2=Q_{s0}^2$ (left plot) and respectively  $Q^2=5Q_{s0}^2$ (right plot), and for $\eta\le 3$.
 The plotted quantity is the ratio
 \beq\label{sigmaz}
 \frac{\sigma_{\rmT}(z)}{\sigma_{\rmT}(z=0.5)}\,=\,\frac{\int \dif^2 \bk\, \mcal{J}_\rmT(\bk, z)}
 {\int \dif^2 \bk\, \mcal{J}_\rmT(\bk, z=0.5)}\,,\eeq
 where $\sigma_{\rmT}(z)$ is merely a short-hand notation for the respective
 differential cross-section per unit of $z$, that is ${\dif\sigma}_{\rmT}/{\dif z}$.
As compared to  Fig.~\ref{fig:int},  we now implicitly concentrate on a more limited range of values for
 the ratio $\bar Q^2/Q_{s0}^2$, which are more realistic from the viewpoint of the phenomenology.
This new way of plotting the results is interesting in that it emphasises their strong dependence
upon $z$ in the vicinity of $z=1$, a feature that could be observed in the experiments.

This strong $z$--dependence can be understood from our analytic approximations 
in Eqs.~\eqref{intJlow} and \eqref{intJhigh}. These expressions show that,
when increasing $z$ towards 1, $\sigma_{\rmT}(z)$ increases like $\ln\frac{1}{1-z}$ so long as
${\bar{Q}^2}\ll {Q_s^2(\eta)}$ and (roughly) like the power $1/(1-z)^\gamma$ when
${\bar{Q}^2}\gg {Q_s^2(\eta)}$. For $z\gtrsim 0.8$, the two plots in Fig.~\ref{fig:zdep} 
are representative for these two limiting kinematical regimes. And indeed, all the curves there show a rather
pronounced increase with increasing $z$ above 0.8. This increase is stronger in Fig.~\ref{fig:zdep_b} 
(``high $\bar Q^2$'') than in Fig.~\ref{fig:zdep_a}  (``low $\bar Q^2$''), and it becomes less pronounced
when increasing $\eta$, thus reflecting the onset of an anomalous dimension $\gamma < 1$.

\subsection{$R_{pA}$ ratios for the integrated cross-section}

In Sect.~\ref{sec:cronin} we observed that a suggestive way to visualise the nuclear modifications
is by plotting the ratio $R_{pA}$ between the SIDIS cross-sections in $eA$ collisions and in $ep$ collisions
(times the nucleon number $A$), respectively. Previously, we studied this ratio for 
the case of the differential cross-section in $k_\perp$, but a similar ratio can of course be defined
for the cross-sections integrated over $k_\perp$.

\begin{figure}
  \centering
  \begin{subfigure}[t]{0.47\textwidth}
    \includegraphics[width=\textwidth]{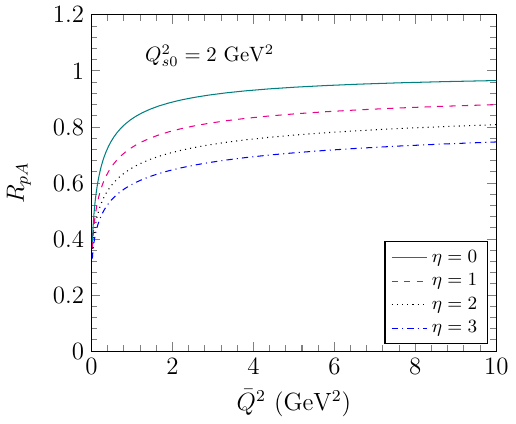}
    \caption{}
    \label{fig:Rpaint_a}
  \end{subfigure}
  \hfill
  \begin{subfigure}[t]{0.49\textwidth}
    \includegraphics[width=\textwidth]{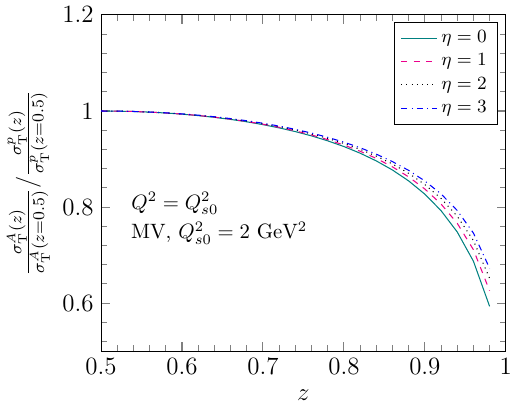}
    \caption{}
    \label{fig:Rpaint_b}
  \end{subfigure}
  \caption{\small (a) The ratio $R_{pA}$ for the  SIDIS cross-section integrated over $\bk$, 
  as defined in Eq.~\eqref{RpAz}, and its rapidity evolution of up to $\eta=3$.
   The ratio is always smaller than 1 in contrast to the unintegrated case (cf.~Fig.~\ref{fig:Cronin}). (b) 
   The nuclear modification factor for the normalised cross-section introduced  in \eqn{sigmaz} 
   plotted as a function of $z$, for fixed $Q^2$ and for various values of $\eta\le 3$. As $z$ approaches 
   1 the ratio drops significantly below 1, due to the different $z$-dependences observed in 
   Figs.~\ref{fig:zdep_a} and \ref{fig:zdep_b}, respectively.}\label{fig:Rpaint}
\end{figure}

Specifically, in  Fig.~\ref{fig:Rpaint_a} we display the ratio
\beq\label{RpAz}
 R_{pA}(z)\,\equiv\,
 \frac{\int \dif^2 \bk\,\mcal{J}_\rmT^A(z,\bk)}{A^{1/3}\int \dif^2 \bk\,\mcal{J}_\rmT^p(z,\bk)}\,,
 \eeq
 as a function of $\bar Q^2=z(1-z)Q^2$ up to $\bar Q^2=5Q_{s0}^2(A)=10$~GeV$^2$ and
 for different rapidities up to $\eta=3$. This ratio would be
 equal to one in the absence of nuclear effects, but by inspection of  Fig.~\ref{fig:Rpaint_a}
 one sees that this is clearly not the case: the ratio is strictly smaller than one and monotonously
 increasing with $\bar Q^2$ (albeit almost flat for  $\bar Q^2 \gtrsim Q_{s0}^2$) within the whole 
  range under consideration.
   Moreover, the nuclear suppression is significantly increasing with $\eta$ at any value for $\bar Q^2$.
 
 This behaviour can be easily understood, at least qualitatively. At the level of the MV model,
 it is well known that the effect of gluon saturation in a large nucleus is to push the gluon
 distribution towards large transverse momenta 
 \cite{Kharzeev:2003wz,Iancu:2004bx}: the low momentum modes at $k_\perp< Q_s$
 get depleted and the gluons which are missing at low $k_\perp$ are redistributed in the modes
 with larger momenta  $k_\perp\gtrsim Q_s$.
Hence, if one integrates the  $k_\perp^2$-distribution up to some value $\bar Q^2$, the total
number of gluons thus counted is smaller than it would have been in the absence of saturation.
Since the saturation effects are much smaller in a proton, the 
$R_{pA}$ ratio for the integrated gluon distributions is smaller than one, but it rises towards
one when increasing $\bar Q^2$. With increasing energy (or $\eta$), the gluon distribution rises
faster in the proton than in the nucleus, hence the $R_{pA}$ ratio  becomes even smaller.

This is precisely the kind of behaviour one sees in Fig.~\ref{fig:Rpaint_a}. Strictly speaking, this plot does
not apply to the integrated gluon distribution in the target, but rather to the integrated
cross-section for quark production in DIS. Yet, these two quantities are closely related to each other.
This is visible 
e.g. at the level of  \eqn{JTinlow}, where the Fourier transform of the dipole $S$-matrix is
recognised as the ``dipole'' version of the unintegrated gluon distribution in the target 
(the ``dipole TMD'') \cite{Marquet:2009ca,Dominguez:2011wm}.

Furthermore, Fig.~\ref{fig:Rpaint_b} shows the nuclear modification factor for the quantity depicted in 
Fig.~\ref{fig:zdep}, which is itself a ratio, cf. \eqn{sigmaz}. That is, the quantity depicted 
in Fig.~\ref{fig:zdep_a} for the case of a nucleus is divided by the same quantity computed for 
a proton\footnote{Note that, in forming this ratio, there is no  factor of  $A$
in the denominator. Indeed, the global proportionality with $A$ of the cross-section
corresponding to $eA$ collisions has already cancelled in the ratio \eqref{sigmaz}.}.
In both calculations, one uses $Q^2=Q_{s0}^2(A)=2$~GeV$^2$. Recalling that $Q_{s0}^2(A)=
A^{1/3}Q_{s0}^2(p)$ and $A^{1/3}=6$, it is easy
to see that the quantity appearing in the denominator of this ratio
 (the one corresponding to the proton) should behave similarly
to that depicted  Fig.~\ref{fig:zdep_b}. In other terms, each of the curves 
shown in Fig.~\ref{fig:Rpaint_b} can be thought of as the ratio of the 2 corresponding curves 
in Fig.~\ref{fig:zdep_a} and 
Fig.~\ref{fig:zdep_b},  respectively. From this perspective, the behaviour visible in  Fig.~\ref{fig:Rpaint_b} 
is easy to understand: when $Q^2=Q_{s0}^2(A)=6Q_{s0}^2(p)$,
one probes a dense gluon distribution in the nuclear, but a dilute one in the proton. Hence, 
the cross-section $\sigma_{\rmT}(z)$ increases much slower when increasing $z$ towards 1
for the nucleus than for the proton (as also visible by comparing the two plots in Fig.~\ref{fig:zdep}).
Accordingly, the nucleus-to-proton ratio is decreasing, as shown in Fig.~\ref{fig:Rpaint_b}.
Interestingly, this suppression is quite strong already for not too large values
of $z$ (say $z\simeq 0.8$) and thus it might be a robust signal of saturation in the EIC data.

 \section{Inclusive DIS: symmetric vs. aligned-jet configurations}
\label{sect:sym-al}

In most of the previous developments in this paper, we assumed that the longitudinal momentum fraction
$z$ of the measured quark is very close to one, in such a way that $\bar{Q}^2 \equiv z(1-z)Q^2 \ll Q_s^2$.
Asymmetric dipole configurations, such that
$z$ is either very close to one, $1-z\ll 1$, or very small $z\ll 1$, have often been
discussed in a different context, namely in relation with the so-called 
``aligned jet'' contributions to the inclusive DIS cross-section
$\sigma_\rmT\equiv 
\sigma_{\gamma_{\rmT}^*A} (x,Q^2)$, cf. \eqn{sigmat}. So, it is natural to wonder what is the
relation between these two physical regimes, if any. In this section, we would like to explain that
these situations are in fact different and do not overlap with each other. 


Before we discuss the inclusive DIS cross-section, in which the longitudinal fraction
$z$ is integrated over, let us first summarise
the interesting kinematical regimes for fixed values of $Q^2$, $z$ and $Q_s^2$:
 
 \texttt{(i)} Low  virtuality $Q^2\ll Q_s^2$: in this case, one clearly has $\bar{Q}^2\ll Q_s^2$ for any value of $z$.
 In this regime, saturation effects are certainly important, but in practice they may mix with 
 non-perturbative effects like vector meson fluctuations of the virtual photon, due to the fact that
 such low values for $Q^2$ are close to the QCD confinement scale $\Lambda^2$.
 
 \texttt{(ii)} High virtuality $Q^2\gg Q_s^2$, but such that $z$ is sufficiently small, or sufficiently close to one,
 in order to have  $\bar{Q}^2 \ll Q_s^2$. This is the kinematical regime that we almost exclusively considered 
 throughout this paper, in the particular context of SIDIS and for $z\simeq 1$. 

 \texttt{(iii)} High virtuality  $Q^2\gg Q_s^2$ and $z$ values such that $\bar Q^2 \gg Q_s^2$ as well. 
 Clearly, this regime is not very interesting for studies of gluon saturation (except marginally, via effects
 like geometric scaling and the BK anomalous dimension discussed in Sect.~\ref{sec:highbarQ}).
In turn, this regime includes two different situations, depending upon the values of $z$: (a) aligned-jet configurations, which are such that $1\gg z(1-z)\gg Q_s^2/Q^2$,  and (b) symmetric configurations, 
for which $z$
is neither very close to zero, nor very close to 1; we shall succinctly refer to them as $z\sim 1/2$.

It is generally stated that the aligned-jet configurations 
give the dominant contribution to $\sigma_\rmT$ at high $Q^2$,
but as we shall shortly see this conclusion can be modified by the high-evolution evolution
and the associated anomalous dimension, cf. \eqn{tbkdil}.

%
%

 We now present leading-order, parametric,  estimates for the inclusive DIS
 cross-section $\sigma_\rmT$. For convenience, we shall use the simple notation
 $ \tilde\sigma_\rmT$ for the cross-section in \eqn{sigmaT} {\it without} the factor $R^2_A N_c \alpha_{\rm em} e_q^2$, that is, without the constant geometrical, colour and electromagnetic factors. Still, we shall take into account the numerical constant and the $z$-dependent factor. We successively consider the low-virtuality and the high-virtuality regimes and
distinguish between contributions from various regions in $z$.
 

\texttt{(i)} When $Q^2 \ll Q_s^2$, one can use the ``black-disk'' approximation in \eqn{intJlow} for any value of $z$,
to deduce
\begin{align}
        \label{sigmalow}
         \tilde\sigma_\rmT \simeq 
        \int\limits_0^1 \dif z \left[z^2 + (1-z)^2 \right]
        \left[ \ln \frac{Q_s^2}{Q^2} + \ln \frac{1}{z(1-z)}
        \right]
        \simeq \frac{2}{3}\ln \frac{Q_s^2}{Q^2} 
        \qquad \textrm{for} \quad Q^2 \ll Q_s^2,
\end{align}
where only the first term in the square brackets is important to logarithmic accuracy. Clearly, this
dominant contribution comes from the symmetric configurations with  $z\sim 1/2$.

 \texttt{(ii)} When $Q^2\gg Q_s^2$, but $z$ is restricted to the corners in the phase-space where
 $z(1-z)\lesssim Q_s^2/Q^2$, we can still rely on \eqn{intJlow}, but with appropriate limits on the
integral over $z$; namely, instead of \eqn{sigmalow}, one should write
 \begin{align}
        \label{sigmahighsat}
         \tilde\sigma_\rmT \simeq 
        2 \int\limits_0^{Q_s^2/Q^2} \dif z 
        \left[ \ln \frac{1}{z} - \ln \frac{Q^2}{Q^2_s}
        \right]
        =\, \frac{2 Q_s^2}{Q^2} 
        \qquad \textrm{for} \quad Q^2 \gg Q_s^2 \gtrsim \bar{Q}^2,
\end{align}
where we integrated over small values $z \le Q_s^2/Q^2$ and multiplied the result by a factor of 2
to account for the region at $1-z\le Q_s^2/Q^2$. (Notice however that the upper limit in Eq.~\eqref{sigmahighsat} should involve a fudge factor of order $\mcal{O}(1)$, thus the proportionality constant is not really under control.) This is the contribution to the inclusive 
cross-section from the SIDIS configurations which have been our main focus in this paper --- those
where the dipole fluctuations have very large sizes $r\sim 1/\bar Q\gtrsim 1/Q_s$ and therefore
suffer strong scattering.

\texttt{(iii)} When both $Q^2 \gg Q_s^2$ and $\bar Q^2 \gg Q_s^2$, the dipole scattering is weak
($r\sim 1/\bar Q\ll 1/Q_s$) and we can use the results for ${\dif\sigma_\rmT}/{\dif z}$
from Sect.~\ref{sec:highbarQ},  but with the subsequent integration over $z$ restricted to
$z(1-z) \gg Q_s^2/Q^2$. 

Consider first the  ``low-energy'' case, as described by the MV model; using
 \eqn{intJhighMV2}, one finds
 \begin{align}
        \label{sigmahighMV}
         \tilde\sigma_\rmT \simeq 
        \frac{2}{3}\, \frac{Q_\rmA^2}{Q^2}
        \int\limits_{Q_s^2/Q^2}
        \frac{\dif z}{z} \left(\ln \frac{Q^2}{\Lambda^2} + \ln z \right)
        \simeq \frac{2}{3}\, 
        \frac{Q_\rmA^2}{Q^2}\, \ln\frac{Q^2}{\Lambda^2}\,
        \ln \frac{Q^2}{Q_s^2}\,
        \quad \textrm{for} \quad Q^2 \gg Q_s^2 \quad \textrm{(MV model)},
\end{align}
 where in the final result we kept only the term
generated by the logarithmic integration over $z$, which
is enhanced by the large logarithm $\ln ({Q^2}/{Q_s^2})$. (A similar contribution from $z\sim 1$ 
has been included via a factor of 2.) 
Clearly, this dominant contribution comes from the {\it aligned-jet configurations},
as previously defined --- that is, $z$ is small, but not {\it too} small (and similarly for $1-z$), in such
a way that $1\gg z(1-z)\gg Q_s^2/Q^2$.

%

Consider now the high-energy regime $\abar\eta \gtrsim 1$,
where  the BK evolution becomes important. Using Eq.~\eqref{intJhigh}
and neglecting proportionality constants, we obtain
\begin{align}
        \label{sigmahigh}
        \!\!\!\tilde \sigma_\rmT \approx
        2 \left(\frac{Q_s^2}{Q^2}\right)^\gamma\!\!\!
        \int\limits_{Q_s^2/Q^2}^{1}\!\! \dif z\, 
        \frac{z^2+(1-z)^2}{[z(1-z)]^\gamma} 
        \left[ \ln \frac{Q^2}{Q_s^2} + \ln z(1-z)\right]
        \approx
        \left(\frac{Q_s^2}{Q^2}\right)^\gamma \!\ln \frac{Q^2}{Q_s^2} 
        \quad \textrm{for} \quad Q^2 \gg Q_s^2.
\end{align}
Since $\gamma < 1$, the above integral is dominated by large values\footnote{Unlike in
\eqn{sigmahighMV}, the logarithm $ \ln({Q^2}/{Q_s^2}) $ which appears in
 the final result in \eqn{sigmahigh} 
is not generated by the integral over $z$, rather it comes from the transverse logarithm 
originally present in the BK solution for the dipole amplitude at high energy, cf.
 \eqn{tbkdil}.} $z\sim 1/2$ (in particular, the
lower limit at $z= Q_s^2/Q^2\ll 1$ played no role for the final result), i.e.~by {\it symmetric} dipole
configurations, with very small sizes $r\sim 1/Q$. This change of behaviour w.r.t. the MV model
calculation in Eq.~\eqref{sigmahighMV} is due to the
fact that, when $\gamma < 1$, the dipole amplitude for a single
scattering rises more slowly with the dipole size $r$ 
than in the regime of ``colour transparency'' ($\gamma=1$).

The results in Eqs.~\eqref{sigmahighMV} and \eqref{sigmahigh} confirm our earlier statement
in this section, namely the fact that the aligned-jet configurations dominate the DIS
cross-section $\sigma_\rmT$ at high $Q^2$ only so long as one can neglect the onset
of the BK anomalous dimension.
This is the actual physical situation at not too high energies (in particular, in the MV model and,
more generally, in the regime where the DGLAP evolution is the dominant QCD dynamics).
It is furthermore the case at {\it any} energy, provided $Q^2$ is sufficiently high $Q^2$  ---
 larger than the upper limit of the geometric scaling region (which, via
the high-energy evolution, grows with $\eta$ even faster than $Q_s^2$)
 \cite{Iancu:2002tr,Mueller:2002zm,Iancu:2004bx}).

{The above results also show that, irrespective of the precise value of $\gamma\le 1$, 
inclusive DIS at high $Q^2\gg Q_s^2$ is controlled by relatively small dipoles undergoing weak scattering
($r\ll 1/Q_s$). The corresponding contribution of the large dipoles with $r\gtrsim 1/Q_s$,
as estimated in \eqn{sigmahighsat}, is parametrically smaller, in the sense that it lacks
the logarithm enhancement visible in Eqs.~\eqref{sigmahighMV} and \eqref{sigmahigh}.
We thus conclude that the SIDIS configurations which are most interesting for
studies of saturation yield a relatively small contribution to the inclusive cross-section,
which is however sizeable --- it is only logarithmically suppressed compared to the
dominant contribution.}

Before we conclude this section, let us notice that, when $\gamma < 1$,
 the parametric estimate shown in \eqn{sigmahigh} for $\sigma_\rmT$ also holds for the
 longitudinal cross-section $\sigma_\rmL$; in that case too, this result is generated by 
 symmetric dipole configurations with $z\sim 1/2$ and $r\sim 1/Q$.  But unlike $\sigma_\rmT$,
 the longitudinal cross-section is controlled by symmetric  configurations even 
when $\gamma=1$, so in that case $\sigma_\rmL$ is lacking the large
logarithm $ \ln({Q^2}/{Q_s^2}) $ visible in \eqn{sigmahighMV} and thus is suppressed relative
to $\sigma_\rmT$.

\section{Conclusions}
\label{sec:conc}

In this paper, we have identified a new physical regime that should be interesting for studies
of unitarity corrections and gluon saturation in electron-nucleus deep inelastic scattering at the 
future Electron-Ion Collider. In the dipole picture for DIS at small Bjorken $x$, this regime
corresponds to very asymmetric quark-antiquark fluctuations, such that one of the two quarks
carries a large fraction $z\simeq 1$ (meaning $1-z\ll 1$) of the longitudinal momentum of its
parent virtual photon. This longitudinal asymmetry implies that the dipole fluctuation has a
very large transverse size $r\sim 1/\bar Q\gg 1/Q$, with $\bar Q^2=z(1-z)Q^2$. When moreover
one has also has $\bar Q^2\lesssim Q_s^2$, with $Q_s$ the saturation momentum in the
nuclear target, then the dipole-nucleus scattering is strong and thus probes high gluon density
effects, like gluon saturation, in the target wavefunction. 

We have proposed several observables in view of experimental studies of this regime.
They all refer to SIDIS experiments in which one measures the jet (or hadron) produced by the
final-state evolution of the quark carrying the large longitudinal momentum fraction $z$.

For experiments which measure both the longitudinal fraction $z$ and the transverse momentum
$k_\perp$ of the final jet (or hadron), we have argued that the most interesting kinematics
for a study of saturation is at $\bar Q^2\ll k_\perp^2\lesssim Q_s^2$.  We identified two
interesting phenomena in this regime. 

For sufficiently low transverse momenta $k_\perp^2\ll Q_s^2$, we found that
the scattering probes the ``black disk limit'', i.e. the regime
where the dipole amplitude reaches the unitarity limit $T=1$. In this case,
the inelastic scattering is suppressed by unitarity, hence the total SIDIS cross-section
is controlled by its  elastic (or coherent)  component, as given by the ``shadow'' of the 
strong scattering. In our dipole picture,
this corresponds to configurations where the splitting of the virtual photon
into a $q\bar q$ pair occurs {\it after} the photon has crossed the nucleus without
interacting with it. 

On the other hand, for larger transverse momenta $k_\perp^2\sim Q_s^2$,
the dominant component is the inelastic scattering of the measured quark, leading to
transverse momentum broadening, that is, to a Gaussian distribution 
which peaks at $k_\perp\sim Q_s$. A similar, ``Cronin'', peak could also be seen in the $R_{pA}$ ratio
between the SIDIS cross-sections in $eA$ and in $ep$ collisions, respectively, albeit only for 
extremely small values of the ratio $\bar Q^2/Q^2_s$, that might be difficult to reach in practice.
Yet, the deviation of $R_{pA}$ from unity is found to be significantly large including
for more realistic kinematical conditions, so the nuclear effects should be easy to observe.
From the viewpoint of the target, the multiple scattering of the measured quark 
probes the nuclear gluon distribution at, or near, saturation, as measured by a special unintegrated
gluon distribution known as the ``dipole TMD".

We have also studied the effects of the high energy evolution, as described by the collinearly-improved
version of the BK equation (with running coupling) recently proposed in 
Refs.~\cite{Ducloue:2019ezk,Ducloue:2019jmy}. We found that, in general, these effects are quite small,
due to the slowing down of the evolution by the various (collinear and running-coupling) resummations,
and also to the fact that the we have only considered relatively small rapidity evolutions, in line
with the kinematics expected at the EIC. The only observable for which even a small rapidity evolution 
may have a qualitative effect is the $R_{pA}$ ratio: the Cronin peak disappears after just one unit
of rapidity and is replaced by uniform nuclear suppression  
up to relatively large transverse momenta, well above $Q_s$.

Since the transverse momentum of the tagged jet (or hadron) may be difficult to measure
in practice, especially at very forward rapidities, we also study the saturation effects on the 
SIDIS cross-section integrated over  $k_\perp$. Once again, the interesting regime for saturation is at
$\bar Q^2\ll Q^2_s$. In this regime, we find that the (integrated) elastic and inelastic cross-sections
are comparable with each other --- they are even identical to leading logarithmic accuracy.
Moreover, they are both of order $\ln(Q^2_s/\bar Q^2)$, with the logarithm coming from 
integrating over the wavefunction of the virtual photon, that is, over the $k_\perp$-distribution
of the $q\bar q$ pair produced by the decay of the virtual photon. For the elastic scattering, this is
also the {\it final} distribution. For the inelastic channel, the final distribution is generated via 
multiple scattering, so the integral over $k_\perp$ is controlled by values $k_\perp\sim Q_s$.

To better emphasise the distinguished nature of this behaviour at low $\bar Q^2\ll Q^2_s$, we
have also computed the integrated SIDIS cross-section in the weak scattering regime at 
$\bar Q^2\gg Q^2_s$. We thus found the expected ``leading-twist'' suppression by the factor
$Q^2_s/\bar Q^2$ (possibly modified by an anomalous dimension introduced by the high
energy evolution). A particularly suggestive way to present our results is by plotting the
integrated cross-section as a function of $z$ for a fixed and relatively large value of the virtuality
 $Q^2\gg Q_s^2$ (cf. Fig.~\ref{fig:zdep}):  in the weak scattering 
regime at $\bar Q^2\gg Q_s^2$ one sees a rapid, power-like, growth when increasing $z$ towards 1, 
which is eventually tamed (when $\bar Q^2\lesssim Q^2_s$) by saturation. This change of behaviour
can perhaps be better visualised by plotting the corresponding $R_{pA}$ ratio,
as shown in Fig.~\ref{fig:Rpaint_b}.
 
We have finally considered the inclusive DIS cross-section, with the purpose
of clarifying the relation between the special, large-$z$, SIDIS configurations that we have 
investigated in this paper and the ``aligned jet'' configurations traditionally discussed in relation
with DIS at large $Q^2\gg Q_s^2$. 
Although both types of configurations are associated with dipole fluctuations
which are very asymmetric in their $z$ sharing (i.e. $z(1-z)\ll 1$) and hence relatively large
($r\sim 1/\bar Q\gg 1/Q$), they nevertheless  correspond to
different physical regimes, which do not overlap with each other.
The ``aligned jets'' correspond to dipoles which are still small enough ($r\ll 1/Q_s$) to scatter
only weekly; these are the typical configurations which control inclusive DIS at high $Q^2$. 
On the contrary, the ``large-$z$'' SIDIS configurations which are interesting for 
saturation are sufficiently large to suffer strong scattering ($r\gtrsim 1/Q_s$)
even at high $Q^2\gg Q_s^2$. Their contribution to the inclusive cross-section is
parametrically suppressed, but only logarithmically,  so it should represent a sizeable fraction
of the total cross-section measured in the experiments.

Throughout this analysis, we restricted ourselves to the leading-order pQCD approximation, 
where the particle ``measured'' in SIDIS at large $z$ is a bare, on-shell, quark. 
In general, the quark emerging from the scattering with the hadronic target will have some 
time-like virtuality, so it will evolve via final-state radiation, thus giving rise to a jet. 
In an actual experiment, one could measure either a jet with a large value of $z$ (defined as
the total longitudinal momentum fraction carried by the jet constituents),
or a single,  ``leading'', hadron, which itself has a large value of $z$.

 Leaving aside the
non-perturbative effects, like hadronisation or higher-twist corrections, it is important
to observe that already the higher-order pQCD corrections are expected to be quite different in
the two cases (jet production and single hadron production, respectively). Indeed, for
an exclusive process like hadron production, the limit $z\to 1$ introduces a strong constraint
on the radiation in the final state: the total energy fraction carried by the unmeasured emissions cannot
exceed the small value $1-z$. Such a constraint is well known to introduce special higher-order
corrections, where each power of $\alpha_s$ is accompanied by the double logarithm $\ln^2\frac{1}{1-z}$.
(See App.~\ref{sec:jet} for an explicit, albeit admittedly sketchy, calculation at next-to-leading order.)
Physically, these large corrections express the fact that it is very unlikely to observe a final state
in which most of the total energy of the incoming photon is carried by a hadron {\it alone}.
The existence of these double-logarithmic corrections spoils the validity of any fixed-order 
approximation --- {\it a fortiori}, of our present estimates at leading order.

However, the situation becomes much simpler if in the final state one measures a jet.  Unmeasured emissions
which are relatively energetic (with energy fractions larger than $1-z$) are now allowed, so long as they
belong to the forward jet. The final state is now composed with more likely configurations and, as a result,
there are no double-logarithmic corrections anymore. (Technically, this is ensured by an efficient compensation
between ``real'' and ``virtual'' emissions; see App.~\ref{sec:jet} for an example at NLO.) In view of this,
our present results can be seen as a {\it bone fidae}  leading-order approximation to the SIDIS cross-section
for forward jets with large $z\simeq 1$. On the other hand, much more work --- including all-order resummations
of the double-logarithmic corrections --- is needed to give a reliable estimate for the single hadron production
in SIDIS at large $z$.

Still for the case of forward jet production, one can also worry about the importance of the higher-twist corrections 
--- the non-perturbative corrections suppressed by inverse powers of the jet virtuality. As already stressed,
both final jets have relatively small transverse momenta $k_\perp^2 \simeq (1-z)Q^2\ll Q^2$ (when $z\simeq 1$).
If this transverse scale were to be representative for the jet virtuality, then indeed the twist-expansion
would be questionnable. However, this is not the case for the measured jet, with a large energy fraction $z\sim 1$.
By inspection of the kinematics, one can deduce that this jet should be time-like and  relatively hard, with
an invariant mass squared $M^2_z \equiv k^\mu k_\mu$ of the order of the initial photon virtuality:
 $M^2_z \sim Q^2$. 
This follows from the appropriate generalisation of the kinematical constraint \eqref{xg},
as obtained by replacing\footnote{For a right-moving particle with invariant mass $M^2=
2k^+k^--k_\perp^2$, the light-cone energy reads $k^-=(k_{\perp}^2+M^2)/2k^+$.}
 $k_{1\perp}^2\to k_{1\perp}^2+M^2_z$: after also using $k_{1\perp}^2\simeq k_{2\perp}^2 \simeq
 (1-z)Q^2$ when $z\simeq 1$, one sees that the limit $z\to 1$ puts no special constraint on the jet
 invariant mass  $M^2_z$, which therefore is of order $Q^2$. 
 
  The fact that the produced jet is hard when $Q^2\gg Q_s^2$ also ensures that
 the SIDIS process at large $z$ is leading-twist in the sense of the QCD parton picture
 --- it measures a genuine quark distribution in the hadronic target --- even when
 $\bar Q^2 = z(1-z)Q^2$ is low enough, $\bar Q^2\lesssim Q_s^2$, for the saturation
 effects to be important.  In other terms, the saturation effects that we discussed 
 in this paper can be characterised as {\it leading-twist quark shadowing}.
 This may appear as a surprise since in our
 calculations the saturation  effects manifest themselves as multiple scattering ---
 a phenomenon generally associated with higher-twist corrections. Recall however that
 the standard twist expansion for DIS is formulated in a special frame 
 (a target infinite momentum frame, such as the Bjorken frame) and a special gauge
 (the target light-cone gauge, that is, $A^-=0$ in the present conventions),
 which are not the frame and gauge employed by the dipole picture that we have used
 throughout. 
%
It would be conceptually interesting to repeat
 some of the previous calculations in the traditional QCD parton picture, to verify our above
 statements about the twist expansion and, especially, to understand the role of
 gluon saturation and of the unitarity corrections in that framework.


The previous remarks clearly show that our present study is only preliminary. 
Further theoretical studies are needed,  especially for the case of forward hadron production.
On the experimental side, one still need to ascertain the feasibility of hadron or jet measurements 
at very forward rapidities and semihard transverse momenta. 
We hope that our present analysis will give an incentive for 
further developments and will inspire new studies of gluon saturation in 
deep inelastic scattering at the EIC.

\section*{Acknowledgements} 
The work of E.I. is supported in part by the Agence Nationale de la Recherche project 
 ANR-16-CE31-0019-01.   The work of A.H.M.
is supported in part by the U.S. Department of Energy Grant \# DE-FG02-92ER40699. 

\appendix
 
\section{Reducing the number of integrations}
\label{sec:app1}

Here we shall show how to analytically reduce the number of integrations to be done when calculating all the possible $\bk$-dependent cross sections, i.e.~longitudinal, transverse, elastic, inelastic or total ones. 

We shall see first that is is possible to do exactly one of the two 2-dimensional integrations over
the dipole sizes $\br$ and $\br'$. Regarding the elastic cross sections, cf.~for example the transverse one in Eq.~\eqref{JTel}, this is trivial since (after making use of $T^*(\br) = T(-\br)$) it can be written as the modulus squared of a 2-dimensional integration. Now, let's consider the longitudinal cross section in Eq.~\eqref{JL}. The first two terms (denoted by labels $a$ and $b$ just below in Eq.~\eqref{ilab}), those proportional to $T(\br)$ and $T(-\br')$, contribute equally and using the integral
\begin{align}
	\label{int1}
	\int \frac{\dif^2 \br'}{(2\pi)^2}\, \rme^{\rmi \bk \cdot \br'} \rmK_0(\bar{Q}r')  = 
	\frac{1}{2\pi} \int_0^{\infty} \dif r'r' \rmJ_0(k_\perp r') \rmK_0(\bar{Q}r') = 
	\frac{1}{2\pi}\frac{1}{k_\perp^2 + \bar{Q}^2}, 
\end{align}
we readily get
\begin{align}
\label{ilab}
	\mcal{J}_{\rmL}^{a+b} = 
	\int \frac{\dif^2 \br}{(2\pi)^2}\,
	\rme^{-\rmi \bk \cdot \br} 
	\frac{\bar{Q}^2}{k_\perp^2 + \bar{Q}^2}\,\rmK_0(\bar{Q}r)
	T(\br).
\end{align}
For the third term, we write $T(\br - \br')$ in terms of its Fourier transform, i.e.
\begin{align}
	\label{tft}
	T(\br - \br') = 
	\int \dif^2 \bl\, \rme^{\rmi \bl \cdot (\br - \br')}
	\tilde{T}(\bl)
\end{align}
and then using Eq.~\eqref{int1} it is easy to perform the two integrations over both $\br$ and $\br'$  to get
\begin{align}
\label{ilc}
	\mcal{J}_{\rmL}^{c} = 
	-\frac{\bar{Q}^2}{4 \pi}
	\int \frac{\dif^2 \bl\, \tilde{T}(\bl)}{[(\bk-\bl)^2+\bar{Q}^2]^2}.
\end{align}
Now we rewrite back $\tilde{T}(\bl)$ as an inverse Fourier transform and subsequently let $\bl \to \bl + \bk$ to find
\begin{align}
\label{ilc2}
\hspace*{-0.1cm}
	\mcal{J}_{\rmL}^{c} =
	-\frac{1}{4 \pi} 
	\int \frac{\dif^2\br}{(2\pi)^2}\, 
	\rme^{-\rmi \bk \cdot \br} T(\br)
	\int \dif^2\bl\, \rme^{-\rmi \bl \cdot \br}
	\frac{\bar{Q}^2}{[\ell_\perp^2+\bar{Q}^2]^2}
	= - \int \frac{\dif^2 \br}{(2\pi)^2}\,
	\rme^{-\rmi \bk \cdot \br} 
	\frac{\bar{Q} r}{4}\,\rmK_1(\bar{Q}r)
	T(\br).
	\end{align}
So putting together Eqs.~\eqref{ilab} and \eqref{ilc2} one finds
\begin{align}
\label{JL2}
	\mcal{J}_\rmL = 
	\int \frac{\dif^2 \br}{(2\pi)^2}\,
	\rme^{-\rmi \bk \cdot \br} 
	\left[ \frac{\bar{Q}^2}{k_\perp^2 + \bar{Q}^2}\,\rmK_0(\bar{Q}r) 
	- \frac{\bar{Q} r}{4}\, \rmK_1(\bar{Q}r)\right] 
	T(\br).  
\end{align}	
Similarly for the transverse sector we need the vector integral
\begin{align}
	\label{int2}
	\int \frac{\dif^2 \br'}{(2\pi)^2}\, \rme^{\rmi \bk \cdot \br'}\,
	\frac{\br'}{r'}\, \rmK_1(\bar{Q}r')  = 
	\frac{\rmi}{2\pi} \frac{\bk}{k_\perp}
	\int_0^{\infty} \dif r'r' \rmJ_1(k_\perp r') \rmK_1(\bar{Q}r') = 
	\frac{\rmi}{2\pi}
	\frac{\bk}{\bar{Q}}
	\frac{1}{k_\perp^2 + \bar{Q}^2}, 
\end{align}
so that for the terms involving $T(\br)$ and $T(\br')$ we have
\begin{align}
\label{itab}
	\mcal{J}_{\rmT}^{a+b} = 
	\int \frac{\dif^2 \br}{(2\pi)^2}\,
	\rme^{-\rmi \bk \cdot \br}\,
	\frac{\rmi \bk \cdot \br}{\bar{Q}r}
	\frac{\bar{Q}^2}{k_\perp^2 + \bar{Q}^2}\,\rmK_1(\bar{Q}r)
	T(\br).
\end{align}
For the third term, we introduce the Fourier transform of $T(\br - \br')$ to obtain 
\begin{align}
\label{iltc}
	\mcal{J}_{\rmT}^{c} = 
	-\frac{1}{4 \pi}
	\int \frac{\dif^2 \bl\, (\bk-\bl)^2\, \tilde{T}(\bl)}{[(\bk-\bl)^2+\bar{Q}^2]^2}.
\end{align}
Going back to the inverse Fourier transform and letting $\bl \to \bl + \bk$ we get
\begin{align}
\label{itc2}
	\mcal{J}_{\rmT}^{c} &=
	-\frac{1}{4 \pi} 
	\int \frac{\dif^2\br}{(2\pi)^2}\, 
	\rme^{-\rmi \bk \cdot \br} T(\br)
	\int \dif^2\bl\, \rme^{-\rmi \bl \cdot \br}
	\frac{\ell_\perp^2}{[\ell_\perp^2+\bar{Q}^2]^2}
	\nonumber \\
	& = - \int \frac{\dif^2 \br}{(2\pi)^2}\,
	\rme^{-\rmi \bk \cdot \br} 
	\left[\frac{1}{2}\,\rmK_0(\bar{Q}r)-\frac{\bar{Q} r}{4}\,\rmK_1(\bar{Q}r)\right]
	T(\br).
	\end{align}
Again, putting together Eqs.~\eqref{itab} and \eqref{itc2}, one finds
\begin{align}
\label{JT2}
	\mcal{J}_\rmT = 
	\int \frac{\dif^2 \br}{(2\pi)^2}
	\rme^{-\rmi \bk \cdot \br } 
	\bigg\{
	\frac{\rmi \bk \!\cdot\!\br}{\bar{Q}r} 
	\frac{\bar{Q}^2}{k_\perp^2+\bar{Q}^2}
	\rmK_1(\bar{Q}r) -\frac{1}{2}
	\bigg[\rmK_0(\bar{Q}r) - \frac{\bar{Q}r}{2} 
	\rmK_1(\bar{Q}r) \bigg]
	\bigg\}
	T(\br).  
\end{align}

Eqs.~\eqref{JL2} and \eqref{JT2} (and also Eq.~\eqref{JTel} which is the modulus squared of a 2-dimensional integration as already said) are valid for any $T(\br)$. For all our purposes in this paper, and for the vast majority of the studies related to gluon saturation, the amplitude does not depend on the orientation of the dipole. Thus one can easily do the angular integration in terms of the Bessel functions $\rmJ_0(k_\perp r)$ and $\rmJ_1(k_\perp r)$ by using Eqs.~\eqref{int1}, \eqref{int2} and the additional integral 
\begin{align}
	\label{int3}
	\int \frac{\dif^2 \br}{(2\pi)^2}\, \rme^{-\rmi \bk \cdot \br} \bar{Q}r\,\rmK_1(\bar{Q}r)  = 
	\frac{1}{2\pi} \int_0^{\infty} \dif r\,r \rmJ_0(k_\perp r) \,
	\bar{Q}r\,\rmK_1(\bar{Q}r) = 
	\frac{1}{\pi}\frac{\bar{Q}^2}{(k_\perp^2 + \bar{Q}^2)^2}.
\end{align}
We shall not write here the resulting expressions, since they are trivially obtained from Eqs.~\eqref{JL2}, \eqref{JT2} and Eq.~\eqref{JTel}. Therefore, one is finally left with a single integration to perform numerically, the one over a dipole size.

\section{Elastic cross section in the MV model}
\label{sec:elastic}

In this Appendix we will give an analytic expression for the elastic cross section in the MV model and in the general case $\bar{Q}^2 \ll k_\perp^2, Q_s^2$. This means we can expand the Bessel function $\rmK_1(\bar{Q}r)$ for small argument, since the integration is practically restricted to $r_{\rm max} \sim 1/k_{\perp}$, so that $\bar{Q} r_{\rm max} \sim \bar{Q}/k_{\perp} \ll 1$. Thus, the starting expression for the quantity of interest is given in Eq.~\eqref{JTelMV} and here we shall calculate directly the integral which gives the derivative of the Weisz\"acker-Williams
unintegrated gluon distribution, i.e.
\begin{align}
\label{W}
	\nabla_{\bk} \mathcal{W} = -\rmi  
	\int \frac{\dif^2 \br}{(2\pi)^2}\,
	\frac{\br}{r^2}\,
	\rme^{-\rmi \bk \cdot \br}\,
	T(r) 
	= -\frac{1}{2\pi}\,
	\frac{\bk}{k_\perp}  
	\int_0^{1/\Lambda}\!\! \dif r\, 
	\rmJ_1(k_\perp r) \,
	T(r).
\end{align}
We have performed the angular integration using the first equality in Eq.~\eqref{int2}. In the above $T=1-S$ is the MV model amplitude, with $S$ given in \eqref{SMV}. We shall follow the procedure developed in \cite{Iancu:2004bx} (cf.~Sect.~2.2 and Appendix A there), which is in fact applicable to a wide class of ``observables'' in the MV model. Using Eq.~\eqref{QsMV} to replace $Q_A$ in terms of $Q_s$ and after doing simple algebraic manipulations we can rewrite $S$ as
\begin{align}
\label{MVexpand}
S = \exp\left(-\frac{r^2 Q_s^2}{4}\right)
\exp\left(-\frac{r^2 Q_s^2}{4}\, \frac{\ln (1/r^2Q_s^2)}{\ln (Q_s^2/\Lambda^2)} \right)
\simeq 	
\exp\left(-\frac{r^2 Q_s^2}{4}\right)
\left(1 - \frac{r^2 Q_s^2}{4}\, \frac{\ln (1/r^2Q_s^2)}{\ln (Q_s^2/\Lambda^2)} \right).
\end{align}
In writing the second, approximate, equality we have done an expansion to first order in $1/(\ln Q_s^2/\Lambda^2)$ which is a small number, roughly equal to the QCD running coupling evaluated at the saturation momentum. Such an expansion is clearly valid for $r \lesssim 1/Q_s$ and the only potential dangerous regime could be when $r \gg 1/Q_s$, that is, very close to the unitarity limit. However in that regime the $S$-matrix is anyway close to zero and thus the error in our expansion is innocuous. It is important to emphasize that this is not a twist expansion, since the first factor clearly contains multiple scattering to all orders. Rather it should be viewed as an expansion in $r$ around $1/Q_s$ in a well-defined and controlled approximation scheme. The two terms in Eq.~\eqref{MVexpand} give rise to two contributions to $\nabla_{\bk} \mathcal{W}$, more precisely
\begin{align}
\label{W2}
	\nabla_{\bk} \mathcal{W}
	= \nabla_{\bk} \mathcal{W}\big|_{\rm sat} + 
	\nabla_{\bk} \mathcal{W}\big|_{\rm twist}.
\end{align}
These saturation and twist pieces, with the names to be shortly justified, correspondingly read
\begin{align}
\label{Wsat}
	\nabla_{\bk} \mathcal{W}\big|_{\rm sat}
	= -\frac{1}{2\pi}\,
	\frac{\bk}{k_\perp}  
	\int_0^\infty \dif r\, 
	\rmJ_1(k_\perp r) \, \left[1 - \exp\left(-\frac{r^2 Q_s^2}{4}\right) \right]
\end{align}
and
\begin{align} 
\label{Wtwist}
	\nabla_{\bk} \mathcal{W}\big|_{\rm twist}
	= -\frac{1}{2\pi}\,
	\frac{\bk}{k_\perp}  
	\int_0^\infty \dif r\, 
	\rmJ_1(k_\perp r)\,
	\frac{r^2 Q_s^2}{4}\, \frac{\ln (1/r^2Q_s^2)}{\ln (Q_s^2/\Lambda^2)}\,
	\exp\left(-\frac{r^2 Q_s^2}{4}\right).
\end{align}
The integration in the saturation contribution is easily done and we get
\begin{align}
\label{Wsat2}
	\nabla_{\bk} \mathcal{W}\big|_{\rm sat}
	= -\frac{1}{2\pi}\,
	\frac{\bk}{k_\perp^2}\,  
\exp\left(-k_\perp^2/Q_s^2\right),
\end{align}
while the twist contribution can be given in terms of the exponential integral function ${\rm Ei}(x) = {\rm PV} \int_{-\infty}^{x}
 \dif t\, {\rm e}^t/t$, where PV stands for the Principal Value, and reads  
\begin{align}
\label{Wtwist2}
	\nabla_{\bk} \mathcal{W}\big|_{\rm twist}
	= \frac{1}{2\pi}\,
	\frac{\bk}{k_\perp^2}\, 
	\frac{1}{\ln Q_s^2/\Lambda^2}\, 
\left\{1 + \exp\left(-k_\perp^2/Q_s^2\right) 
\left[\frac{k_\perp^2}{Q_s^2}\ln \frac{4k_\perp^2}{Q_s^2} - 1
- \frac{k_\perp^2}{Q_s^2}\, {\rm Ei}(k_\perp^2/Q_s^2)
\right]
\right\}.
\end{align}
It is very instructive to study Eq.~\eqref{Wtwist2} in the various kinematic regimes and we have
\begin{align}
\label{Wtwistlim}
	\nabla_{\bk} \mathcal{W}\big|_{\rm twist}
	\simeq \frac{1}{2\pi}\,
	\frac{\bk}{k_\perp^2}\, 
	\frac{1}{\ln Q_s^2/\Lambda^2}
	\begin{cases}\ln(4 \rme^{1-\gamma_{\rm E}})
		\displaystyle{\frac{k_\perp^2}{Q_s^2}}
		&\quad \textrm{for} \quad k_{\perp}^2 \ll Q_s^2,
		\\*[0.3cm]
		\mcal{O}(1)
		&\quad \textrm{for} \quad k_{\perp}^2 \sim  Q_s^2,
		\\*[0.3cm]
	 \left(-\displaystyle{\frac{Q_s^2}{k_{\perp}^2}}\right)	
	&\quad \textrm{for} \quad k_{\perp}^2 \gg Q_s^2.
	\end{cases}
\end{align}
We point out that the above piecewise limiting expressions could have been easily obtained directly from Eq.~\eqref{Wtwist}. Let us now focus on Eqs.~\eqref{Wsat2} and \eqref{Wtwistlim}. When $k_{\perp}^2 \ll Q_s^2$ the twist part is power suppressed when compared to the saturation part, while it is still suppressed by the factor $1/(\ln Q_s^2/\Lambda^2)$ when $k_{\perp}^2 \sim Q_s^2$. Thus, the saturation part dominates everywhere in the regime $k_{\perp}^2 \lesssim Q_s^2$. On the contrary, since the saturation part is ``compact'', the power-law tail of the twist part dominates when $k_{\perp}^2 \gg Q_s^2$. Notice also that this analysis completely justifies our notation a posteriori. Putting everything together (and using also Eq.~\eqref{QsMV} to simplify the final expression at high-$k_\perp$) we arrive at    
\begin{align}
\label{Wlim}
	\nabla_{\bk} \mathcal{W}
	\simeq -\frac{1}{2\pi}\,
	\frac{\bk}{k_\perp^2}\,
	\begin{cases}
	\exp\left(-k_\perp^2/Q_s^2\right)
		&\quad \textrm{for} \quad k_{\perp}^2 \lesssim Q_s^2,
		\\*[0.3cm]
	 \displaystyle{\frac{Q_A^2}{k_{\perp}^2}}	
	&\quad \textrm{for} \quad k_{\perp}^2 \gg Q_s^2.
	\end{cases}
\end{align}
Finally, using the left equation in \eqref{JTelMV} we get for the elastic cross section in the MV model
\begin{align}
\label{Wlim1}
	\mcal{J}_{\rmT,\rm el}
	\simeq \frac{1}{4\pi k_\perp^2}\,
	\begin{cases}
	\exp\left(-2 k_\perp^2/Q_s^2\right)
		&\quad \textrm{for} \quad k_{\perp}^2 \lesssim Q_s^2,
		\\*[0.3cm]
	 \displaystyle{\frac{Q_A^4}{k_{\perp}^4}}	
	&\quad \textrm{for} \quad k_{\perp}^2 \gg Q_s^2.
	\end{cases}
\end{align}
The upper case in the above is exactly Eq.~\eqref{JTelGBW} which was also discussed in the main text (recall that $\bar{Q}^2 \ll k_\perp^2, Q_s^2$ in this Appendix). The lower case confirms the strong power-law suppression of the elastic cross section in the hard regime when compared to the inelastic one given in Eq.~\eqref{JTsingle}.

\section{High-momentum tail in the MV model}
\label{sec:mv}

Here we would like to give the details in the derivation of Eq.~\eqref{JTsingle} for the high-momentum limit of the transverse cross section in the MV model case. In the main text it was explained that the dominant, logarithmically enhanced, contribution arises from dipoles such that $\rho \ll r \ll 1/\bar{Q}$, where $\bm{\rho} = \br - \br '$. Here we shall be a little more general and we will take the logarithmic limit only at the end of the calculation. As shown in Appendix \ref{sec:app1}, the integration over the one transverse coordinate can be done exactly to give Eq.~\eqref{JT2}, and where in the current notation we must let $\br \to \bm{\rho}$, that is
\begin{align}
\label{JTmv1}
	\mcal{J}_\rmT \simeq 
	-\frac{1}{2}
	\int \frac{\dif^2 \br}{(2\pi)^2}\,
	\rme^{-\rmi \bk \cdot \bm{\rho}}\, 
	\rmK_0(\bar{Q}\rho)\,T(\bm{\rho}).  
\end{align}
We have kept only the term proportional to $\rmK_0$, since the others can be shown not to lead to a logarithmic behavior. For the MV model amplitude, we shall not rely on the ``harmonic approximation'' in Eq.~\eqref{MVT0}, rather we shall take a step back and write \cite{Iancu:2002xk} 
\begin{align}
	\label{MVT0int} 
	T(\bm{\rho})= \frac{Q_\rmA^2}{\pi}
	\int_\Lambda \frac{\dif^2\bm{q}}{q_\perp^4} \,
	\big(1- \rme^{\rmi \bq \cdot \bm{\rho}}\big).
\end{align}
(Expanding the above to second order for small $\rho$ indeed leads to Eq.~\eqref{MVT0}.) Inserting the above into Eq.~\eqref{JTmv1}, reversing the order of integrations and using Eq.~\eqref{int1} we find 
\begin{align}
\label{JTmv2}
	\mcal{J}_\rmT \simeq 
	\frac{Q_\rmA^2}{4 \pi^2}
	 \int_\Lambda \frac{\dif^2\bm{q}}{q_\perp^4} \,
	 \left[\frac{1}{(\bk - \bq)^2 +\bar{Q}^2} - \frac{1}{k_\perp^2 + \bar{Q}^2} \right]
	 \,\simeq \,
	 \frac{Q_\rmA^2}{4 \pi^2 k_\perp^2}
	 \int_\Lambda \frac{\dif^2\bm{q}}{q_\perp^4} \,
	 \frac{2\bk \cdot \bq - q_\perp^2}{(\bk - \bq)^2 +\bar{Q}^2},
\end{align}
where in writing the second approximate equality we have just combined the two fractions and used $k_\perp^2 \gg \bar{Q}^2$. The above contains two logarithmic regimes of integrations. The ``harder'' one is for $\bar{Q} \ll p_\perp \ll k_\perp$, with $\bp \equiv \bq-\bk$ and from only the first term in the first equality above we find    
\begin{align}
\label{JTmvlog1}
	\mcal{J}_\rmT \simeq 
	\frac{Q_\rmA^2}{4 \pi k_\perp^4} \ln \frac{k_\perp^2}{\bar{Q}^2}. 
\end{align}
Note that the power $1/k_\perp^4$ has been generated by the exchange (a hard scattering with 
transfer momentum $q_\perp\simeq k_\perp$), whereas the logarithm came from the integration
$\int (\dif^2\bm{p}/\bp^2)$ over the wavefunction of the virtual photon.

The second, ``softer'', regime is for $\Lambda \ll q_\perp \ll k_\perp$ and one must be very careful in expanding the fraction in the integrand after the second equality in Eq.~\eqref{JTmv2} to keep all the terms that eventually lead to $\mcal{O}(1/k_\perp^4)$ contributions in $\mcal{J}_{\rmT}$. Notice that the leading $1/k_{\perp}^2$ contributions from the two terms in the first expression in Eq.~\eqref{JTmv2} cancel each other in the limit of interest, and this is why we prefer to work with the second, equivalent, expression. One has
\begin{align}
	\label{fracexpand}
	\frac{2\bk \cdot \bq - q_\perp^2}{(\bk - \bq)^2 +\bar{Q}^2} 
	\,\simeq\,
	\frac{1}{k_\perp^2} 
	\frac{2\bk \cdot \bq - q_\perp^2}{1  - \frac{2 \bk \cdot \bq}{k_\perp^2}}
	\,\simeq\,
	\frac{1}{k_\perp^2}
	\left[\frac{4 (\bk \cdot \bq)^2 }{k_\perp^2} - q_\perp^2 + 2 \bk \cdot \bq \right].
\end{align}
After doing the angular integration the last term vanishes while the first one acquires an averaging factor of 1/2, so overall  Eq.~\eqref{fracexpand} is equal to $q_\perp^2/k_\perp^2$. The ensuing integral
over $\bq$ generates the Coulomb logarithm characteristic of soft scattering. We finally deduce
the following contribution to $\mcal{J}_\rmT$ in Eq.~\eqref{JTmv2}:
\begin{align}
\label{JTmvlog2}
	\mcal{J}_\rmT \simeq 
	\frac{Q_\rmA^2}{4 \pi k_\perp^4} \ln \frac{k_\perp^2}{\Lambda^2}. 
\end{align}
As it should be clear from the above manipulations, the power $1/k_\perp^4$ now originates
from the original transverse momentum distribution ${1}/({k_\perp^2 + \bar{Q}^2})$ generated by the
decay of the virtual photon --- more precisely, it arises 
as the difference between two such distributions, with one of them shifted with the momentum
$\bq$ transferred by the scattering.

Eqs.~\eqref{JTmvlog1} and \eqref{JTmvlog2} are added to give the final result in Eq.~\eqref{JTsingle}.

\section{Collinearly improved BK equation}
\label{sec:BK}

The BK evolution equation we have used for the numerical calculations in this work reads \cite{Ducloue:2019ezk,Ducloue:2019jmy} (see also \cite{Beuf:2014uia,Iancu:2015vea})
\begin{equation}
	\label{bketa}
	\frac{\del {S}_{\bx\by}(\eta)}{\del \eta}  = 
	\int \frac{\dif^2\bz}{2\pi}\, 
	\bar{\alpha}_{\rm \scriptscriptstyle BLM}\,
	\frac{(\bx-\by)^2}{(\bx-\bz)^2(\bz-\by)^2}
	\left[\frac{r^2}{\bar z^2}\right]^{\pm A_1}\!\!
	\big[{S}_{\bx\bz}(\eta \minus \delta_{\bx\bz;r})
	{S}_{\bz\by}(\eta \minus \delta_{\bz\by;r}) - {S}_{\bx\by}(\eta) \big] 
\end{equation}
and we briefly explain below all the various elements in the above. The gluon is emitted at the transverse position $\bz$ by either the quark or the antiquark of the parent dipole located at $(\bx,\,\by)$ and the indices in the $S$-matrices stand for its dependence on the corresponding dipole coordinates. When compared to the LO BK equation we note various differences which all arise from the resummation of higher order corrections enhanced by large logarithms.

(i) The rapidity shifts in the arguments of the $S$-matrices for the daughter dipoles resum double logarithms associated with the time ordering in the successive emissions and they are given by
\begin{equation}
\label{delta}
  \delta_{\bx\bz;r} \equiv \max \left\{0,\ln\frac{r^2}{|\bx\minus\bz|^2} \right\} 
\end{equation}
with $r\equiv|\bx-\by|$, and similarly for $\delta_{\bz\by;r}$. They
become significant for emissions in which one of the daughter dipoles
is much smaller than the parent one.

(ii) The factor $\left[{r^2}/{\bar z^2}\right]^{\pm A_1}$, where 
$\bar z \equiv \min\{|\bx-\bz|,|\bz-\by|\}$ and with $A_1=11/12$, 
includes the resummation of the first set of single DGLAP logarithms. The sign in the exponent $\pm A_1$ is taken to be plus for ${r^2}<{\bar z^2}$ and minus ${r^2}>{\bar z^2}$, so it is always suppressing the evolution.

(iii) The use of a running coupling resums the corrections related to the one-loop QCD $\beta$-function. When one of the three dipoles is much smaller than the other two, pQCD requires that the scale should be the smallest dipole size, i.e.~one should use $\abar(r_{\rm min})$ in Eq.~\eqref{bketa}, where $r_{\min} \equiv
\min\{|\bx-\by|,|\bx-\bz|,|\bz-\by|\}$. Here, we shall use what is close to a
BLM prescription (called ``fast apparent convergence'' in \cite{Iancu:2015joa}), which minimizes the NLO correction to the BK equation proportional to the one-loop $\beta$-function. It is defined as
\begin{equation}
\label{ablm}
  \bar{\alpha}_{\rm \scriptscriptstyle BLM} = 
  \left[ 
    \frac{1}{\abar(|\bx \minus\by|)}
    + \frac{(\bx\minus\bz)^2 - (\bz\minus\by)^2}{(\bx\minus\by)^2}
    \frac{\abar(|\bx-\bz|)-\abar(|\bz-\by|)}{\abar(|\bx-\bz|)\abar(|\bz-\by|)}
  \right]^{-1}
\end{equation}
and one can easily check that it reduces to $\abar(r_{\min})$ in the limit of very disparate dipoles discussed just above. The running $\abar(r)$ on the right-hand side of Eq.~\eqref{ablm} is dictated by QCD and reads  
\begin{equation}
\label{alpha0}
\abar(r) = \frac{1}{\bar b_0\ln\big[4/(r_*^2\Lambda_{\rm QCD}^2)\big]},
\end{equation}
with $\bar b_0=0.75$ (corresponding to 3 flavors) and
$\Lambda=0.2$~GeV. The Landau pole is avoided by the introduction of $r_*$ according to $r_* = r/\sqrt{1+r^2/r_{\rm max}^2}$ with $r_{\rm max} = 4$ GeV$^{-1}$, which freezes the coupling in the deep IR to the value $\abar(r \gg r_{\rm max}) \simeq 0.728$.

Finally, we point out that \eqn{bketa} is solved as an initial value problem, but one must pay particular attention to the non-locality in $\eta$. The shift in Eq.~\eqref{delta} introduces a dependence to rapidities smaller than $\eta$ and thus when we start the evolution at a rapidity $\eta_0$ (which is taken to vanish throughout the current paper) we need to specify the initial condition for $\eta \le \eta_0$. We shall assume a constant behaviour in $\eta$ (cf.~the discussion in Sect.~9 of
Ref.~\cite{Ducloue:2019ezk}), that is
\beq\label{ICeta}
{S}_{\bx\by}(\eta <\eta_0) ={S}_{\bx\by}(\eta_0)
\eeq
and the latter is simply given by the MV model as written in Eq.~\eqref{SMVnum}.

\section{Useful integrals}
\label{sec:int}

\comment{\begin{align}
\label{intLam}
\int \dif^2\bm{\rho}\, \rme^{-\rmi \bk \cdot \bm{\rho}} \ln\frac{1}{\rho^2\Lambda^2}
\simeq \frac{4\pi}{k_\perp^2}\quad\mbox{for}\quad {k_\perp^2\gg \Lambda^2}
\end{align}}

First, we list some integrals which are useful in calculating the high-momentum tail of the $\bk$-dependent cross sections. Generally, for $-1 < \gamma < -1/2$, one has
\begin{align}
	\label{int4}
	\int_0^{\infty}
	\dif \rho\, \rho^{1+2\gamma} \rmJ_0(\rho) = 
	\frac{4^{\frac{1}{2} +\gamma} \Gamma\left(1+\gamma \right)}
	{\Gamma\left(-\gamma\right)}
\end{align}
and for $-1 < \gamma < 1/4$
\begin{align}
	\label{int5}
	\int_0^{\infty}
	\dif \rho\, \rho^{2\gamma} \rmJ_1(\rho) = 
	\frac{4^{\gamma} \Gamma\left(1 + \gamma \right)}
	{\Gamma\left(1 - \gamma\right)}.
\end{align}
By analytic continuation we shall take the above to be true for values of $\gamma$ outside the aforementioned interval. Differentiating with $\gamma$ we can also calculate integrals involving a power of $\ln \rho^2$. For example we have
\begin{align}
	\label{int6}
	\int_0^{\infty}
	\dif \rho\, \rho^{3} \ln \rho^2 \,\rmJ_0(\rho) = 
	\lim_{\gamma \to 1}
	\frac{\dif }{\dif \gamma}
	\frac{4^{\frac{1}{2} +\gamma} \Gamma\left(1+\gamma \right)}
	{\Gamma\left(-\gamma\right)} = 8.
\end{align}
Second, we give a generic integral involving the modified Bessel function $\rmK_1$ needed for the evaluation of the high momentum tail of the integrated over-$\bk$ cross sections. We have
\begin{align}
	\label{intK1}
	\int_0^\infty \dif \rho\, \rho^{1+2\gamma} \,\rmK_1^2(\rho) = 
	\frac{\sqrt\pi\, \Gamma(\gamma) \Gamma(1 + \gamma) \Gamma(2 + \gamma)}{4 \Gamma(3/2 + \gamma)},
\end{align}
which holds for any $\gamma>0$. In particular for $\gamma=1$ and $\gamma=2$ we respectively get 2/3 and 8/5.

\section{Double-logarithmic corrections: hadron measurement versus jet measurement}
\label{sec:jet}

In the concluding section, we have mentioned an important difference between measuring a quark (``hadron'') and,
respectively, a jet, in a SIDIS experiment where the longitudinal fraction $z$ of the measured system is close to one.
In the first case --- hadron measurement ---, one expects large next-to-leading order (NLO) corrections enhanced by a double-logarithm
$\ln^2\frac{1}{1-z}$, which make the leading-order predictions unreliable. In the second case --- jet measurement ---,
such large corrections are however absent. In this Appendix, we would like to make this statement more precise, by
presenting an admittedly sketchy calculation of the relevant NLO corrections, to the double-logarithmic accuracy of interest.
We plan to present this calculation in full detail in a subsequent publication.

\begin{figure}
  \centering
  \begin{subfigure}[t]{0.45\textwidth}
    \includegraphics[width=\textwidth]{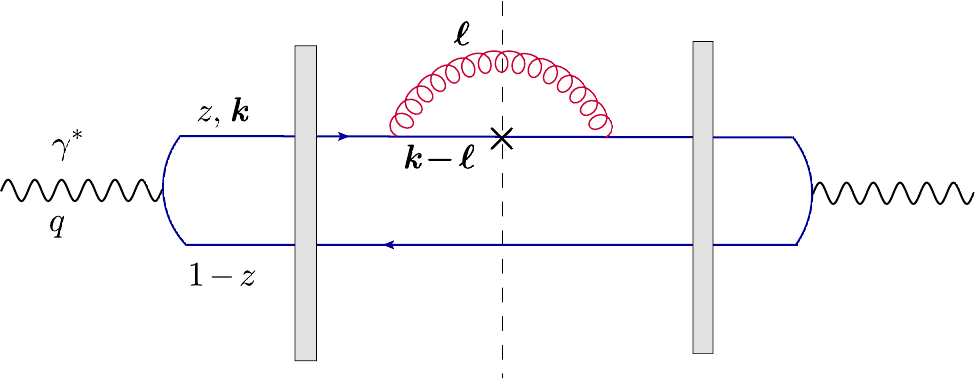}
    \caption{}\label{fig:real}
  \end{subfigure}
  \hfill
  \begin{subfigure}[t]{0.45\textwidth}
    \includegraphics[width=\textwidth]{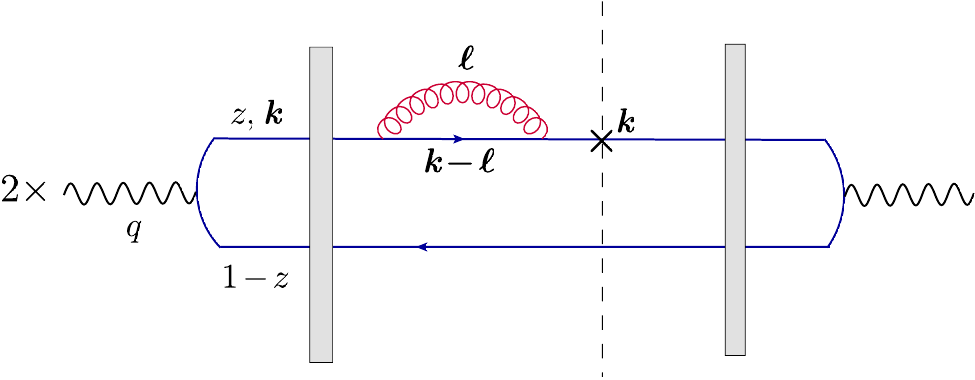}
    \caption{}\label{fig:virt}
  \end{subfigure}
  \caption{\small  The NLO graphs encoding the final-state evolution of the measured quark.
  (a) A real gluon emission. (b) A virtual gluon emission. Graph (b) is multiplied by a factor of 2 to account for
  the identical contribution where the gluon emission is associated with the quark in the complex conjugate
  amplitude.}
    \label{fig:NLO}
    \end{figure}

In the case of a quark measurement, that we shall address first,
the large NLO corrections of order $\alpha_s \ln^2\frac{1}{1-z}$ are generated by late gluon emissions, 
which are a part of the familiar DGLAP evolution of the final quark. More precisely, they are the result of 
a kinematical   mismatch between  ``real'' and ``virtual'' gluon emissions, as
represented by the graphs in Fig.~\ref{fig:real} and Fig.~\ref{fig:virt}, respectively. Namely, the 
momentum of the measured quark is equal to $k$ for the ``virtual'' graph~\ref{fig:virt} and to $k-\ell$ for the
``real'' graph~\ref{fig:real}. Here, $\ell$ is the gluon 4-momentum and is integrated over (since the gluon is not measured in the final state).
In both cases, the longitudinal momentum of the final quark  --- that is, $k^+$ for Fig.~\ref{fig:virt} and
respectively $k^+-\ell^+$ for Fig.~\ref{fig:real} --- is measured to be a fraction $z$ of the initial momentum $q^+$
of the virtual photon.

As generally throughout this paper, we use the (time-orderd) light-cone perturbation theory together with the
projectile LC gauge $A^+=0$. In particular, all the (real or virtual) quanta are on-shell and the lack of energy
conservation at the emission vertices translates into energy denominators which encode the duration
of the quantum emission process. Also, the LC longitudinal momentum of each particle is positive and bound by the 
respective momentum of its parent. For the case of the ``real'' emissions, it is furthermore constrained by the 
kinematics of the final state. 

Accordingly, the longitudinal momentum of the ``virtual'' gluon in Fig.~\ref{fig:virt}
is constrained by $\ell^+\le k^+=zq^+$, whereas that of the ``real'' gluon in Fig.~\ref{fig:real} rather
obeys $\ell^+\le (1-z)q^+$. (Indeed, once the quark is measured with $zq^+$, the total longitudinal momentum
carried by the unmeasured antiquark and gluon is known to be $(1-z)q^+$, and this represents the upper bound
on $\ell^+$.) This kinematical mismatch lies at the origin of the double-logarithmic NLO corrections, as we now explain.

The calculation of the DGLAP-like NLO corrections in the double logarithmic approximation (DLA) being a standard exercice,
here we shall merely exhibit the final result. These corrections factorise, so they can be fully encoded in a factor, $F_{\rm real}$
or $F_{\rm virt}$, multiplying the leading-order SIDIS cross-section. This factor
has a similar structure for the real and virtual graphs: in both cases, it features the integral over the phase-space of the
emitted gluon, in logarithmic units for both the longitudinal ($\ell^+$) and the transverse ($\bl$) momentum. However,
$F_{\rm real}$ and $F_{\rm virt}$ differ from each other in their integration limits over $\ell^+$ (as already discussed) and also
in their global sign: the virtual corrections are negative, as they describe a reduction in the probability to find a ``bare''
quark in the final state.

Specifically, the factor $F_{\rm virt}$ encoding the effect of the ``virtual'' graph in Fig.~\ref{fig:virt} is found as
\begin{align}
\label{Fvirt1}
	F_{\rm virt} = -\frac{\alpha_s C_F}{\pi^2}\int_0^{k^+}\frac{\rmd \ell^+}{\ell^+}\int \frac {\rmd^2 \bl}{\left(\bl - \frac{\ell^+}{k^+}\bk\right)^2
	+m^2\left(\frac{\ell^+}{k^+}\right)^2}\ \Theta\left(\frac{2\ell^+}{\bl^2} - \frac{2q^+}{Q^2} \right)\,.
\end{align}
The longitudinal integration being by now quite obvious, let us give some explanations about the structure of the transverse integration.
First, we have introduced a non-zero quark mass $m$, to screen the would-be collinear divergence in the limit where the
gluon emission angle relative to its parent quark approaches to zero: $\theta_{\ell k}\to 0$ with $\theta_{\ell k}^2\equiv\left(\frac{\bl}{\ell^+}-
\frac{\bk}{k^+}\right)^2$. A non-zero mass $m$ limits this angle to a minimal value $\theta_{\ell k} \ge m/k^+$, via the well-known
``dead cone effect'': gluon radiation within an angle smaller than $m/k^+$ around the direction of the emitter is forbidden
by the kinematics. The other important ingredient in \eqn{Fvirt1} is the step function enforcing the formation time $\tau_\ell
= {2\ell^+}/{\bl^2}$ of the gluon to be larger than the coherence time $\tau_q=2q^+/Q^2$ of the virtual photon. This is the
condition that the gluon emissions under consideration correspond indeed to final-state radiation. One can check that early
radiation, with $\tau_\ell < \tau_q$, does not generate collinear divergences, hence it does not contribute to DLA.  After a change 
in the integration variable, $\bl \to \bl + \frac{\ell^+}{k^+}\bk$, and to the
DLA accuracy of interest, \eqn{Fvirt1} can be further simplified to
\begin{align}
\label{Fvirt2}
	F_{\rm virt} = -\frac{\alpha_s C_F}{\pi}\int_0^{q^+}\frac{\rmd \ell^+}{\ell^+}\int_{m^2(\ell^+/q^+)^2}^{Q^2(\ell^+/q^+)}
	 \frac {\rmd \ell_\perp^2}{\ell_\perp^2}
	\,.
\end{align}
As compared to  \eqn{Fvirt1}, we have also approximated $k^+=zq^+\simeq q^+$ in the integration limits, in view of the
fact that $1-z\ll 1$.

The corresponding ``real'' contribution is given by a similar expression, except for a change of sign and for the much lower
value of the upper limit on $\ell^+$:
\begin{align}
\label{Freal}
	F_{\rm real} = \frac{\alpha_s C_F}{\pi}\int_0^{(1-z)q^+}\frac{\rmd \ell^+}{\ell^+}\int_{m^2(\ell^+/k^+)^2}^{Q^2(\ell^+/q^+)}
	 \frac {\rmd \ell_\perp^2}{\ell_\perp^2}
	\,.
\end{align}

Clearly, the net contribution $F_{\rm quark}\equiv F_{\rm real}  + F_{\rm virt} $
 reads (the lower script ``quark'' is meant to recall that we measure a quark in the final state)
\begin{align}
\label{Fquark}
	F_{\rm quark} & = -\frac{\alpha_s C_F}{\pi}\int_{(1-z)q^+}^{q^+}\frac{\rmd \ell^+}{\ell^+}\int_{m^2(\ell^+/k^+)^2}^{Q^2(\ell^+/q^+)}
	 \frac {\rmd \ell_\perp^2}{\ell_\perp^2} \nonumber \\*[0.2cm]
	 & = -\frac{\alpha_s C_F}{\pi}\ln\frac{1}{1-z}\left[\ln\frac{Q^2}{m^2} + \frac{1}{2}\ln\frac{1}{1-z}\right]
	\,.
\end{align}
The final result in \eqn{Fquark}
exhibits the anticipated double-logarithmic correction, which becomes arbitrarily large in the limit $z\to 1$.
Notice that this correction is independent of the transverse momentum $\bk$ of the measured quark,
so it would be the same for both the double differential cross-section 
$\frac{\dif\sigma}{\dif z\, \dif^2\bk}$ and the cross-section $\frac{\dif\sigma}{\dif z}$ integrated over $\bk$. 

To summarise, this large NLO correction to the SIDIS cross-section for quark (more generally, hadron) production
occurs because of the large disparity between the longitudinal phase-spaces for ``real'' and respectively ``virtual''
gluon emissions. In turn, this disparity is a consequence of the fact that the ``real'' gluon is not measured in the final
state, so its longitudinal momentum must be very low, $\ell^+\le (1-z)q^+$, in order for the measured quark to carry a large
longitudinal fraction $z\simeq 1$ by itself. However, if instead a quark, we measure a jet, then the ``real'' gluon can be
a part of that jet and its longitudinal momentum is not constrained anymore. This will be explained in what follows.

When measuring jets in the final state, we would like to be able to distinguish between a hard jet with 
longitudinal fraction $z\simeq 1$ and a soft one, with fraction $1-z\ll 1$. To that aim, the angular opening of the hard
jet must be limited by the angle $\delta$ between the 2 jets, that can be estimated as
\beq\label{deltajet}
\delta\simeq\frac{k_\perp}{q^+-k^+}=\frac{k_\perp}{q^+(1-z)}\,.
\eeq
Concerning the ``real'' gluon emissions, it is convenient to distinguish between soft emissions, with  $\ell^+ < (1-z)q^+$,
and hard ones, with $\ell^+ > (1-z)q^+$. (The intermediate values with $\ell^+\sim (1-z)q^+$ do not contribute at DLA.)
The soft emissions can have relatively large emission angles, hence for them the jet cone condition
$\theta_{\ell k}< \delta$ may be quite restrictive. However, such soft emissions do not count for the jet energy balance,
for none of the 2 jets, so in practice they are not observable and can be neglected. The hard emissions on the other
hand must belong to the hard jet. Recalling that $\theta_{\ell k}=\ell_\perp/\ell^+$ (we use the $\ell_\perp$ variable 
occurring in  \eqn{Fvirt2}), we have the condition
\beq\label{jetcond}
\frac{\ell_\perp} {\ell^+}< \delta \quad \mbox{or}\quad \ell_\perp^2 <  (\ell^+)^2 \frac{k_\perp^2}{[q^+(1-z)]^2}\,.\eeq
What we would like to show is that this constraint is in fact automatically satisfied by the ``real'' emissions with
$\ell^+> (1-z)q^+$. Indeed, the upper limit on $\ell_\perp^2$ in the second inequality above is in fact  larger than the
upper limit on the respective integral in \eqn{Freal}, as we now verify: the inequality that we need, namely,
\beq
Q^2 \frac{\ell^+}{q^+} \, <\, (\ell^+)^2 \frac{k_\perp^2}{[q^+(1-z)]^2}
\eeq
is equivalent to
\beq
Q^2  \, <\, \frac{k_\perp^2}{(1-z)}\,\frac{\ell^+}{q^+(1-z)}\,,
\eeq
which is obviously satisfied since  $k_\perp^2\ge Q^2 z(1-z) \simeq Q^2 (1-z)$, and 
 $\ell^+\gg (1-z)q^+$.
Accordingly, when going from a quark to a jet measurement, the integral over ``real'' gluon emissions extends up to $k^+\simeq
q^+$, so the virtual contributions (which are still given by \eqn{Fvirt2}) are exactly cancelled by real emissions of gluons
lying within the jet cone \eqref{deltajet} --- at least to the double-logarithmic accuracy of interest. In other terms, they
double-logarithmic corrections found in \eqn{Fquark} for the case of quark production are absent in the case
of jet production.

\bigskip

\begin{thebibliography}{10}
\expandafter\ifx\csname url\endcsname\relax
  \def\url#1{\texttt{#1}}\fi
\expandafter\ifx\csname urlprefix\endcsname\relax\def\urlprefix{URL }\fi
\expandafter\ifx\csname href\endcsname\relax
  \def\href#1#2{#2} \def\path#1{#1}\fi

\bibitem{Accardi:2012qut}
A.~Accardi, et~al., {Electron Ion Collider: The Next QCD Frontier}:
  {Understanding the glue that binds us all}, Eur. Phys. J. A 52~(9) (2016)
  268.
\newblock \href {http://arxiv.org/abs/1212.1701} {\path{arXiv:1212.1701}},
  \href {http://dx.doi.org/10.1140/epja/i2016-16268-9}
  {\path{doi:10.1140/epja/i2016-16268-9}}.

\bibitem{Aschenauer:2017jsk}
E.~Aschenauer, S.~Fazio, J.~Lee, H.~Mantysaari, B.~Page, B.~Schenke,
  T.~Ullrich, R.~Venugopalan, P.~Zurita, {The electron\textendash{}ion
  collider: assessing the energy dependence of key measurements}, Rept. Prog.
  Phys. 82~(2) (2019) 024301.
\newblock \href {http://arxiv.org/abs/1708.01527} {\path{arXiv:1708.01527}},
  \href {http://dx.doi.org/10.1088/1361-6633/aaf216}
  {\path{doi:10.1088/1361-6633/aaf216}}.

\bibitem{Gelis:2002nn}
F.~Gelis, J.~Jalilian-Marian, {From DIS to proton nucleus collisions in the
  color glass condensate model}, Phys. Rev. D 67 (2003) 074019.
\newblock \href {http://arxiv.org/abs/hep-ph/0211363}
  {\path{arXiv:hep-ph/0211363}}, \href
  {http://dx.doi.org/10.1103/PhysRevD.67.074019}
  {\path{doi:10.1103/PhysRevD.67.074019}}.

\bibitem{Marquet:2009ca}
C.~Marquet, B.-W. Xiao, F.~Yuan, {Semi-inclusive Deep Inelastic Scattering at
  small x}, Phys. Lett. B 682 (2009) 207--211.
\newblock \href {http://arxiv.org/abs/0906.1454} {\path{arXiv:0906.1454}},
  \href {http://dx.doi.org/10.1016/j.physletb.2009.10.099}
  {\path{doi:10.1016/j.physletb.2009.10.099}}.

\bibitem{Dominguez:2011wm}
F.~Dominguez, C.~Marquet, B.-W. Xiao, F.~Yuan, {Universality of Unintegrated
  Gluon Distributions at small x}, Phys. Rev. D83 (2011) 105005.
\newblock \href {http://arxiv.org/abs/1101.0715} {\path{arXiv:1101.0715}},
  \href {http://dx.doi.org/10.1103/PhysRevD.83.105005}
  {\path{doi:10.1103/PhysRevD.83.105005}}.

\bibitem{Mantysaari:2019hkq}
H.~M\"antysaari, N.~Mueller, F.~Salazar, B.~Schenke, {Multigluon Correlations
  and Evidence of Saturation from Dijet Measurements at an Electron-Ion
  Collider}, Phys. Rev. Lett. 124~(11) (2020) 112301.
\newblock \href {http://arxiv.org/abs/1912.05586} {\path{arXiv:1912.05586}},
  \href {http://dx.doi.org/10.1103/PhysRevLett.124.112301}
  {\path{doi:10.1103/PhysRevLett.124.112301}}.

\bibitem{Marquet:2007vb}
C.~Marquet, {Forward inclusive dijet production and azimuthal correlations in
  pA collisions}, Nucl. Phys. A796 (2007) 41--60.
\newblock \href {http://arxiv.org/abs/0708.0231} {\path{arXiv:0708.0231}},
  \href {http://dx.doi.org/10.1016/j.nuclphysa.2007.09.001}
  {\path{doi:10.1016/j.nuclphysa.2007.09.001}}.

\bibitem{Albacete:2010pg}
J.~L. Albacete, C.~Marquet, {Azimuthal correlations of forward di-hadrons in
  d+Au collisions at RHIC in the Color Glass Condensate}, Phys. Rev. Lett. 105
  (2010) 162301.
\newblock \href {http://arxiv.org/abs/1005.4065} {\path{arXiv:1005.4065}},
  \href {http://dx.doi.org/10.1103/PhysRevLett.105.162301}
  {\path{doi:10.1103/PhysRevLett.105.162301}}.

\bibitem{Stasto:2011ru}
A.~Stasto, B.-W. Xiao, F.~Yuan, {Back-to-Back Correlations of Di-hadrons in dAu
  Collisions at RHIC}, Phys.Lett. B716 (2012) 430--434.
\newblock \href {http://arxiv.org/abs/1109.1817} {\path{arXiv:1109.1817}},
  \href {http://dx.doi.org/10.1016/j.physletb.2012.08.044}
  {\path{doi:10.1016/j.physletb.2012.08.044}}.

\bibitem{Metz:2011wb}
A.~Metz, J.~Zhou, {Distribution of linearly polarized gluons inside a large
  nucleus}, Phys. Rev. D 84 (2011) 051503.
\newblock \href {http://arxiv.org/abs/1105.1991} {\path{arXiv:1105.1991}},
  \href {http://dx.doi.org/10.1103/PhysRevD.84.051503}
  {\path{doi:10.1103/PhysRevD.84.051503}}.

\bibitem{Lappi:2012nh}
T.~Lappi, H.~M{\"a}ntysaari, {Forward dihadron correlations in deuteron-gold
  collisions with the Gaussian approximation of JIMWLK}, Nucl.Phys. A908 (2013)
  51--72.
\newblock \href {http://arxiv.org/abs/1209.2853} {\path{arXiv:1209.2853}},
  \href {http://dx.doi.org/10.1016/j.nuclphysa.2013.03.017}
  {\path{doi:10.1016/j.nuclphysa.2013.03.017}}.

\bibitem{Iancu:2013dta}
E.~Iancu, J.~Laidet, {Gluon splitting in a shockwave}, Nucl.Phys. A916 (2013)
  48--78.
\newblock \href {http://arxiv.org/abs/1305.5926} {\path{arXiv:1305.5926}},
  \href {http://dx.doi.org/10.1016/j.nuclphysa.2013.07.012}
  {\path{doi:10.1016/j.nuclphysa.2013.07.012}}.

\bibitem{Kotko:2015ura}
P.~Kotko, K.~Kutak, C.~Marquet, E.~Petreska, S.~Sapeta, A.~van Hameren,
  {Improved TMD factorization for forward dijet production in dilute-dense
  hadronic collisions}, JHEP 09 (2015) 106.
\newblock \href {http://arxiv.org/abs/1503.03421} {\path{arXiv:1503.03421}},
  \href {http://dx.doi.org/10.1007/JHEP09(2015)106}
  {\path{doi:10.1007/JHEP09(2015)106}}.

\bibitem{Dumitru:2015gaa}
A.~Dumitru, T.~Lappi, V.~Skokov, {Distribution of Linearly Polarized Gluons and
  Elliptic Azimuthal Anisotropy in Deep Inelastic Scattering Dijet Production
  at High Energy}, Phys. Rev. Lett. 115~(25) (2015) 252301.
\newblock \href {http://arxiv.org/abs/1508.04438} {\path{arXiv:1508.04438}},
  \href {http://dx.doi.org/10.1103/PhysRevLett.115.252301}
  {\path{doi:10.1103/PhysRevLett.115.252301}}.

\bibitem{Altinoluk:2015dpi}
T.~Altinoluk, N.~Armesto, G.~Beuf, A.~H. Rezaeian, {Diffractive Dijet
  Production in Deep Inelastic Scattering and Photon-Hadron Collisions in the
  Color Glass Condensate}, Phys. Lett. B 758 (2016) 373--383.
\newblock \href {http://arxiv.org/abs/1511.07452} {\path{arXiv:1511.07452}},
  \href {http://dx.doi.org/10.1016/j.physletb.2016.05.032}
  {\path{doi:10.1016/j.physletb.2016.05.032}}.

\bibitem{Hatta:2016dxp}
Y.~Hatta, B.-W. Xiao, F.~Yuan, {Probing the Small- x Gluon Tomography in
  Correlated Hard Diffractive Dijet Production in Deep Inelastic Scattering},
  Phys. Rev. Lett. 116~(20) (2016) 202301.
\newblock \href {http://arxiv.org/abs/1601.01585} {\path{arXiv:1601.01585}},
  \href {http://dx.doi.org/10.1103/PhysRevLett.116.202301}
  {\path{doi:10.1103/PhysRevLett.116.202301}}.

\bibitem{Marquet:2016cgx}
C.~Marquet, E.~Petreska, C.~Roiesnel, {Transverse-momentum-dependent gluon
  distributions from JIMWLK evolution}, JHEP 10 (2016) 065.
\newblock \href {http://arxiv.org/abs/1608.02577} {\path{arXiv:1608.02577}},
  \href {http://dx.doi.org/10.1007/JHEP10(2016)065}
  {\path{doi:10.1007/JHEP10(2016)065}}.

\bibitem{vanHameren:2016ftb}
A.~van Hameren, P.~Kotko, K.~Kutak, C.~Marquet, E.~Petreska, S.~Sapeta,
  {Forward di-jet production in p+Pb collisions in the small-x improved TMD
  factorization framework}, JHEP 12 (2016) 034.
\newblock \href {http://arxiv.org/abs/1607.03121} {\path{arXiv:1607.03121}},
  \href {http://dx.doi.org/10.1007/JHEP12(2016)034}
  {\path{doi:10.1007/JHEP12(2016)034}}.

\bibitem{Albacete:2018ruq}
J.~L. Albacete, G.~Giacalone, C.~Marquet, M.~Matas, {Forward dihadron
  back-to-back correlations in $pA$ collisions}, Phys. Rev. D 99~(1) (2019)
  014002.
\newblock \href {http://arxiv.org/abs/1805.05711} {\path{arXiv:1805.05711}},
  \href {http://dx.doi.org/10.1103/PhysRevD.99.014002}
  {\path{doi:10.1103/PhysRevD.99.014002}}.

\bibitem{Dumitru:2018kuw}
A.~Dumitru, V.~Skokov, T.~Ullrich, {Measuring the Weizs\"acker-Williams
  distribution of linearly polarized gluons at an electron-ion collider through
  dijet azimuthal asymmetries}, Phys. Rev. C 99~(1) (2019) 015204.
\newblock \href {http://arxiv.org/abs/1809.02615} {\path{arXiv:1809.02615}},
  \href {http://dx.doi.org/10.1103/PhysRevC.99.015204}
  {\path{doi:10.1103/PhysRevC.99.015204}}.

\bibitem{Salazar:2019ncp}
F.~Salazar, B.~Schenke, {Diffractive dijet production in impact parameter
  dependent saturation models}, Phys. Rev. D 100~(3) (2019) 034007.
\newblock \href {http://arxiv.org/abs/1905.03763} {\path{arXiv:1905.03763}},
  \href {http://dx.doi.org/10.1103/PhysRevD.100.034007}
  {\path{doi:10.1103/PhysRevD.100.034007}}.

\bibitem{Kolbe:2020tlq}
I.~Kolb\'e, K.~Roy, F.~Salazar, B.~Schenke, R.~Venugopalan, {Inclusive prompt
  photon-jet correlations as a probe of gluon saturation in electron-nucleus
  scattering at small $x$}\href {http://arxiv.org/abs/2008.04372}
  {\path{arXiv:2008.04372}}.

\bibitem{Mantysaari:2020lhf}
H.~M\"antysaari, K.~Roy, F.~Salazar, B.~Schenke, {Gluon imaging using azimuthal
  correlations in diffractive scattering at the Electron-Ion Collider}\href
  {http://arxiv.org/abs/2011.02464} {\path{arXiv:2011.02464}}.

\bibitem{Mueller:2013wwa}
A.~Mueller, B.-W. Xiao, F.~Yuan, {Sudakov double logarithms resummation in hard
  processes in the small-x saturation formalism}, Phys. Rev. D 88~(11) (2013)
  114010.
\newblock \href {http://arxiv.org/abs/1308.2993} {\path{arXiv:1308.2993}},
  \href {http://dx.doi.org/10.1103/PhysRevD.88.114010}
  {\path{doi:10.1103/PhysRevD.88.114010}}.

\bibitem{Zheng:2014vka}
L.~Zheng, E.~Aschenauer, J.~Lee, B.-W. Xiao, {Probing Gluon Saturation through
  Dihadron Correlations at an Electron-Ion Collider}, Phys. Rev. D 89~(7)
  (2014) 074037.
\newblock \href {http://arxiv.org/abs/1403.2413} {\path{arXiv:1403.2413}},
  \href {http://dx.doi.org/10.1103/PhysRevD.89.074037}
  {\path{doi:10.1103/PhysRevD.89.074037}}.

\bibitem{Iancu:2003xm}
E.~Iancu, R.~Venugopalan, {The color glass condensate and high energy
  scattering in QCD}\href {http://arxiv.org/abs/hep-ph/0303204}
  {\path{arXiv:hep-ph/0303204}}.

\bibitem{Gelis:2010nm}
F.~Gelis, E.~Iancu, J.~Jalilian-Marian, R.~Venugopalan, {The Color Glass
  Condensate}, Ann.Rev.Nucl.Part.Sci. 60 (2010) 463--489.
\newblock \href {http://arxiv.org/abs/1002.0333} {\path{arXiv:1002.0333}},
  \href {http://dx.doi.org/10.1146/annurev.nucl.010909.083629}
  {\path{doi:10.1146/annurev.nucl.010909.083629}}.

\bibitem{Kovchegov:2012mbw}
Y.~V. Kovchegov, E.~Levin, {Quantum chromodynamics at high energy}, {Cambridge
  University Press}, 2012.

\bibitem{Balitsky:1995ub}
I.~Balitsky, {Operator expansion for high-energy scattering}, Nucl. Phys. B463
  (1996) 99--160.
\newblock \href {http://arxiv.org/abs/hep-ph/9509348}
  {\path{arXiv:hep-ph/9509348}}, \href
  {http://dx.doi.org/10.1016/0550-3213(95)00638-9}
  {\path{doi:10.1016/0550-3213(95)00638-9}}.

\bibitem{JalilianMarian:1997jx}
J.~Jalilian-Marian, A.~Kovner, A.~Leonidov, H.~Weigert, {The BFKL equation from
  the Wilson renormalization group}, Nucl. Phys. B504 (1997) 415--431.
\newblock \href {http://arxiv.org/abs/hep-ph/9701284}
  {\path{arXiv:hep-ph/9701284}}, \href
  {http://dx.doi.org/10.1016/S0550-3213(97)00440-9}
  {\path{doi:10.1016/S0550-3213(97)00440-9}}.

\bibitem{JalilianMarian:1997gr}
J.~Jalilian-Marian, A.~Kovner, A.~Leonidov, H.~Weigert, {The Wilson
  renormalization group for low x physics: Towards the high density regime},
  Phys.Rev. D59 (1998) 014014.
\newblock \href {http://arxiv.org/abs/hep-ph/9706377}
  {\path{arXiv:hep-ph/9706377}}, \href
  {http://dx.doi.org/10.1103/PhysRevD.59.014014}
  {\path{doi:10.1103/PhysRevD.59.014014}}.

\bibitem{Kovner:2000pt}
A.~Kovner, J.~G. Milhano, H.~Weigert, {Relating different approaches to
  nonlinear QCD evolution at finite gluon density}, Phys. Rev. D62 (2000)
  114005.
\newblock \href {http://arxiv.org/abs/hep-ph/0004014}
  {\path{arXiv:hep-ph/0004014}}, \href
  {http://dx.doi.org/10.1103/PhysRevD.62.114005}
  {\path{doi:10.1103/PhysRevD.62.114005}}.

\bibitem{Iancu:2000hn}
E.~Iancu, A.~Leonidov, L.~D. McLerran, {Nonlinear gluon evolution in the color
  glass condensate. I}, Nucl. Phys. A692 (2001) 583--645.
\newblock \href {http://arxiv.org/abs/hep-ph/0011241}
  {\path{arXiv:hep-ph/0011241}}, \href
  {http://dx.doi.org/10.1016/S0375-9474(01)00642-X}
  {\path{doi:10.1016/S0375-9474(01)00642-X}}.

\bibitem{Iancu:2001ad}
E.~Iancu, A.~Leonidov, L.~D. McLerran, {The renormalization group equation for
  the color glass condensate}, Phys. Lett. B510 (2001) 133--144.
\newblock \href {http://arxiv.org/abs/hep-ph/0102009}
  {\path{arXiv:hep-ph/0102009}}, \href
  {http://dx.doi.org/10.1016/S0370-2693(01)00524-X}
  {\path{doi:10.1016/S0370-2693(01)00524-X}}.

\bibitem{Ferreiro:2001qy}
E.~Ferreiro, E.~Iancu, A.~Leonidov, L.~McLerran, {Nonlinear gluon evolution in
  the color glass condensate. II}, Nucl. Phys. A703 (2002) 489--538.
\newblock \href {http://arxiv.org/abs/hep-ph/0109115}
  {\path{arXiv:hep-ph/0109115}}, \href
  {http://dx.doi.org/10.1016/S0375-9474(01)01329-X}
  {\path{doi:10.1016/S0375-9474(01)01329-X}}.

\bibitem{Kovchegov:1999yj}
Y.~V. Kovchegov, {Small-x F2 structure function of a nucleus including multiple
  pomeron exchanges}, Phys. Rev. D60 (1999) 034008.
\newblock \href {http://arxiv.org/abs/hep-ph/9901281}
  {\path{arXiv:hep-ph/9901281}}, \href
  {http://dx.doi.org/10.1103/PhysRevD.60.034008}
  {\path{doi:10.1103/PhysRevD.60.034008}}.

\bibitem{Mueller:1999wm}
A.~H. Mueller, {Parton saturation at small x and in large nuclei}, Nucl. Phys.
  B 558 (1999) 285--303.
\newblock \href {http://arxiv.org/abs/hep-ph/9904404}
  {\path{arXiv:hep-ph/9904404}}, \href
  {http://dx.doi.org/10.1016/S0550-3213(99)00394-6}
  {\path{doi:10.1016/S0550-3213(99)00394-6}}.

\bibitem{McLerran:1993ni}
L.~D. McLerran, R.~Venugopalan, {Computing quark and gluon distribution
  functions for very large nuclei}, Phys. Rev. D49 (1994) 2233--2241.
\newblock \href {http://arxiv.org/abs/hep-ph/9309289}
  {\path{arXiv:hep-ph/9309289}}, \href
  {http://dx.doi.org/10.1103/PhysRevD.49.2233}
  {\path{doi:10.1103/PhysRevD.49.2233}}.

\bibitem{McLerran:1994vd}
L.~D. McLerran, R.~Venugopalan, {Green's functions in the color field of a
  large nucleus}, Phys. Rev. D50 (1994) 2225--2233.
\newblock \href {http://arxiv.org/abs/hep-ph/9402335}
  {\path{arXiv:hep-ph/9402335}}, \href
  {http://dx.doi.org/10.1103/PhysRevD.50.2225}
  {\path{doi:10.1103/PhysRevD.50.2225}}.

\bibitem{Arsene:2004ux}
I.~Arsene, et~al., {On the evolution of the nuclear modification factors with
  rapidity and centrality in d + Au collisions at s(NN)**(1/2) = 200-GeV},
  Phys.Rev.Lett. 93 (2004) 242303.
\newblock \href {http://arxiv.org/abs/nucl-ex/0403005}
  {\path{arXiv:nucl-ex/0403005}}, \href
  {http://dx.doi.org/10.1103/PhysRevLett.93.242303}
  {\path{doi:10.1103/PhysRevLett.93.242303}}.

\bibitem{Adams:2006uz}
J.~Adams, et~al., {Forward neutral pion production in p+p and d+Au collisions
  at s(NN)**(1/2) = 200-GeV}, Phys.Rev.Lett. 97 (2006) 152302.
\newblock \href {http://arxiv.org/abs/nucl-ex/0602011}
  {\path{arXiv:nucl-ex/0602011}}, \href
  {http://dx.doi.org/10.1103/PhysRevLett.97.152302}
  {\path{doi:10.1103/PhysRevLett.97.152302}}.

\bibitem{Kharzeev:2002pc}
D.~Kharzeev, E.~Levin, L.~McLerran, {Parton saturation and N(part) scaling of
  semihard processes in QCD}, Phys. Lett. B 561 (2003) 93--101.
\newblock \href {http://arxiv.org/abs/hep-ph/0210332}
  {\path{arXiv:hep-ph/0210332}}, \href
  {http://dx.doi.org/10.1016/S0370-2693(03)00420-9}
  {\path{doi:10.1016/S0370-2693(03)00420-9}}.

\bibitem{Baier:2003hr}
R.~Baier, A.~Kovner, U.~A. Wiedemann, {Saturation and parton level Cronin
  effect: Enhancement versus suppression of gluon production in p-A and A-A
  collisions}, Phys. Rev. D 68 (2003) 054009.
\newblock \href {http://arxiv.org/abs/hep-ph/0305265}
  {\path{arXiv:hep-ph/0305265}}, \href
  {http://dx.doi.org/10.1103/PhysRevD.68.054009}
  {\path{doi:10.1103/PhysRevD.68.054009}}.

\bibitem{Albacete:2003iq}
J.~L. Albacete, N.~Armesto, A.~Kovner, C.~A. Salgado, U.~A. Wiedemann, {Energy
  dependence of the Cronin effect from nonlinear QCD evolution}, Phys.Rev.Lett.
  92 (2004) 082001.
\newblock \href {http://arxiv.org/abs/hep-ph/0307179}
  {\path{arXiv:hep-ph/0307179}}, \href
  {http://dx.doi.org/10.1103/PhysRevLett.92.082001}
  {\path{doi:10.1103/PhysRevLett.92.082001}}.

\bibitem{Kharzeev:2003wz}
D.~Kharzeev, Y.~V. Kovchegov, K.~Tuchin, {Cronin effect and high p(T)
  suppression in pA collisions}, Phys.Rev. D68 (2003) 094013.
\newblock \href {http://arxiv.org/abs/hep-ph/0307037}
  {\path{arXiv:hep-ph/0307037}}, \href
  {http://dx.doi.org/10.1103/PhysRevD.68.094013}
  {\path{doi:10.1103/PhysRevD.68.094013}}.

\bibitem{Iancu:2004bx}
E.~Iancu, K.~Itakura, D.~Triantafyllopoulos, {Cronin effect and high
  p-perpendicular suppression in the nuclear gluon distribution at small x},
  Nucl.Phys. A742 (2004) 182--252.
\newblock \href {http://arxiv.org/abs/hep-ph/0403103}
  {\path{arXiv:hep-ph/0403103}}, \href
  {http://dx.doi.org/10.1016/j.nuclphysa.2004.06.033}
  {\path{doi:10.1016/j.nuclphysa.2004.06.033}}.

\bibitem{Blaizot:2004wu}
J.~P. Blaizot, F.~Gelis, R.~Venugopalan, {High-energy pA collisions in the
  color glass condensate approach. 1. Gluon production and the Cronin effect},
  Nucl.Phys. A743 (2004) 13--56.
\newblock \href {http://arxiv.org/abs/hep-ph/0402256}
  {\path{arXiv:hep-ph/0402256}}, \href
  {http://dx.doi.org/10.1016/j.nuclphysa.2004.07.005}
  {\path{doi:10.1016/j.nuclphysa.2004.07.005}}.

\bibitem{Ducloue:2019ezk}
B.~Duclou{\'e}, E.~Iancu, A.~H. Mueller, G.~Soyez, D.~N. Triantafyllopoulos,
  {Non-linear evolution in QCD at high-energy beyond leading order}, JHEP 04
  (2019) 081.
\newblock \href {http://arxiv.org/abs/1902.06637} {\path{arXiv:1902.06637}},
  \href {http://dx.doi.org/10.1007/JHEP04(2019)081}
  {\path{doi:10.1007/JHEP04(2019)081}}.

\bibitem{Ducloue:2019jmy}
B.~Duclou\'e, E.~Iancu, G.~Soyez, D.~Triantafyllopoulos, {HERA data and
  collinearly-improved BK dynamics}, Phys. Lett. B 803 (2020) 135305.
\newblock \href {http://arxiv.org/abs/1912.09196} {\path{arXiv:1912.09196}},
  \href {http://dx.doi.org/10.1016/j.physletb.2020.135305}
  {\path{doi:10.1016/j.physletb.2020.135305}}.

\bibitem{Beuf:2014uia}
G.~Beuf, {Improving the kinematics for low-x QCD evolution equations in
  coordinate space}, Phys.Rev. D89 (2014) 074039.
\newblock \href {http://arxiv.org/abs/1401.0313} {\path{arXiv:1401.0313}},
  \href {http://dx.doi.org/10.1103/PhysRevD.89.074039}
  {\path{doi:10.1103/PhysRevD.89.074039}}.

\bibitem{Iancu:2015vea}
E.~Iancu, J.~Madrigal, A.~Mueller, G.~Soyez, D.~Triantafyllopoulos, {Resumming
  double logarithms in the QCD evolution of color dipoles}, Phys.Lett. B744
  (2015) 293--302.
\newblock \href {http://arxiv.org/abs/1502.05642} {\path{arXiv:1502.05642}},
  \href {http://dx.doi.org/10.1016/j.physletb.2015.03.068}
  {\path{doi:10.1016/j.physletb.2015.03.068}}.

\bibitem{Iancu:2015joa}
E.~Iancu, J.~D. Madrigal, A.~H. Mueller, G.~Soyez, D.~N. Triantafyllopoulos,
  {Collinearly-improved BK evolution meets the HERA data}, Phys. Lett. B750
  (2015) 643--652.
\newblock \href {http://arxiv.org/abs/1507.03651} {\path{arXiv:1507.03651}},
  \href {http://dx.doi.org/10.1016/j.physletb.2015.09.071}
  {\path{doi:10.1016/j.physletb.2015.09.071}}.

\bibitem{Lappi:2015fma}
T.~Lappi, H.~M{\"a}ntysaari, {Direct numerical solution of the coordinate space
  Balitsky-Kovchegov equation at next to leading order}, Phys.Rev. D91~(7)
  (2015) 074016.
\newblock \href {http://arxiv.org/abs/1502.02400} {\path{arXiv:1502.02400}},
  \href {http://dx.doi.org/10.1103/PhysRevD.91.074016}
  {\path{doi:10.1103/PhysRevD.91.074016}}.

\bibitem{Lappi:2016fmu}
T.~Lappi, H.~M{\"a}ntysaari, {Next-to-leading order Balitsky-Kovchegov equation
  with resummation}, Phys. Rev. D93~(9) (2016) 094004.
\newblock \href {http://arxiv.org/abs/1601.06598} {\path{arXiv:1601.06598}},
  \href {http://dx.doi.org/10.1103/PhysRevD.93.094004}
  {\path{doi:10.1103/PhysRevD.93.094004}}.

\bibitem{Albacete:2015xza}
J.~L. Albacete, {Resummation of double collinear logs in BK evolution versus
  HERA data}, Nucl. Phys. A957 (2017) 71--84.
\newblock \href {http://arxiv.org/abs/1507.07120} {\path{arXiv:1507.07120}},
  \href {http://dx.doi.org/10.1016/j.nuclphysa.2016.07.008}
  {\path{doi:10.1016/j.nuclphysa.2016.07.008}}.

\bibitem{Beuf:2020dxl}
G.~Beuf, H.~H\"anninen, T.~Lappi, H.~M\"antysaari, {Color Glass Condensate at
  next-to-leading order meets HERA data}, Phys. Rev. D 102 (2020) 074028.
\newblock \href {http://arxiv.org/abs/2007.01645} {\path{arXiv:2007.01645}},
  \href {http://dx.doi.org/10.1103/PhysRevD.102.074028}
  {\path{doi:10.1103/PhysRevD.102.074028}}.

\bibitem{Stasto:2000er}
A.~M. Stasto, K.~J. Golec-Biernat, J.~Kwiecinski, {Geometric scaling for the
  total gamma* p cross-section in the low x region}, Phys. Rev. Lett. 86 (2001)
  596--599.
\newblock \href {http://arxiv.org/abs/hep-ph/0007192}
  {\path{arXiv:hep-ph/0007192}}, \href
  {http://dx.doi.org/10.1103/PhysRevLett.86.596}
  {\path{doi:10.1103/PhysRevLett.86.596}}.

\bibitem{Iancu:2002tr}
E.~Iancu, K.~Itakura, L.~McLerran, {Geometric scaling above the saturation
  scale}, Nucl. Phys. A708 (2002) 327--352.
\newblock \href {http://arxiv.org/abs/hep-ph/0203137}
  {\path{arXiv:hep-ph/0203137}}, \href
  {http://dx.doi.org/10.1016/S0375-9474(02)01010-2}
  {\path{doi:10.1016/S0375-9474(02)01010-2}}.

\bibitem{Mueller:2002zm}
A.~Mueller, D.~Triantafyllopoulos, {The Energy dependence of the saturation
  momentum}, Nucl.Phys. B640 (2002) 331--350.
\newblock \href {http://arxiv.org/abs/hep-ph/0205167}
  {\path{arXiv:hep-ph/0205167}}, \href
  {http://dx.doi.org/10.1016/S0550-3213(02)00581-3}
  {\path{doi:10.1016/S0550-3213(02)00581-3}}.

\bibitem{Mueller:2001fv}
A.~H. Mueller, {Parton saturation: An Overview} (2001) 45--72\href
  {http://arxiv.org/abs/hep-ph/0111244} {\path{arXiv:hep-ph/0111244}}.

\bibitem{Iancu:2002xk}
E.~Iancu, A.~Leonidov, L.~McLerran, {The Color glass condensate: An
  Introduction} (2002) 73--145\href {http://arxiv.org/abs/hep-ph/0202270}
  {\path{arXiv:hep-ph/0202270}}.

\bibitem{Kovchegov:2003dm}
Y.~V. Kovchegov, L.~Szymanowski, S.~Wallon, {Perturbative odderon in the dipole
  model}, Phys. Lett. B586 (2004) 267--281.
\newblock \href {http://arxiv.org/abs/hep-ph/0309281}
  {\path{arXiv:hep-ph/0309281}}, \href
  {http://dx.doi.org/10.1016/j.physletb.2004.02.036}
  {\path{doi:10.1016/j.physletb.2004.02.036}}.

\bibitem{Hatta:2005as}
Y.~Hatta, E.~Iancu, K.~Itakura, L.~McLerran, {Odderon in the color glass
  condensate}, Nucl. Phys. A760 (2005) 172--207.
\newblock \href {http://arxiv.org/abs/hep-ph/0501171}
  {\path{arXiv:hep-ph/0501171}}, \href
  {http://dx.doi.org/10.1016/j.nuclphysa.2005.05.163}
  {\path{doi:10.1016/j.nuclphysa.2005.05.163}}.

\end{thebibliography}

\end{document}